\documentclass[journal]{IEEEtran}
\usepackage{colortbl}
\usepackage{ifpdf}
\usepackage{cite}
\usepackage{amsmath,amssymb,amsfonts}
\usepackage{algorithmic}
\usepackage{graphicx, adjustbox}
\usepackage{array}
\usepackage{enumitem}
\usepackage{breqn}
\usepackage{ dsfont }
\usepackage{ upgreek }
\usepackage{textcomp}
\usepackage{multirow}
\usepackage{algorithm}
\usepackage{algorithmic}
\usepackage{caption}
\usepackage{pgfplots}
\usepackage{tkz-euclide,subfigure}
\usepackage{tikz}

\begin{document}
	
	\title{Blockchain for 5G and Beyond Networks: \\  A State of the Art Survey}
	
	\author{Dinh C. Nguyen,~\IEEEmembership{Student Member,~IEEE,}
		Pubudu N. Pathirana,~\IEEEmembership{Senior Member,~IEEE,}
		Ming Ding,~\IEEEmembership{Senior Member,~IEEE,}
		Aruna Seneviratne,~\IEEEmembership{Senior Member,~IEEE}
		
		\thanks {*This work was supported in part by the CSIRO Data61, Australia.}
		\thanks{Dinh C. Nguyen is with School of Engineering, Deakin University, Waurn Ponds, VIC 3216, Australia, and also with the Data61, CSIRO, Docklands, Melbourne, Australia  (e-mail: cdnguyen@deakin.edu.au).}% <-this % stops a space
		\thanks{Pubudu N. Pathirana is with School of Engineering, Deakin University, Waurn Ponds, VIC 3216, Australia (email: pubudu.pathirana@deakin.edu.au).}
		\thanks{Ming Ding is with Data61, CSIRO, Australia (email: ming.ding@data61.csiro.au).}% <-this % stops a space
		\thanks{Aruna Seneviratne is with School of Electrical Engineering and Telecommunications, University of New South Wales (UNSW), NSW, Australia (email: a.seneviratne@unsw.edu.au).}
	}
	
	\markboth{IEEE COMMUNICATIONS SURVEYS \& TUTORIALS}%
	{Shell \MakeLowercase{\textit{et al.}}: Bare Demo of IEEEtran.cls for IEEE Journals}

	\maketitle
	
	\begin{abstract}
	The fifth generation (5G) wireless networks are on the way to be deployed around the world. The 5G technologies target to support diverse vertical applications by connecting heterogeneous devices and machines with drastic improvements in terms of high quality of service, increased network capacity and enhanced system throughput. Despite all these advantages that 5G will bring about, there are still major challenges to be addressed, including decentralization, transparency, risks of data interoperability, network privacy and security vulnerabilities. Blockchain, an emerging disruptive technology, can offer innovative solutions to effectively solve the challenges in 5G networks. Driven by the dramatically increased capacities of the 5G networks and the recent breakthroughs in the blockchain technology, blockchain-based 5G services are expected to witness a rapid development and bring substantial benefits to future society. In this paper, we provide a state-of-art survey on the integration of blockchain with 5G networks and beyond. In this detailed survey, our primary focus is on the extensive discussions on the potential of blockchain for enabling key 5G technologies, including cloud computing, edge computing, Software Defined Networks, Network Function Virtualization, Network Slicing, and D2D communications. We then explore and analyse the opportunities that blockchain potentially empowers important 5G services, ranging from spectrum management, data sharing, network virtualization, resource management to interference management, federated learning, privacy and security provision. The recent advances in the applications of blockchain in 5G Internet of Things are also surveyed in a wide range of popular use-case domains, such as smart healthcare, smart city, smart transportation, smart grid and UAVs. The main findings derived from the comprehensive survey on the cooperated blockchain-5G networks and services are then summarized, and possible research challenges with open issues are also identified. Lastly, we complete this survey by shedding new light on future directions of research on this newly emerging area.
	\end{abstract}
	
	
	\IEEEpeerreviewmaketitle

	\section{Introduction}
\begin{figure*}
	\centering
	\includegraphics[height=8.2cm, width=18cm]{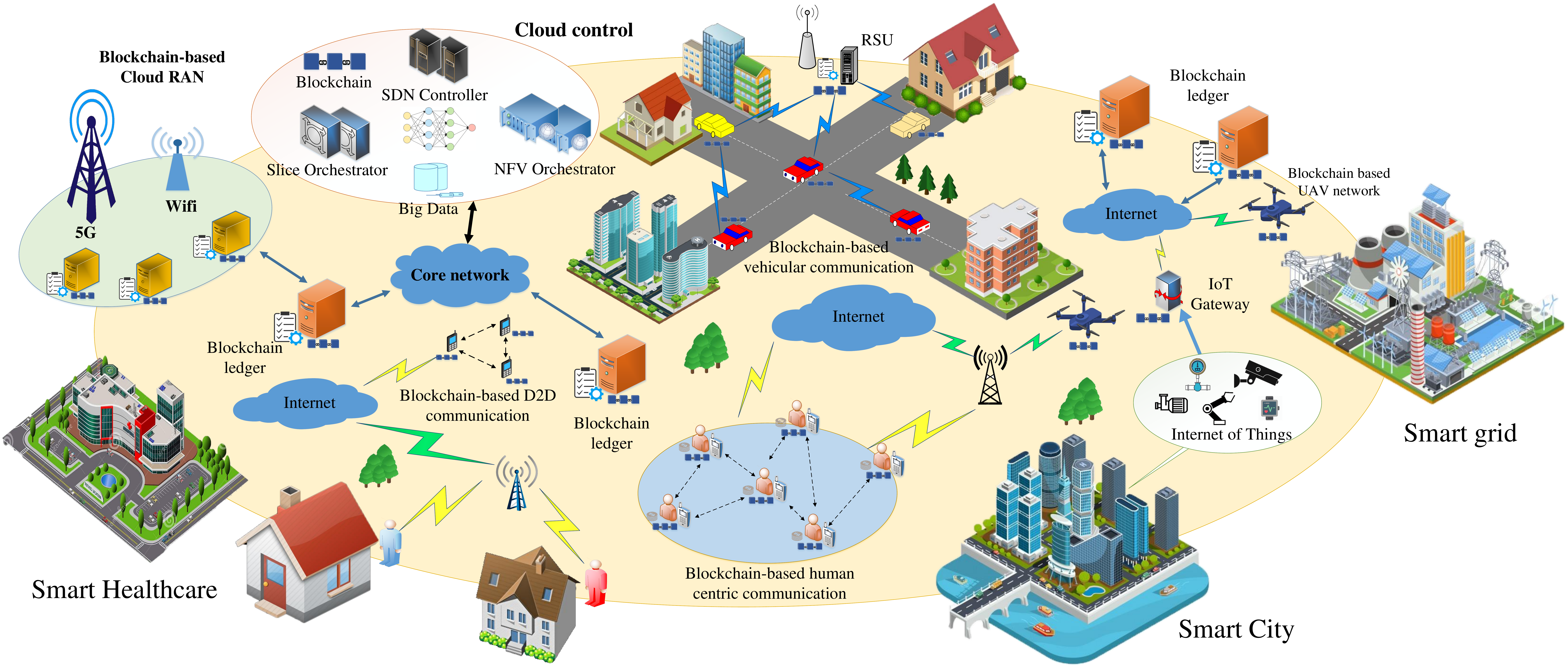}
	\caption{The convergence of blockchain and 5G.  }
	\vspace{-0.17in}
\end{figure*}
The fifth generation 5G technology, referred to as beyond 2020 communications systems, represents the next important phase of the global telecommunication evolution, with recent successful deployments in several areas across almost all the continents\footnote{{https://www.speedtest.net/ookla-5g-map}}. The 5G networks are characterized by three major features with its ability to support Enhanced Mobile Broadband, Massive Machine Type Communication and the provisioning of Ultra-reliable Low Latency Communication services \cite{1}. Driven by the explosion of smart mobile devices and the rapid advances of communication technologies, 5G could be a technical enabler for a plethora of new innovative business opportunities and industrial applications, and facilitates the seamless collaboration across domains by interconnecting billions of devices. The 5G mobile networks promise to revolutionize global industries and provide immediate impacts on customers and business stakeholders. The main vision of future 5G services is to provide a customized and advanced user-centric value, enabling connection of nearly all aspects of the human life to communication networks to meet the ever growing demands of user traffic and emerging services \cite{2}. To achieve these objectives, several underlying wireless technologies have been proposed to enable future 5G networks, including cloud computing, edge computing, Software Defined Networking (SDN), Network Function Virtualization (NFV), Network Slicing, and D2D communication \cite{3}. However, the rapid surge and breakneck expansion of 5G wireless services in terms of scale, speed, and capacity also pose new security challenges such as network reliability, data immutability, privacy \cite{4} that must be considered and solved before wide deployments. 

Many security solutions have been used in the previous generations of communication networks (i.e., 2G, 3G and 4G) [48]. For example, in the physical layer of 2G-4G networks, Hybrid Automatic Repeat reQuest (HARQ) techniques, combining Forward Error Correction (FEC) channel codes and Automatic Repeat reQuest (ARQ) have been used widely, which can detect and rectify wrong data bits in supporting data authentication. Moreover, for detecting errors in data communications, data storage, and data compression, error-detection techniques such as cyclic redundancy check (CRC) have been leveraged in the radio link control (RLC) layer for data reliability guarantees. However, these security techniques and architectures used in the previous generations (2G-4G), apparently, will not suffice for 5G due to the following reasons.
	
\begin{itemize}
	\item A critical reason is that the above security techniques used in 2G-4G are powerless to deal with the problem of data tampering, such as deletion, injection, alternation in 5G networks.
	\item Another reason is the dynamics of new technologies and services in 5G networks, which pose new requirements on security and privacy beyond protecting data integrity.
\end{itemize}
In particular, the emerging 5G technologies such as SDN, NFV, network slicing and D2D communications in 5G will support new service delivery models and thus further exacerbate the security challenges. Unlike the legacy cellular networks, 5G wireless networks are going to be decentralized and ubiquitous service-oriented which have a special emphasis on security and privacy requirements from the perspective of services. In particular, the security management in 5G is more complex due to various types of and a massive number of devices connected. How to provide an open data architecture for flexible spectrum sharing, data sharing, multiuser access, for example, to achieve ubiquitous 5G service provisions while ensuring high data immutability and transparency is a critical issue. Succinctly, the security architectures of the previous generations lack the sophistication needed to secure 5G networks.

In the 5G/6G era, immutability, decentralization and transparency are crucial security factors that ensure the successful roll-out of new services such as IoT data collection, driverless cars, Unmanned Aerial Vehicles (UAVs), Federated Learning (FL). Among the existing technologies, blockchain is the most promising one to meet these new security requirements and reshape the 5G communication landscape \cite{5}, \cite{6}. Hence, 5G needs blockhain for its wide 5G service deployments.  
From the technical perspective, blockchain is a distributed ledger technology that was firstly used to serve as the public digital ledger of cryptocurrency Bitcoin \cite{7} for economic transactions. The blockchain is basically a decentralized, immutable and transparent database. The concept of blockchain is based on a peer-to-peer network architecture in which transaction information is managed flexibly by all network participants and not controlled by any single centralized authority. In particular, the blockchain technology boasts a few desirable characteristics of decentralization, immutability, accountability, and truly trustless database storage which significantly improve network security and save operational costs \cite{8}. The rapid development and the adoption of blockchain as a disruptive technology are paving the way for the next generation of financial and industrial services.  Currently, blockchain technology has been investigated and applied in various applications, such as Internet of Things (IoT) \cite{9}, \cite{10}, edge computing \cite{11}, smart city \cite{12}, vehicular networks \cite{13}, and industries \cite{14}.

For the inherent superior properties, blockchain has the potential to be integrated with the 5G ecosystems to empower mobile networks and services as shown in Fig. 1. Due to the advanced technical capabilities to support future network services, blockchain was regarded as one of the key technical drivers for 6G at the 2018 Mobile World Congress (MWC) \cite{15}. It is also predicted that blockchains would be a key technology in reaping real benefits from 5G networks, for giving birth to novel applications from autonomous resource sharing, ubiquitous computing to reliable content-based storage and intelligent data management \cite{16}.

The combination of blockchain and 5G is also expected to pave the way for emerging mobile services \cite{17}. In fact, 5G is all about connecting heterogeneous devices and complex networks interconnecting more than 500 billion mobile devices by 2030 \cite{18}. Besides, the emerging Internet of Things (IoT), and Massive Machine Communications (MMC) are predicted to create over 80 billion connections by 2020 \cite{19}. In such a context, the ultra-dense small cell networks, a fundamental component of 5G infrastructure, will provide connections and energy efficiencies of radio links with high data rates and low latencies. However, it introduces trust and secure interoperability concerns among complex sub-networks. Therefore, providing a reliable cooperation among heterogeneous devices is vitally important for 5G mobile networks. In this regard, blockchain with its immutable and decentralized transaction ledgers can enable distributed massive communication with high security and trustworthiness \cite{20}. Moreover, network slicing associated with other emerging technologies such as cloud/ edge computing, SDN, NFV, and D2D communication are also key enablers for future 5G networks and services. A big challenge for current 5G platforms is the need to guarantee an open, transparent, and secure system among the extraordinary number of resources and mobile users. Blockchain with its innovative concepts of decentralized operation can provide a high level of data privacy, security, transparency, immutability for storage of 5G heterogeneous data \cite{21}, \cite{22}. Blockchain is thus expected to be an indispensable tool to fulfill the performance expectations for 5G systems with minimal costs and management overheads. 

\textit{\textbf{Related survey works and Contributions:}}
\begin{figure*}
	\centering
	\includegraphics [height=15.5cm, width=14cm]{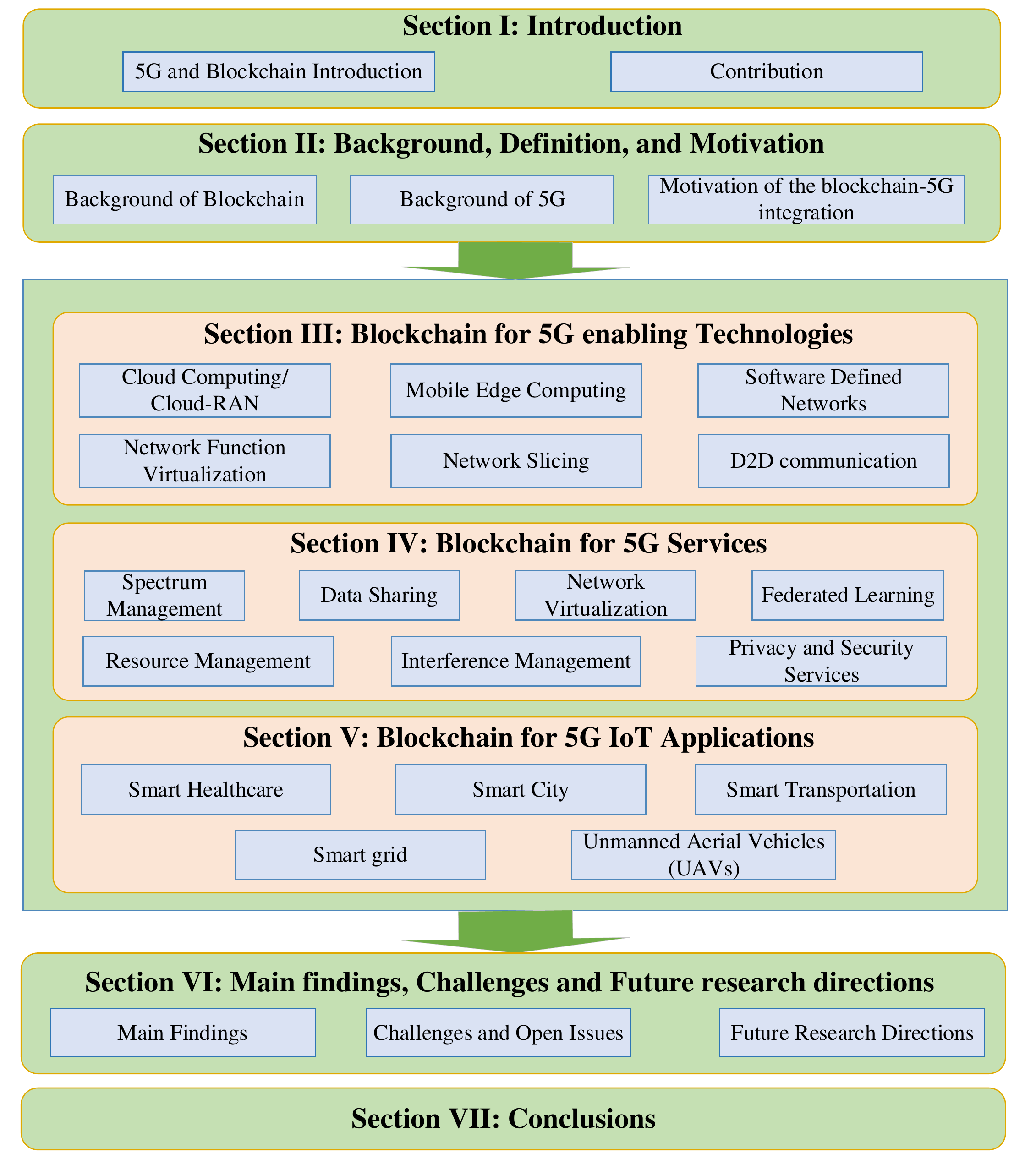}
	\caption{The structure of the paper.   }
	\vspace{-0.17in}
\end{figure*}
Blockchains have gained momentum in the academia, with a number of surveys published in \cite{9}, \cite{10}, \cite{11}, \cite{12}, \cite{13}, \cite{14}, which have discussed many aspects such as architecture, concepts, technologies and application domains. The 5G systems have also attracted attention \cite{1}, \cite{2}, \cite{3}, \cite{4}. Despite growing interest in blockchain and 5G, the focus of existing survey works is on each of the specific technologies. There have been no surveys that emphasize the integration of blockchain and 5G. The authors in \cite{23} only provided a brief introduction of the blockchain adoption in secure 5G resource management and reliable network orchestration. The survey in \cite{24} provided a short survey on the potential of blockchain for 5G networks in Industry 4.0. Similarly, the studies in \cite{25}, \cite{26} presented a brief review on the benefits of blockchain for 5G-based industrial IoTs. 
 
Thus, to our best knowledge, there is no comprehensive survey on the integrated use of blockchain and 5G technologies and services.  In this paper, we provide an extensive survey on the integration of blockchain and 5G technologies for providing services, including cloud computing, edge computing, Software Defined Networks, Network Function Virtualization, Network Slicing, and D2D communication. We also detail the use of blockchain for supporting important 5G services, ranging from spectrum management, data sharing, network virtualization, resource management to mitigating interference, federated learning, privacy and security attacks. The potential of blockchain in 5G IoT networks is also discussed through a number of use-case domains, such as smart healthcare, smart city, smart transportation, smart grid and UAVs. Besides, we highlight the research challenges and open issues, and point out the promising future research directions related to the blockchain-5G integrations. The main contributions of this survey article can be summarized as follows:
\begin{enumerate}
	\item We conduct a state-of-art survey on the convergence of blockchain and 5G, starting with an analysis on the background, definitions as well as highlighting the motivations of the integration of these two emerging technologies. 
	\item We provide a review on the adoption of blockchain for enabling key 5G technologies, with a particular focus on cloud computing, edge computing, Software Defined Networks, Network Function Virtualization, Network Slicing, and D2D communication.
	\item	We present an in-depth discussion on opportunities that blockchain brings to 5G services, including spectrum management, data sharing, network virtualization, resource management, interference management, federated learning, privacy and security services.
	\item We investigate the potential of leveraging blockchains in 5G IoT networks and review the latest developments of the integrated blockchain-5G IoT applications in a number of domains, ranging from smart healthcare, smart city, smart transportation to smart grid and UAVs. 
	\item Based on the comprehensive survey, we summarize the main findings, highlight research challenges and open issues, and point out several future research directions. 
\end{enumerate}

\textit{\textbf{Structure of this survey:}} The structure of this survey is shown as Fig. 2. Section II presents an overview of blockchain and 5G networks, and then highlight the motivations for the integration of blockchains in 5G networks and services. In Section III, we present a state-of-art survey on the convergence of blockchain and key 5G technologies, namely cloud computing, edge computing, Software Defined Networks, Network Function Virtualization, Network Slicing, and D2D communication. We also provide a comprehensive discussion on the use of blockchain for supporting fundamental 5G requirements, ranging from spectrum management, data sharing, network virtualization, resource management to interference management, federated learning privacy and security services in Section IV. The benefits of blockchain for 5G IoT applications are analysed in details in Section V, with a focus on popular applications such as smart healthcare, smart city, smart transportation, smart grid and UAVs. We summarize the key main findings in Section VI, and the potential research challenges and future research directions are also outlined. Finally, Section VII concludes the paper. A list of acronyms used throughout the paper is presented in TABLE I.

\begin{table}
	\caption{List of key acronyms.}
	\label{table}
	\scriptsize
	\centering
	\captionsetup{font=scriptsize}
	\setlength{\tabcolsep}{5pt}
	\begin{tabular}{p{2.5cm}|p{5cm}}
		\hline
		\textbf{Acronyms}& 
		\textbf{Definitions}
		\\
		\hline
		3GPP& Third Generation Partnership Project
		\\
		MWC& Mobile World Congress
		\\
		NGMN&Next Generation Mobile Networks
		\\
		ETSI&European Telecommunications Standards Institute
		\\
		MNO&Mobile Network Operator
		\\
		MVNO &Mobile Virtual Network Operator
		\\
		ML & Machine learning
		\\
		UAVs & Unmanned Aerial Vehicles
		\\
		SDN& Software-Defined Networking
		\\
		SDI& Software-Defined Infrastructure
		\\
		NFV&Network Functions Virtualisation
		\\
		VNFs& Virtual Network Functions 
		\\
		D2D&Device-to-Device
		\\
		VM& Virtual	Machine 
		\\
		Cloud-RANs& Cloud Radio Access Networks
		\\
		BBU&Baseband Unit
		\\
		IoT&Internet of Thing
		\\
		MEC& Mobile Edge Computing
		\\
		ESPs &Edge Service Providers
		\\
		VANETs & Vehicular ad-hoc Networks
		\\
		MANO& Management and Network Orchestration
		\\
		SFC& Service Function Chaining
		\\
		VMOA& Virtual Machine Orchestration Authentication
		\\
		V2V & Vehicle-to-Vehicle
		\\
		RSU & Roadside Units
		\\
		CCN& Content Centric Networking
		\\
		SLA& Service-Level Agreement 
		\\
		IPFS & Inter-Planetary File System
		\\
		DoS & Denial-of-Service
		\\
		QoS& Quality of Services
		\\
		QoE&Quality of Experience
		\\
		CSI & Channel State Information
		\\
		FUEs & Femtocell Users
		\\
		PoW&Proof of Work
		\\
		PBFT & Practical Byzantine Fault Tolerance 
		\\
		EHRs & Electronic Health Records
		\\
		MaaS & Mobility-as-a-Service
		\\
		TPAs & Third Party Auditors 
		\\
		ITS & Intelligent Transportation System
		\\
		V2G & Vehicle-to-Grid
		\\
		EVs & Electric Vehicles
		\\
		\hline
	\end{tabular}
	\label{tab1}
	\vspace{-0.2in}
\end{table}
\section{Blockchain and 5G: background, definition and motivation}	
\subsection{	Blockchain }
Blockchain is mostly known as the technology underlying the cryptocurrency Bitcoin \cite{7}. The core idea of a blockchain is decentralization. This means that blockchain does not store any of its database in a central location. Instead, the blockchain is copied and spread across a network of participants (i.e. computers). Whenever a new block is added to the blockchain, every computer on the network updates its blockchain to reflect the change. This decentralized architecture ensures robust and secure operations on blockchain with the advantages of tamper resistance and no single-point failure vulnerabilities. In particular, blockchain can be accessible for everyone and is not controlled by any network entity. This is enabled by a mechanism called consensus which is a set of rules to ensure the agreement among all participants on the status of the blockchain ledger. The general concept on how blockchain operates is shown in Fig. 3. 

In general, blockchains can be classified as either a public (permission-less) or a private (permissioned) blockchain \cite{29}. A public blockchain is accessible for everyone and anyone can join and make transactions as well as participate in the consensus process. The best-known public blockchain applications include Bitcoin and Ethereum. Private blockchains on the other hand are an invitation-only network managed by a central entity. A participant has to be permissioned using a validation mechanism. In order to realize the potential of blockchain in 5G networks, it is necessary to understand the operation concept, main properties of blockchain, and understand how blockchain can bring opportunities to 5G applications. In this section, we first present the main components of a blockchain network. Next, we discuss the key characteristics of blockchains in terms of immutability, decentralization, transparency, security and privacy, which can benefit for 5G networks and services. 

\subsubsection{Main components of blockchain}
Blochain features several key components which are summarized as the following.

- \textit{Data block:} Blockchain is essentially a chain of blocks, a linear structure beginning with a so-called genesis block and continuing with every new block linked to the chain. Each block contains a number of transactions and is linked to its immediately-previous block through a hash label. In this way, all blocks in the chain can be traced back to the previous one, and no modification or alternation to block data is possible. Specially, a typical structure of data block includes two main components, including transaction records and a blockchain header \cite{27}. Here, transaction records are organized in a Merkle tree based structure where a leaf node represents a transaction of a blockchain user. For example, a user can make a request with associated metadata (i.e. transferred money or contract) to establish a transaction that is also signed with the private key of user for trust guarantees. Meanwhile, the block header contains the following information: 1) hash of the block for validation, 2) Merkle root to store a group of transactions in each block, 3) nonce value which is a number that is generated by consensus process to produce a hash value below a target difficulty level, and 4) timestamp which refers to the time of when the block is created. A typical blockchain structure is illustrated in Fig. 4.

- \textit{Distributed ledger (database):} Distributed ledger is a type of database which is shared and replicated among the entities of a peer-to-peer network. The shared database is available for all network participants within the blockchain ecosystem. Distributed ledger records transactions similar to the process of data exchange among the members of the network. Participants of the network can achieve on the agreement by a consensus mechanism in a distributed environment where no third party is required to perform the transaction. For example, if a person joins the Bitcoin application, then he has to abide by all rules and guidelines which are established in the programming code of the Bitcoin application. He can make transactions to exchange currency or information with other members automatically without a third party such as a financial institution. In the distributed ledger, every record has a unique cryptographic signature associated with timestamp which makes the ledger auditable and immutable. 

\begin{figure}
	\centering
	\includegraphics [height=4.3cm,width=8.8cm]{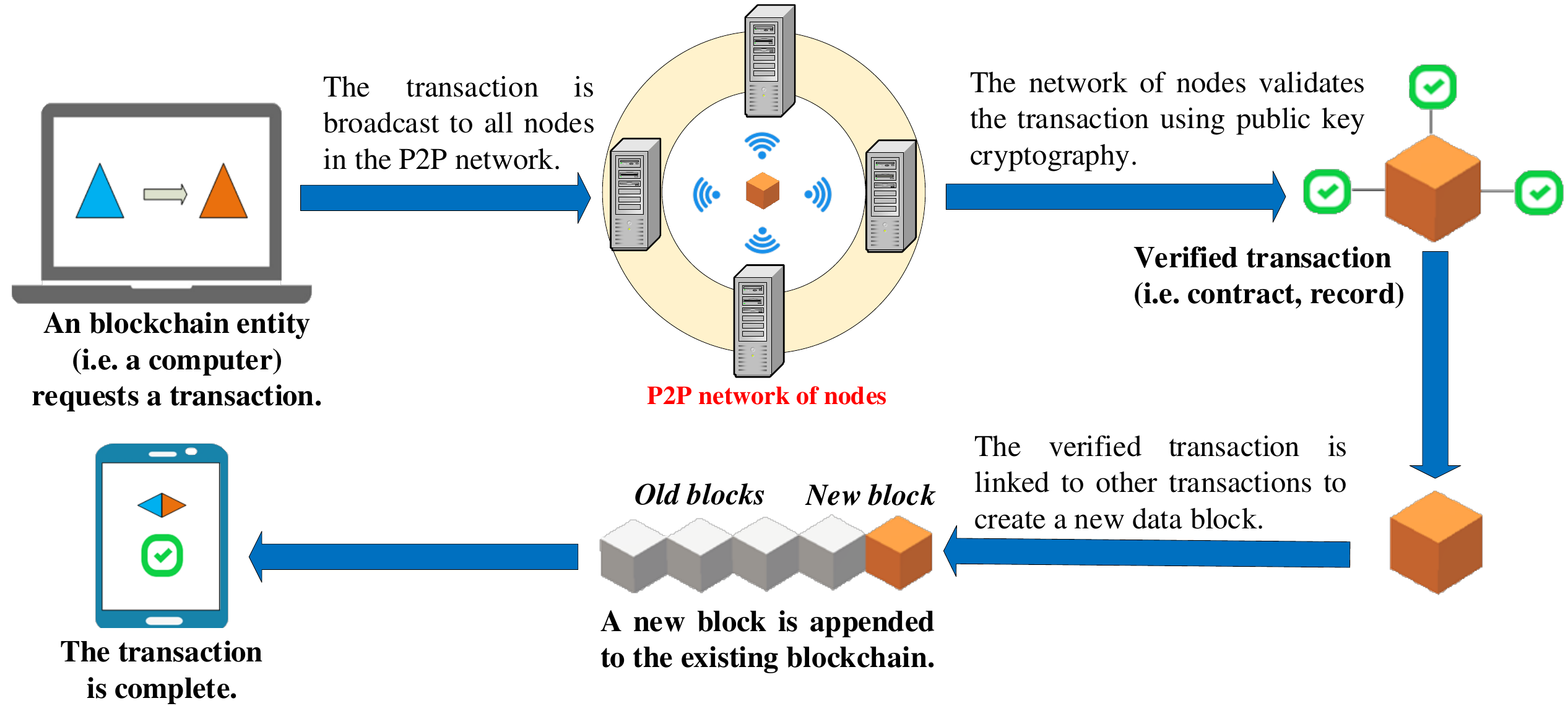}
	\caption{The concept of blockchain operation.}
	\label{fig11}
	\vspace{-0.15in}
\end{figure}

- \textit{Consensus algorithms:} When nodes start to share or exchange data on a blockchain platform, there is no centralized parties to regulate transaction rules and preserve data against security threats. In this regard, it is vitally necessary to validate the block trustfulness, keep track the data flow and guarantee safe information exchange to avoid fraud issues, such as double-spending attacks \cite{28}. These requirements can be met by using validation protocols called as consensus algorithms. In the blockchain context, a consensus algorithm is a process used to reach agreement on a single data block among multiple unreliable nodes. An example of consensus applications is in Bitcoin blockchain. Bitcoin adopts a Proof of Work algorithm (PoW) \cite{7} as an enabling consensus mechanism run by miners to ensure security in a untrusted network. Software on the network of miners uses their computation resources to solve complex mathematical puzzles. The first miner solving the puzzle to create a new block will receive a reward as an encouragement for future mining contributions. However, a critical drawback of PoW is its high resource consumption which would be unsustainable in the future. As a result, other efficient consensus algorithms appears as strong alternatives, such as Proof-of-stake (PoS), Byzantine Faulty Tolerant (BFT). Details of conceptual features and related technical issues of such consensus algorithms can be referenced to previous excellent surveys \cite{5}, \cite{29}.

- \textit{Smart contracts:} A smart contract is a programmable application that runs on a blockchain network. Since the first smart contract platform known as Ethereum \cite{5} was released in 2015, smart contracts have increasingly become one of the most innovative topics in the blockchain area. When we talk about smart contracts, the natural question is: What makes smart contracts so smart? This is due to their self-executing nature which means the codes will execute automatically the contractual clauses defined in the contract once the conditions have been met. For example, when a person signs a smart contract to transfer his funds, the funds will transfer automatically themselves over the blockchain network. Then the transfer information will be recorded as a transaction which is kept on the blockchain as an immutable ledger. Such a type of self-executing agreement relying on the code makes smart contracts unalterable and resistant to external attacks \cite{30}.

In addition to the capability of defining the operational rules and penalties around an agreement similar to the way a traditional contract does, smart contracts are capable of automatically enforcing their obligations to manage transactions. Particularly, smart contracts allow the performance of credible transactions without requiring the involvement of middlemen or third-party intermediaries \cite{31}. This property is particularly useful because it significantly reduces the issues of confliction and saves operation time as well as system costs. Therefore, smart contracts can provide cheaper, faster and more efficient options compared to the traditional systems in which contract conditions are always enforced physically by a central authority, enforcement mechanism or guidance system. With its programmable and automatic features, smart contracts offer a wide range of new applications to solve real-world problems, such as financial services and insurance, mortgage transactions, supply chain transparency, digital identity and records management \cite{31}.
\begin{figure}
	\centering
	\includegraphics [height=4.3cm,width=6.9cm]{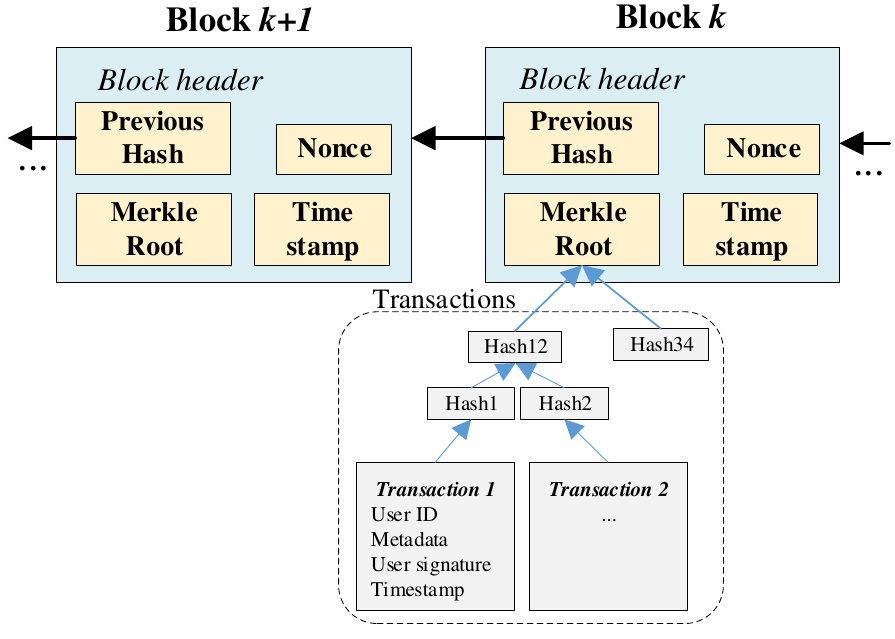}
	\caption{The data block structure.}
	\label{fig11}
	\vspace{-0.15in}
\end{figure}

\begin{table*}[ht]
	\centering
	\caption{Main characteristics of blockchain and their potentials to 5G. }
	\label{table}
	
	\setlength{\tabcolsep}{5pt}
	\begin{tabular}{|p{2.5cm}|p{3.5cm}|p{10cm}|}
		\hline
		\centering \textbf{Key characteristics of blockchain}& 
		\centering \textbf{Description}&	
		\textbf{Potential applications to 5G networks and services}
		\\
		\hline
		\textbf{Decentralization} &No central authority or trusted third party is needed to perform transactions. Users have full control on their own data.&Eliminate the need of trusted external authorities in 5G ecosystems, i.e. spectrum licenses, band managers, and database managers in spectrum management; central cloud/edge service manager in mobile computing and D2D networks; UAV control center in 5G UAV networks; and complex cryptographic primitives in 5G IoT systems. Decentralizing 5G networks potentially eliminates single-point failures, ensures data availability and enhance service delivery efficiency.  
		\\
		\hline
		\textbf{Immutability}&It is very difficult to modify or change the data recorded in the blockchain.&Enable high immutability for 5G services. Spectrum sharing, data sharing, virtualized network resource provisions, resource trading can be recorded immutably into the only-appended blockchain. Besides, D2D communications, ubiquitous IoT networking and large-scale human-centric interconnections can be achieved via peer-to-peer networks of ubiquitous blockchain nodes without being modified or changed. The high immutability is very useful for 5G networks to performing accounting tasks, i.e. logging of session statistics and usage information for billing, resource utilization, and trend analysis.
		\\
		\hline
		\textbf{Transparency} &All information of transactions on blockchain (i.e. public ledgers) can be viewable to all network participants. & Provide better localized visibility into 5G service usage. The same copy of records of blockchain spreads across a large network for public verifiability. This enables service providers and users to fully access, verify and track transaction activities over the network with equal rights. Also, blockchains potentially offer transparent ledger solutions for truly open 5G architectures (i.e. decentralized network virtualization, distributed edge computing, distributed IoT networks). Blockchain ledgers also support fair service trading applications (i.e. resource trading, payment) under the control of all network entities. 
		\\
		\hline
		\textbf{Security and privacy} & Blockchain employs asymmetric cryptography for security with high authentication, integrity, and nonrepudiation. Smart contracts available on blockchain can support data auditability, access control and data provenance for privacy. & Provide high security for 5G networks involved in decentralized ledgers. Blockchain helps secure the 5G networks by providing distributed trust models with high access authentication, in turn enabling 5G systems to protect themselves and ensure data privacy. By storing data information (i.e. IoT metadata) across a network of computers, the task of compromising data becomes much more difficult for hackers. Besides, smart contracts, as trustless third parties, potentially support 5G services, such as data authentication, user verification, and preservation of 5G resource against attacks. 
		\\
		\hline
	\end{tabular}
\end{table*}
\subsubsection{Main characteristics of blockchain}
As a general-purpose database technology, in theory blockchain can be applied to any data-related context. However, the efficiency of distributed ledgers come with costs. Blockchain technology may be not the best solution for every scenario. The important step in assessing the potential benefits of blockchain in 5G is to ask whether its characteristics such as decentralization, immutability, transparency, security and privacy are useful for 5G networks and services. We will briefly review such key properties as follows.

\textit{Immutability:} It is the ability for a blockchain ledger to keep transaction data unchangeable over time. Technically, transactions are timestamped after being verified by the blockchain network and then included into a block which is secured cryptographically by a hashing process. It links to and incorporates the hash of the previous block. This mechanism connects multiple blocks together and builds a chronological chain. Particularly, the hashing process of a new block always contains metadata of the hash value of previous block, which makes the chain data strongly unalterable. This property of blockchain supports secure data storage and sharing in 5G scenarios, i.e. secure spectrum sharing, D2D communication or privacy-preserved network virtualization. Further, by deploying immutable transaction ledgers, the network operators can establish secure communications to perform heterogeneous networking and computing, such as large-scale IoT collaborations or mobile edge/cloud computing over the trustless IoT environments. 

\textit{	Decentralization:} The decentralized nature of blockchain means that it does not rely on a central point of control to manage transactions. Instead of depending on a central authority or third party to perform transactions between network users, blockchain adopts consensus protocols to validate transactions in a reliable and incorruptible manner. This exceptional property brings promising benefits, including eliminating single point failure risks due to the disruption of central authority, saving operational costs and enhancing trustworthiness.

\textit{Transparency:} The transparency of a blockchain stems from the fact that all information of transactions on blockchains (i.e. permission-less ones) is viewable to all network participants. In other words, the same copy of records of blockchain spreads across a large network for public verifiability. As a result, all blockchain users can fully access, verify and track transaction activities over the network with equal rights. Such transparency also helps to maintain the integrity of the blockchain-based systems by reducing risks of unauthorized data alternations. This feature is particularly suitable for 5G ecosystems where the openness and fairness are required. In the cooperative network slicing, for instance, the blockchains can offer transparent ledger solutions to support open and secure data delivery and payment such that the resource providers and slice customers can trace and monitor transactions. Moreover, service trading applications (i.e. mobile resource trading in 5G IoT) can be performed automatically on blockchain by triggering smart contracts, which ensures transparent and reliable data exchange among different service providers and IoT users. 

\textit{Security and privacy:} One of the most appealing aspects of blockchain is the degree of security and privacy that it can provide. The key aspect of security in blockchains is the use of private and public keys. Blockchain systems use asymmetric cryptography to secure transactions between members. These keys are generated randomly with strings of numbers so that it is mathematically impossible for an entity to guess the private key of other users from their public key. This preserves blockchain records against potential attacks and reduces data leakage concerns \cite{32}. Additionally, the privacy service provided by blockchain and smart contract gives the data provenance rights to users. In other words, this ability enables data owners to manage the disclosure of their information on blockchain. Specially, by setting access rules on self-executing smart contracts, blockchain guarantees data privacy and data ownership of individuals. Malicious access is validated and removed by user identification and authorization of smart contract.

\textbf{Remark:} Transparency implies open data, while privacy concerns whether it is possible to infer private and sensitive information from such open data. How to protect people's privacy in open data is a hot topic. A typical example in this area is the face blurring used in the open-access Google Street service. In the context of blockchains, privacy-preserving data provenance based on smart contracts is a promising technique to realize privacy protection in open data [10]. 

From the above high-level analysis, blockchain technology would be a promising candidate for 5G networks and services by providing a number of technical benefits. We summarize the potential applications that blockchain can provide to 5G in TABLE II.

\subsection{5G networks }
The next generations of mobile network (5G and beyond) have revolutionized industry and society by providing an unimaginable level of innovation with significant network and service performance improvements. In this subsection, we present an overview of the 5G networks. Also, 5G design principles are highlighted to provide insights into integrating blockchain in future networks and services. 

\subsubsection{Overview of 5G networks}
Over the past few decades, the world has seen a steady development of communication networks, initializing from the first generation and moving towards the fourth generation. The global communication traffic has shown a drastic increase in recent years and is expected to continue, which triggers the appearance of the forthcoming generation of telecommunication networks, namely 5G, aiming to address the limitations of previous cellular standards and scope with such ever-increasing network capacity. The 5G network can outperform earlier versions of wireless communication technology and provide diverse service abilities as well as encourage full networking among countries globally \cite{33}, \cite{34}. 5G networks also provide solutions for efficient and cost-effective launch of a multitude of new services, tailored for different vertical markets with a wide range of service requirements. In particular, the advances in 5G communication are envisioned as opening up new applications in various domains with great impacts on nearly aspects of our life, such as IoT \cite{35}, smart healthcare \cite{36}, vehicular networks \cite{37}, smart grid \cite{38}, smart city \cite{39}.  Particularly, according to 3GPP and IMT-2020 vision \cite{40}, \cite{41}, the 5G technology is able to provide the following key capabilities:

\begin{itemize}
	\item Provide 1-10Gbps connections to end points in the field and can reach up to 20Gbps in certain scenarios. 
	\item	Provide ultra-low latency services (1ms or less than 1ms).
	\item	Achieve high mobility in the network (up to 500km/h). 
	\item	Enable massive machine-type communication and support high dense network.
	\item	Enable Perception of 99.999\% availability and 90\% reduction in network energy usage.
	\item	Enable 10-100x number of connected devices with the ability to achieve ten year battery life for low power, machine-type devices.
	\item	Enable 1000x bandwidth per unit area.
	
\end{itemize}
In order to achieve such promising performance targets, the 5G networks  leverage a number of underlying technologies, such as cloud/ edge computing, Software-Defined Networking (SDN), Network functions virtualisation (NFV), network slicing, Device-to-Device Communications, Millimeter wave communication \cite{3}. 
\begin{itemize}
	\item \textit{Cloud/edge computing:} Cloud computing has been introduced to meet the increasing demands for resource management, data storage, and mobile sensing in the 5G era. In specific, cloud computing paradigms with resourceful virtual computation centers can well support 5G services such as mobility/network management, resource offloading, and sensing services in various application domains \cite{42}. Meanwhile, as an extension of cloud computing, edge computing has emerged as the promising technology to empower 5G ecosystems. It provides computing services at the edge of the mobile network, with a close proximity to IoT devices, which enables computation and storage services with much lower transmission delays. 
	\item \textit{Software defined networking (SDN):} Using software defined networks, it is possible to run the network using software rather than hardware. It also considers a split between control and data planes, thereby introducing swiftness and flexibility in 5G networks \cite{3}.
	\item 	\textit{Network functions virtualisation (NFV):} When using software defined networks, it is possible to run the different network functions purely using software. NFV enables decoupling the network functions from proprietary hardware appliances so they can run on standardized hardware \cite{3}. The key purpose of NFV is to transform the way networks are built and services are delivered. With NFV, any 5G service operators can simplify a wide array of network functions, as well as maximize efficiencies and offer new revenue-generating services faster and easier than ever before \cite{3}.
	\item \textit{Network slicing:} As 5G will require very different types of networks for the different applications, a scheme known as network slicing has been devices. By using SDN and NFV, it will be possible to configure the type of network that an individual user will require for his application. In this way the same hardware using different software can provide a low latency level for one user, whilst providing voice communications for another using different software and other users may want other types of network performance and each one can have a slice of the network with the performance needed.
	\item \textit{Device-to-Device (D2D) communication:} It allows IoT devices in close proximity to communicate together using a direct link rather than long signal transmissions via traditional base stations. By using D2D communication, 5G heterogeneous data can be transferred quickly between mobile devices in short range, which promises ultra-low latency for communication among users. Moreover, D2D connectivity will make 5G operators more flexible in terms of offloading traffic from the core network, improve spectral efficiency and eliminate unnecessary energy loss due to long data transmissions \cite{43}. 
	\item \textit{Millimeter wave (mmWave) communication:} The mmWave communication technology gives new facilities with a tremendous amount of spectrum to 5G mobile communication networks to supply mobile data demands. It comes with a number of advantages including huge bandwidth, narrow beam, high transmission quality, and strong data access ability to overcome shortcomings caused by the explosive growth in mobile traffic volumes, unprecedented connected devices, and diversified use cases \cite{44}.
\end{itemize}

In the 5G networks, these above technologies will be used to meet the demands of diverse applications from the ongoing traffic explosion of connected devices. For example, the combination of cloud/edge computing and Software Defined Networking and Network Function Virtualization (NFV) is regarded as the potential facilitators for flexible network deployment and operation. Moreover, the network slicing and D2D communication will enable ultra-reliable, affordable broadband access and intelligent use of network data to facilitate the optimal use of network resources with extremely low latency and high-speed device connection \cite{4}, \cite{5}.  The proliferation of 5G networks was initially shaped by the Next Generation Mobile Networks (NGMN) alliance \cite{45} with a 5G initiative for enabling emerging services and business demands with the time target of 2020 and beyond.

\subsubsection{5G design principles}
The rapid advances of new 5G technologies provide an impetus for new fundamental design principles toward 5G networks. The 5G design principle was outlined by the NGMN alliance \cite{46}	as shown in Fig. 5. Specifically, 5G systems can employ software and virtualisation to achieve the service objectives on flexibility, configurability, and scalability. Particularly, one of the key design concepts behind the 5G networks will be network slicing which separates the user and control planes and enables dynamic network function placement \cite{3} for a ubiquitous flexible and extensible infrastructure for all types of communication services on top of which a dynamic service and business environment can involve. The vision of 5G lies in providing smart services with very high data rates, extremely low network latency, manifold increase in base station density and capacity, and brings about significant improvements in the quality of services, quality of user experience, compared to 4G systems. It provides a convergence of pervasive broadband, sensing, and intelligence to establish a greater scale for the fourth industrial revolution that will stimulate the development of society and industrial markets. 
\begin{figure*}
	\centering
	\includegraphics [height=6.7cm, width=15cm]{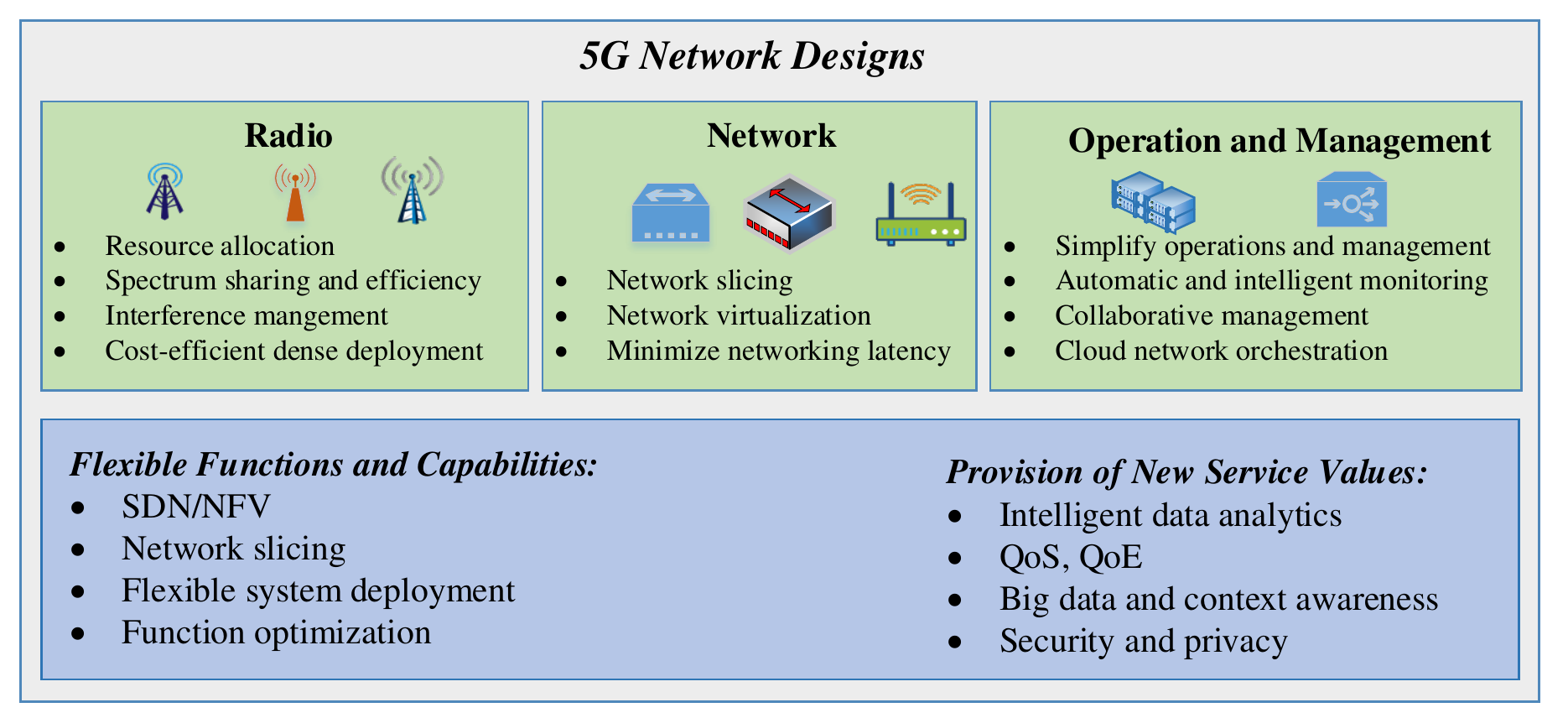}
	\caption{The 5G design principle \cite{46}.   }
	\vspace{-0.17in}
\end{figure*}

The 5G network architecture must support the deployment of security mechanisms and functions (e.g. virtual security firewalls) whenever required in any network perimeter. As presented in Fig. 5, the operation and management need to be simplified. The most prominent technology for simplifying network management is SDN [58]. SDN separates the network control from the data forwarding plane. The control plane is logically centralized to oversee the whole network underneath and control network resources through programmable Application Programming Interfaces (APIs). Network Functions Virtualization (NFV) implements Network Functions (NF) virtually by decoupling hardware appliances (such as firewalls, gateways) from the functions that are running on them to provide virtualized gateways, virtualized firewalls and even virtualized components of the network, leading to the provisions of flexible network functions. Meanwhile, cloud computing/cloud RAN supports unlimited data storage and data processing to cope with the growing IoT data traffic in 5G. The combinations of 5G enabling technologies promise to foster mobile networks with newly emerging services such as intelligent data analytics, big data processing. Specially, different from previous network generations (i.e. 3G/4G), 5G is promising to provide mobile services with extremely low latency, energy savings due to flexibility (i.e. network slicing and proximity of edge computing), all of which will enhance QoS of the network and ensure high QoE for users. 
\subsection{	Motivations of the Blockchain and 5G integration}

In this subsection, we highlight the motivation of the integration which comes from the security challenges of 5G networks and the promising opportunities brought by the incorporation of such two technology families.

\subsubsection{Definition of the integration of Blockchain and 5G}
 To highlight the motivation, we recall the most important properties of both technologies for the integration. Blockchain brings the capability of storing and managing 5G data through its secure distributed ledger. More importantly, blockchain can provide a series of security features such as immutability, decentralization, transparency and privacy, all of which promise to tackle efficiently security issues of current 5G networks. Thus, the main points of blockchain here are its capabilities to support security and network management for 5G networks and applications. On the other side, 5G considered in this paper refers to the latest generation wireless networks which are  envisioned to  provide higher capacity, higher data rate, lower latency, massive device connectivity, enhanced  end-user  quality-of-experience  (QoE), reduced operation cost, and consistent service provisioning. Therefore, the key points of 5G here are its advantages of providing fast and high-quality services and the need for security and networking improvement.

Reviewing the rich and state of the art articles in the field, the motivation behind the integration of blockchain and 5G stems mainly from the promising benefits of blockchain for solving challenges in 5G networks in terms of security, privacy, networking and service management. With the help of innovative blockchain designs, 5G is expected to overcome the existing challenges and open up new opportunities to empower blockchain 5G-based services and applications. In the following, we discuss the motivation of the integration coming from current 5G challenges and then present opportunities brought from the blockchain-5G integrations.

\subsubsection{	Security challenges in 5G networks}
The security associated with 5G technologies has been considered as one of the key requirements related to both 5G and beyond systems. The existing 5G technology infrastructure has remained unsolved challenges in terms of security, networking and computing performance degradation due to its centralized architecture \cite{46}. For example, edge/cloud computing models current rely on centralized service providers (i.e. Amazon cloud), which reveals various security bottlenecks. Indeed, this configuration is vulnerable to single-point failures, which bring threats to the availability of cloud/edge services for on-demand user access. A centralized system does not guarantee seamless provisions of IoT services when multiple users request simultaneously data or servers are disrupted due to software bugs or cyberattacks. 

Moreover, network function virtualization (NFV) and service function chaining in 5G networks, however, also incur new security challenges \cite{47}, \cite{48}. Since end-to-end service function chains may deploy NFVs in an environment involving multiple cloud providers, such data transmissions can be compromised by curious cloud entities, leading to data leakage concerns. Furthermore, in a virtualized scenario, tenants often share the same cloud infrastructure. In this context, the possibility of attacks inside the cloud can increase, which damages the transparency and accountability of service providers. In NFVs, virtualization servers can run on virtual machines (VM) to offer specific functions to execute distinct operating systems such as VM migration or resource allocation using orchestration protocols. However, the security for the communication between the orchestrator and the physical machine VM manager is a real challenge.

The rapid proliferation of mobile data traffic and the increasing user demands on 5G infrastructure also introduce new challenges in terms of security and performance degradation. For example, the increasing requirement for bandwidth-hungry applications for 5G services such as mobile video streaming, big data processing requires a proper 5G spectrum resource management strategy to avoid resource scarcity issues for ensuring continuous service functionalities. Therefore, spectrum sharing between mobile network operators (MNOs) and mobile users is necessary. However, spectrum sharing in such scenarios also raises security concerns and provides a central point of attacks for malicious users \cite{49}. A possible approach is to use certification authorities, providing provide certificates for cognitive radios inside each cell. This approach not only requires infrastructure to be implemented for each cell but also requires a protocol for defence against central-point attacks. Further, it requires greater calculation complexity and longer packet lengths, which increases overhead for spectrum sharing systems and thus reduces the Quality of Services (QoS) of the involved system. Importantly, the use of such centralized architectures also adds single-of-failure bottlenecks when the authority is attacked or out of services, which leads to the disruption of the entire spectrum sharing network. 

In the 5G IoT scenarios such as smart healthcare, smart cities where mobile environments are highly dynamic with the conjunction of ubiquitous IoT devices, heterogeneous networks, largescale data storage, and powerful processing centres such as cloud computing for service provisions, security and privacy issues become much more complex to be solved \cite{50}. In fact, a prohibitively large amount of IoT data will be generated continuously from ubiquitous IoT sensor devices. It is very challenging to immediately identify the objects of interest or detect malicious actions from thousands of data transactions on a large scale. The solution of using a centralized management may be infeasible to such use cases due to long latency, privacy risks due to curious third parties and network congestion. Obviously, how to provide efficient mobile services (i.e. data sharing, data processing, user management) in terms of low latency and increased network throughput while still ensure high degrees of security is a critical challenge. Therefore, there are urgent needs of innovative solutions to overcome the above security and network performance limitations for future 5G networks. 

\subsubsection{	Opportunities brought by blockchain to 5G networks and services}

With its promising security properties, blockchain promises to provide a new set of innovative solutions for 5G networks and services for better security, privacy, decentralization and transform the network management architectures for improved QoS as well as better 5G performances. Therefore, 5G should leverage the benefits of blockchain to accommodate flexibility and security in providing mobile network services and ubiquitous coverage. In short, we highlight the significant opportunities that blockchain can bring to 5G networks and services, with a focus on three main aspects, including security enhancements, system performance improvements, and network simplification. 

\begin{enumerate}
	\item \textit{Security enhancements:} Blockchain promises to enhance the security and privacy of 5G ecosystems, by offering many promising technical properties such as decentralization, privacy, immutability, traceability, and transparency. Blockchain can eliminate the centralized network management concept by decentralizing the network infrastructure where there are no third party authorities needed. As an example, the concept of blockchain-based cloud computing enables decentralization of cloud/edge 5G networks which removes centralized control at the core network and provides a decentralized fair agreement with blockchain consensus platform, which eliminates single point failure bottlenecks and improves significantly system trust. Besides, the security of D2D communication can be achieved by building a peer to peer network via blockchain, which transforms each D2D device as blockchain node to hold a ledge copy with the ability of verifying and monitoring transactions for better system transparency and reliability.  
	
	Especially, different from the conventional database management systems which often use a centralized server to perform access authentication and security mechanisms, blockchain with smart contracts can implement decentralized user access validation by using the computing power of all legitimate network participants. This makes the 5G services (i.e. spectrum sharing, data sharing, resource allocation) strongly resistant to data modifications. Many research works on blockchain [11], [12], [13] demonstrate that the blockchain adoption is beneficial to spectrum 5G management in terms of better verification of spectrum access with blockchain contracts, improved accessibility thanks to the transparency of blockchain. Moreover, the use of blockchain fosters scalable spectrum sharing over the peer-to-peer ledge network where spectrum license holders and band managers are eliminated for high trustworthiness. The ledger services with strong immutability from blockchain also provide a high degree of security and better system protection capability against DoS attacks and threats. Empowered by smart contracts, which provide highly flexible efficient user access control mechanisms via access rules and intelligent coding logics, blockchain potentially introduce new authentication solutions for 5G cellular networks. Instead of relying on external public key infrastructure, contracts can authenticate automatically user access, detect threats and discard malicious access from the networks in an autonomous manner without revealing user information. Besides, by publishing user data to ledger where data is signed by hash functions and appended immutably to blocks, blockchain platforms ensure strong data protection. Blockchain is capable of providing a full control of personal data when sharing over the untrusted network, which is unique from all traditional approaches which hinder users from tracking their data [14]. 
	\item \textit{System performance improvements: }The use of blockchain also potentially improves the performances of 5G systems. In comparison to traditional database platforms such as SQL, blockchain can provide better data storage and management services with low latency data retrieval. In fact, resource requests (i.e. data access) can be verified by decentralized blockchain nodes with the support of intelligent smart contracts without passing a centralized authority, which is promising to reduce network latency. Moreover, motivated by the removal of decentralization, blockchain is able to establish direct communications between 5G service providers and mobile users so that the management cost can be significantly reduced. This would provide a much more flexible and efficient data delivery model for 5G ecosystems but still meet stringent security requirements [12]. For example, blockchain can help establish secure peer-to-peer communication among users (i.e. in D2D communication) using the computing power of all participants to operate the network instead of passing a third party intermediary. This would potentially reduce communication latency, transaction costs, and provide the global accessibility for all users, all of which will enhance the overall system performance. Specially, even when an entity is compromised by malicious attacks or threats, the overall operation of the involved network is still maintained via consensus on distributed ledgers, which in return ensures no single-point failure vulnerabilities for better security. 
	\item \textit{Network simplification:} It is believed that blockchain can simplify the 5G network deployments thanks to its decentralized architectures. Indeed, by leveraging blockchain, the mobile operators now can have no worries about the establishment of centralized control servers. The 5G service delivery can be achieved by the blockchain network where user access, service responses and service trading (i.e. resource trading and payment) can be implemented on the decentralized ledgers among network participants including service providers and mobile users without the need for additional management infrastructure [5]. Therefore, the blockchain adoption potentially reduces network complexity and thus saves significantly operational costs. Furthermore, the transactions for 5G services (i.e. data sharing, spectrum sharing) are controlled by the blockchain network itself where all entities hold the same rights to manage and maintain the network. The capability of exploiting internal resources from participants is also another great advantage that blockchain can provide to simplify the network organization and management for better user experience and facilitation of service transactions, especially in complex mobile environments in the future 5G networks [6]. 
\end{enumerate}

\section{Blockchain for enabling 5G technologies}
Reviewing state-of-art literature works \cite{1}, \cite{3}, \cite{4}, we found that blockchain has mainly cooperated with the key 5G enabling technologies including cloud computing, edge computing, Software Defined Networks, Network Function Virtualization, Network Slicing, and D2D communication. Motivated by this, in this section, we present a review on the integration of blockchain and such 5G technologies. The benefits of blockchain for different 5G use cases and applications empowered from the integration are also analysed in details. 

\subsection{Blockchain for cloud computing/ Cloud RAN}

Cloud computing has drawn significant attention in the last decades thanks to its unlimited resources of storage and computation power, which can provide on-demand, powerful and efficient services with minimum management efforts. Cloud computing has been investigated and integrated extensively with 5G networks, paving the way for the computing-intensive applications involving multi-dimensional massive data processing assisted by the cloud \cite{51}, \cite{52}. In fact, cloud computing paradigms provide a number of technical solutions for realizing 5G services, such as optimizing the communications, processing and storage processes \cite{53}, 5G data content delivery and catching \cite{54}, resource allocation and data transmission management \cite{55}, and cloud-enabled small cell networking for 5G media services \cite{56}. Specially, in order to meet the ever-increasing demand of user association and resource allocation in cellular 5G networks, the architecture of cloud radio access networks (Cloud-RANs) is envisioned as an attractive model that manages the large number of small cells through the centralized cloud controller as baseband unit (BBU) pool \cite{57}. Cloud-RAN is able to offer high-speed interconnection and shared powerful processing to facilitate optimal multicell cooperation and collaborative radio, real-time cloud computing \cite{58}, \cite{59}, which makes Cloud-RAN become a promising candidate of next-generation 5G access networks. 

However, the existing cloud computing models remain unsolved challenges in terms of security, networking and computing performance degradation due to its centralized architecture. Indeed, in the 5G era, the massive data traffic outsourced from IoT devices to the cloud has brought about a series of new security challenges, mainly including data availability, data privacy management, and data integrity \cite{60}. 

\begin{itemize}
	\item \textit{Data availability:} In current cloud network architectures, cloud services are provided and managed centrally by the centralized authority. However, this configuration is vulnerable to single-point failures, which bring threats to the availability of cloud services for on-demand user access. A centralized cloud IoT system does not guarantee seamless provisions of IoT services when multiple users request simultaneously data or cloud servers are disrupted due to software bugs or cyberattacks.
	\item \textit{Privacy management:} Although the centralized cloud 5G networks can provide convenient services, this paradigm raises critical concerns related to user data privacy, considering a large amount of 5G heterogeneous data being collected, transferred, stored and used on the dynamic cloud networks. In fact, IoT users often place their trust in cloud providers managing the applications while knowing very little about how data is transmitted and who is currently using their information \cite{61}. In other words, by outsourcing data protection to the cloud, IoT data owners lose control over their data, which has also adverse impacts on the data ownership of individuals. Moreover, even in the distributed cloud IoT paradigms with multiple clouds, IoT data are not fully distributed but stored in some cloud data centres at high density \cite{62}. In this context, a massive amount of heterogeneous data may be leaked and user privacy is breached if one of the cloud servers is attacked.
	\item \textit{Data integrity:} The storage and analysis of 5G data on clouds may give rise to integrity concerns. Indeed, due to having to place trust on the centralized cloud providers, outsourced data is put at risks of being modified or deleted by third parties without user consent. Moreover, adversaries can tamper with cloud data resources \cite{63}, all of which can breach data integrity. For these reasons, many solutions have been applied to overcome the problem, by using public verification schemes where a third party auditor is needed to perform the integrity verification periodically. This scheme potentially raises several critical issues, including irresponsible verification to generate bias data integrity results or invalidated verification due to malicious auditors. 
	
	\item \textit{Lack of immutability:} The dynamic process of 5G data to clouds and data exchange between cloud providers and mobile users are vulnerable to information modifications and attacks caused by adversaries or third parties. Even entities within the network may be curious about transmitted data over the sharing and unauthorized obtain personal information (i.e. customer data of 5G smart grid or location information of vehicles in vehicular networks). These issues may lead to serious data leakage bottlenecks and consequently damage system immutability.
	
	\item \textit{Lack of transparency:} In the conventional cloud systems, cloud resource providers have full control over outsourced network data (i.e. IoT data) while users are not aware of it and lacks the ability of tracking data after offloading to the cloud. This poses critical challenges on data users to perform verification and monitoring of data flows or usage, especially in the 5G scenarios where transparency among networks members is highly required to ensure fairness and openness, i.e. cloud service providers and slice users in cloud-based network slicing, or between healthcare providers and patients in cloud e-health. 
\end{itemize}

Recently, blockchains have been investigated and integrated in cloud computing to effectively address the above security challenges in the cloud-based 5G networks. For example, the work in \cite{64} takes advantage of blockchain to develop a framework called BlockONet for 5G access scenarios, aiming to improve the network credibility and security in 5G fronthaul. Blockchain is employed to build a verification platform between IoT devices, BBU unit, and manufacturer, where user access information is stored immutably on the chain, while smart contracts are also leveraged to perform automatic user authentication. The benefits from the use of blockchain in Cloud-RAN 5G networks are twofold. First, the concept of blockchain-based Cloud-RAN gets rid of centralized control at the core network and offers a decentralized fair agreement with blockchain consensus platform, which eliminates single point failure bottlenecks and improves significantly system trust. Second, by applying a decentralized blockchain without third parties, the blockchain-based cloud-RAN strategy can achieve optimal resource utilization and save a large amount of signalling and connection costs. In the same direction, the study in \cite{65} applies blockchain to build a trusted authentication architecture for cloud radio access network (Cloud-RAN) in the 5G era. They also show that the proposed schemes can address effectively network access authentication with trusted agreement among service providers and IoT users with reduced operation costs and improved spectrum usage over Cloud-RAN based mobile networks.

Blockchain is also integrated with cloud computing for 5G IoT networks. The study \cite{66} proposed a cloud-centric IoT framework enabled by smart contracts and blockchain for secure data provenance. Blockchain incorporates in cloud computing to build a comprehensive security network where IoT metadata (i.e. cryptographic hash) is stored in blockchain while actual data is kept in cloud storage, which makes it highly scalable for dense IoT deployments. In the system, smart contracts with its autonomous, transparent and immutable properties are also adopted to ensure high cloud data validity. Meanwhile, a secure data sharing architecture was introduced in \cite{67} with attributed based-access control cryptosystem. Its network model consists of four main components: IoT devices, a data owner, a blockchain network and a cloud computing platform. More specific, a permissioned blockchain model is adopted to manage IoT transactions and perform access control for device requests received by cloud, while cloud monitors closely the blockchain network. As a result, such a cloud blockchain integration brings a comprehensive security framework with enhanced privacy preservation, data ownership and secure data sharing. Similarly, a hierarchical access control structure for Cloud blockchain was investigated in \cite{68} with a blockchain-based distributed key management. Especially, the blockchain network topology involves distributed side blockchains deployed at fog nodes and a multi-blockchain operated in the cloud, which would speed up access verification offer flexible storage for scalable IoT networks. In addition, to protect cloud blockchain in security-critical applications, a forensic investigation framework is proposed using decentralized blockchain \cite{69}. Security issues from dynamic interactions between cloud service providers, clients, and IoT devices were considered and analysed with a tamper evident scheme. Blockchain is performed to audit evidence during the investigation of a criminal incident among cloud blockchain entities in a decentralized manner, and therefore avoiding single points of failure on the cloud storage and improving evidence availability.

In addition, blockchain has also incorporated with the cloud federation architectures to further improve the performance of complex 5G-IoT networks in terms of transparent collaboration and interconnected services. As an example, a blockchain framework was proposed on a joint cloud collaboration environment where multiple clouds are interconnected securely by peer-to-peer ledges \cite{70}. The proposed scheme contains three tiers with an IoT sensor network, a federation of multiple clouds, and a service platform. Typically, the blockchain platform can offer many advantages over the schemes based on a single cloud. For instance, since IoT data at each area is stored in a private local cloud in the multi-cloud network, its data security is significantly improved. Further, the single cloud can offer instant services for IoT users through the private blockchain network, which also mitigates risks of malicious attacks on cloud systems \cite{71}. Besides, a cloud blockchain model with micro-clouds was introduced by \cite{72} using blockchain-enabled distributed ledgers. The authors pay special attention to building a joint cloud blockchain to enable secure decentralized collaborative governance services, i.e. immutable data storage, transparent monitoring and resource management for suitable performance on lightweight computing nodes like IoT devices. 

\subsection{	Blockchain for mobile edge computing}
\begin{figure}
	\centering
	\includegraphics[height=7.5cm,width=8cm]{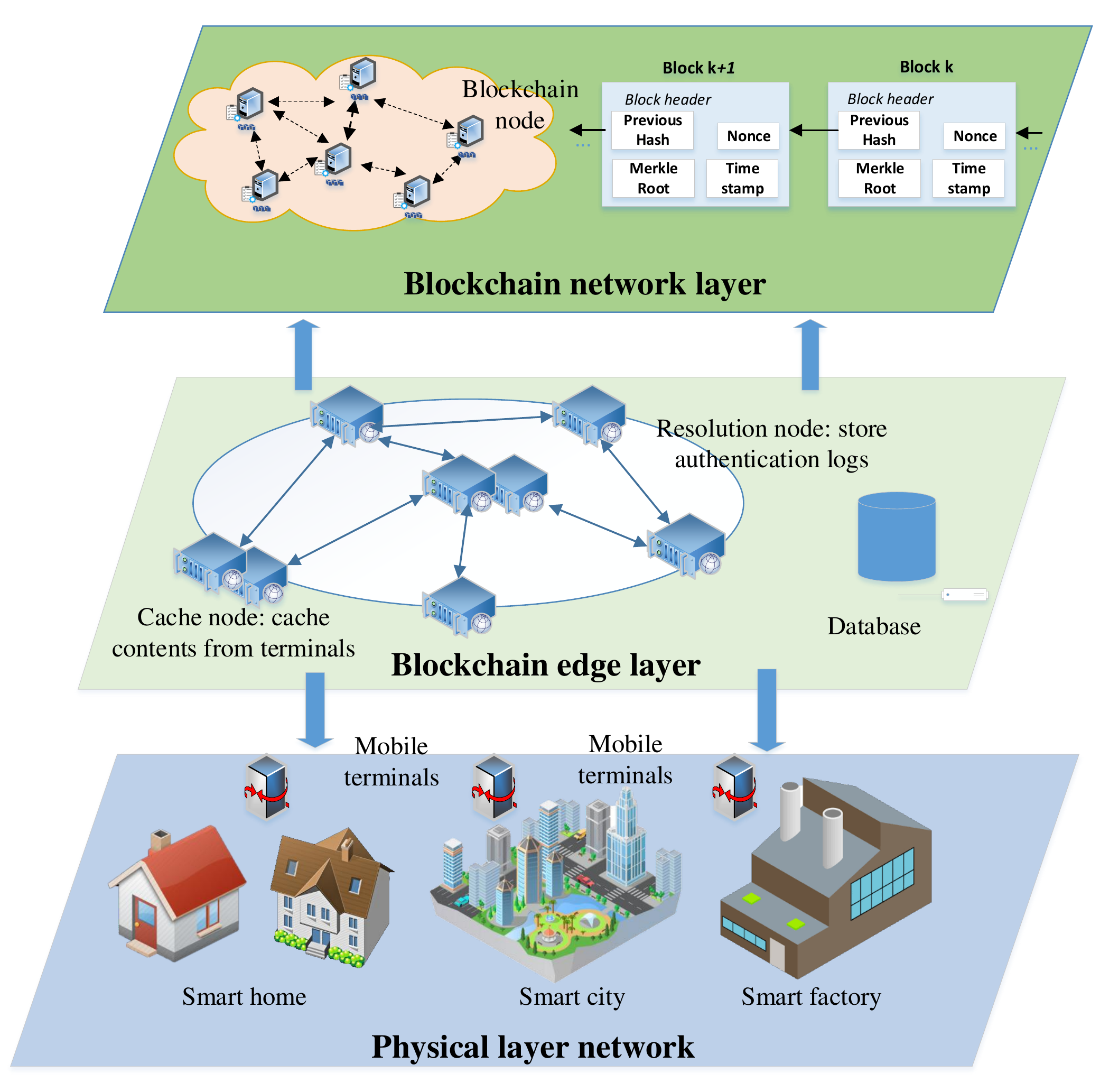}
	\caption{The convergence of blockchain and edge computing for 5G services. }
	\label{fig11}
	\vspace{-0.15in}
\end{figure}
As an extension of cloud computing, mobile edge computing (MEC) has emerged as the promising technology to empower 5G services. Edge computing may have other names such as fog computing, mobile cloud or cloudlet. Similar to the cloud paradigm, edge computing can offer a series of computing services with capabilities of task processing, data storage, heterogeneity support and QoS improvements. In fact, edge servers are less powerful than remote clouds, but they are located at the edge of the network, with a close proximity to IoT devices, which enables highly efficient 5G data computation with much lower transmission delay, compared with the remote cloud \cite{73}. As a result, edge computing can provide instant computing applications to IoT users with low latency and fast service response, which would be particularly useful in the next generation services (i.e. in 5G and beyond). The distributed structure of edge computing also potentially brings numerous benefits, from ubiquitous computing services, scalability improvement to complexity reduction of network management to cope with the explosion of IoT devices and rapid growth of 5G service demands \cite{74}. However, its security is a significant challenge \cite{75}, \cite{76}. Indeed, the migration of 5G services, i.e. data computation, in the dynamic edge computing environments can be vulnerable to malicious attacks (such as jamming attacks, sniffer attacks, denial-of-service attacks, etc.). Further, the setting and configuration information by the edge service providers (ESP) must be trustworthy and secure, but in fact these are actually challenged due to the high dynamism and openness of the MEC system. Another challenge is to ensure data privacy and immutability for outsourced 5G heterogeneous data from external modifications or alternations. Importantly, how to avoid the system disruption caused by the attack on an edge node in the multi-edge computing \cite{75} is of paramount importance for 5G-based edge computing networks. Fortunately, blockchain has come as a promising technical enabler to overcome most of security and networking challenges faced by the existing edge computing architectures. The same decentralization characteristic of both the blockchain and MEC built on the networking, storage, computation, communications makes their combination become natural. The recent research results have demonstrated that blockchain can be applied to the edge computing systems to support a number of services of security and management in edge computing \cite{77}. Generally, the blockchains can support edge computing-based 5G services in three main aspects: networking, storage and computation as shown in Fig. 6.

In fact, with the help of blockchain, the networking capability of edge networks can be optimized. The blockchain is employed in \cite{78} to build a distributed and trusted authentication system to realize reliable authentication and information sharing among different edge-based IoT platforms. In the system, authentication data and user access information can be stored securely on blockchain, which is also capable of automatically tracking activities of mobile terminals (devices) without the need of central authorities. In particular, smart contracts are also utilized to perform trusted content catching in the edge computing network. Meanwhile, the works in \cite{79}, \cite{80} suggest a blockchain-based architecture for vehicular edge computing. Vehicular edge computing is introduced to provide data processing services with low latency, but it also raises privacy concerns since user information can be disclosed during the sharing process. The adaption of blockchain potentially solves such challenges by establishing a secure communication channel empowered by immutable transaction ledgers. Then, this robust and secure concept enables the energy flow and information flow to be protected against external malicious attacks when performing vehicular networking. Furthermore, ensuring security in the transmission process is one of the achievements of blockchain. The authors in \cite{81}, \cite{82} take advantage of blockchain to establish a security mechanism for edge computing-based energy systems where smart contracts are leveraged to build a trusted access control scheme for energy sharing and distribution. Further, the blockchain-based solutions can support efficient conditional anonymity and key management for the privacy-preserving authentication protocol without the need for other complex cryptographic primitives between network users. Moreover, to achieve a trustworthy and efficient edge computing system, the blockchain functionality is applied to the resource management \cite{83}, data sharing \cite{84} or resource allocation \cite{85}, all of which improve edge computing performances while guaranteeing security properties of the network.

In addition, blockchain also provides security features for efficient data storage for edge computing systems. Indeed, blockchain can offer decentralized data storage enabled by the combined storage capacity of a network of peers to store and share contents. The work in \cite{86} proposes a MEC-based sharing economy system by using the blockchain and off-chain framework to store immutable ledgers. Specifically, in a smart vehicular network, blockchain can keep information of the driver and the car profile with the history of maintenance, accident, and other car usage information. The raw vehicular data, i.e. vehicle sensor data, can be captured and processed by the MEC node under the control of the blockchain. Blockchain can also connect the stakeholders of a car through a shared chain and provide help in car-sharing economy scenarios. The work in \cite{87} also proposes a blockchain database to secure communication between the home devices and sensors in the MEC-based smart city. In the sense of the ledger, blockchain can be regarded as a distributed database which keeps data by interconnecting a network of strongly immutable blocks. It is noting that the scalability of blockchain is a critical challenge due to the constrained ledger size, throughput and latency \cite{77}.  In this regard, the on-chain and off-chain storage concept can be very useful. For example,  in the vehicle context, the real-time updates regarding traffic and pollution of nearby roads can be stored locally in a cache unit for autonomous cars, while data hash values can be kept securely in blockchain. Any modifications on the storage unit can be acknowledged by blockchain via decentralized ledgers, improving the trustworthiness of the MEC-based network. Moreover, to facilitate easy access to data in a distrusted MEC blockchain setting, a decentralized big data repository platform, such as Inter-Planetary File System (IPFS) can be necessary for improving storage capability on blockchain \cite{88}. On top of IPFS, several blockchain-based storage platforms such as Filecoin or Storij \cite{10} have been applied as an incentive layer to form an entirely distributed file storage system. These blockchain database systems contain the off-chain service data while providing the on-chain identifier, so that data integrity can be checked by the identifier from the data and hash values in the blockchain and comparing it for monitoring. Such a blockchain platform is integrated with edge computing to solve storage risks caused by dynamic MEC \cite{89}.

Lastly, blockchain can support the computation processes in MEC networks. Specifically, blockchain can provide authentication capability to protect MEC systems. The study in \cite{90} leverages blockchain features such as decentralization, tamper-proofing and consistency to build an authentication layer between edge/fog servers and IoT devices. The main objective is to monitor and verify all computing tasks offloaded to the MEC servers, which preserves edge computing from external attacks. In \cite{91}, smart contracts are employed for MEC to improve the efficiency of IoT computing, i.e. video coding, by providing a self-organized video transcoding and delivery service without a centralized authentication. Blockchain can protect the accuracy, consistency, and origins of the data files in a transparent way. Further, the transactional data are also encrypted and stored on blocks, which has the potential to achieve privacy and security for MEC \cite{92}. 

\subsection{	Blockchain for Software Defined Networking}
Software-Defined Networking (SDN) has gained great attraction over the past years and has been regarded as the key pillar of future 5G networks. SDN is an intelligent networking architecture that envisions to improve the programmability and flexibility of networks. The main concept of SDN is the separation of the control plane outside the network switches and the provisioning of external control of data through a logical software controller, enabling mutual access between different parts of heterogeneous networks \cite{93}. This design architecture not only offers a number of new architecture, management and operation options, but also provides the ability for efficient delivery of user services while exploiting network resources more efficiently. In the 5G context, SDN is developed to make the connectivity services provided by 5G networks programmable, where traffic flows can be dynamically steered and controlled in order to achieve maximum performance benefits. However, despite the obvious advantages that this novel networking paradigm introduces, there remains some non-trivial challenges that hold back its undisputed dominance over legacy solutions, namely security, flexibility and scalability. 
\begin{itemize}
	\item \textit{Security:} In SDN, security is about the authentication in the control plane and mitigation of data modification and leakage in the data plan. In fact, one of the most important shortcomings of SDN is its increased attack surface compared to traditional networking deployments when the controller is modified or compromised. The most fundamental property of the SDN architecture is the decoupling of the control plane and the data plane, but this decoupling also broadens the attack surface of the network and introduces attack bottlenecks for the application layer \cite{94}. Furthermore, the centralized design of the SDN controller is also vulnerable to attacks on the control layer, which can cause controllers, routers, and switches to be maliciously modified, generate and cause loss of flow table information \cite{95}.
	\item \textit{Scalability:} How to build scalable SDN networks to enable multiple SDN controllers to communicate each other and achieve secure information exchanges between them is a challenge. By  providing a distributed network architecture, SDN service providers not only reduce costs and enhance the flexibility to extend the network but also involve the deployment of new services to meet new market requirements \cite{96}.
	\item \textit{Full network decentralization:} The centralized design concept of current SDN models is possibly vulnerable to single-of-failure risks when a network entity is attacked or compromised, which leads to the disruption of the entire network. Therefore, developing a decentralized SDN architecture which can solve this problem and improve quality of services is vitally significant. 
	\item \textit{Network management:} In the multi-SDN environments, SDN devices cannot be interoperable and achieve interconnection and cooperation due to the stringent latency requirements from different 5G service providers. The utilization of network resources requires a centralized repository maintained by all parties for the service provider, but it is challenging to achieve mutual trust between suppliers and the fairness of resource allocation due to the potential conflicts of interest of service providers. How to achieve a trusted network management for an efficient network cooperation multi-SDN networking and perform reliable resource sharing is a challenge \cite{97}. 
	
\end{itemize}

\begin{figure}
	\centering
	\includegraphics [height=6.8cm,width=6cm]{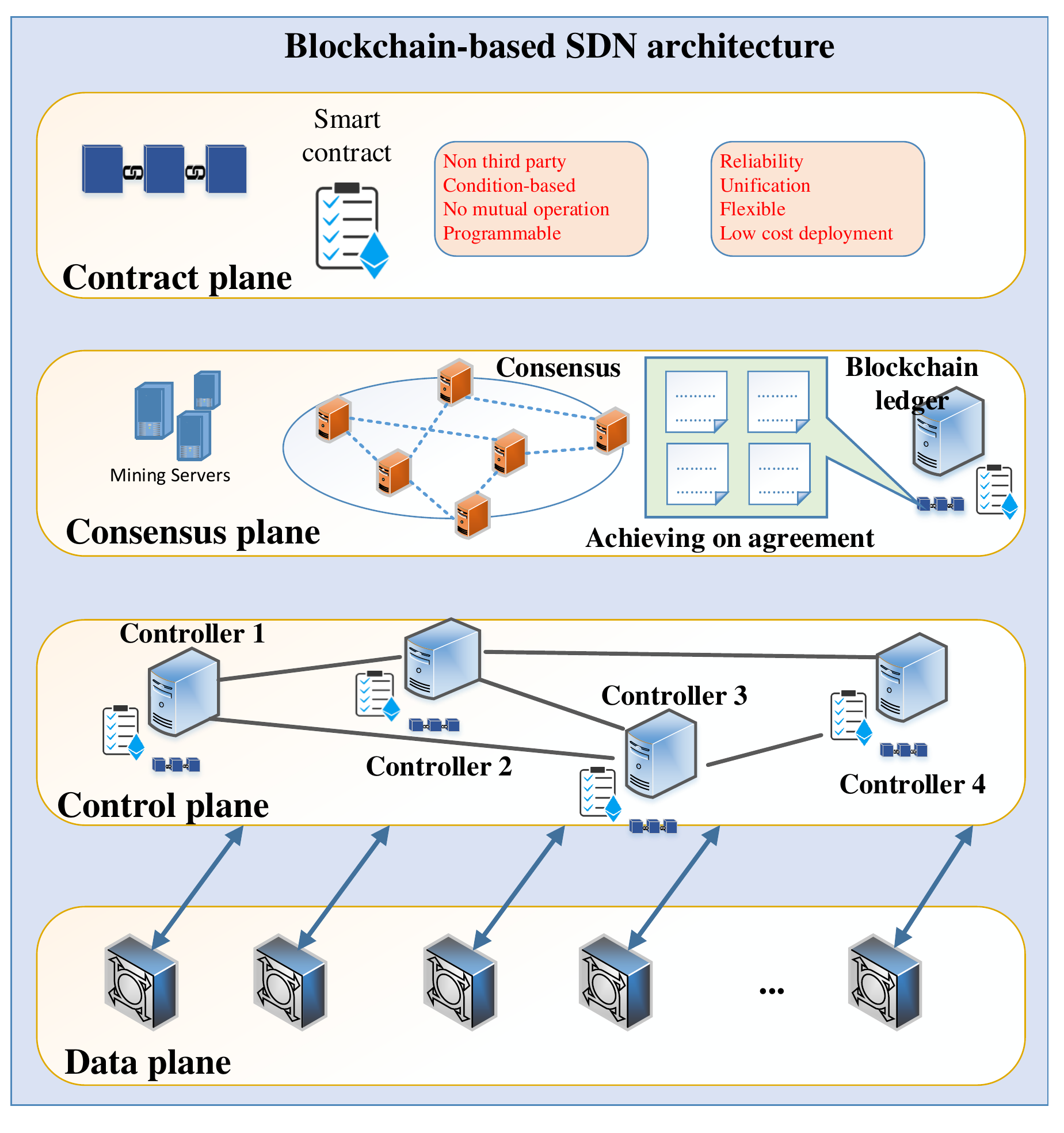}
	\caption{A blockchain-based SDN architecture. }
	\label{fig11}
	\vspace{-0.15in}
\end{figure}
In order to overcome these shortcomings in SDN architectures, many research efforts have been dedicated to research on blockchain as a decentralized security provisioning solution for SDN. The authors in \cite{98} propose blockchain as an authentication solution for SDN-based 5G networks with the objective of eliminating the unnecessary re-authentication in repeated handover among heterogeneous cells. Multiple SDN controllers in this proposed approach can communicate each other and interact with blockchain which enables secure information exchanges between them. Transactions and messages from blockchain can be shared via the dedicated transfer keys to the controller. Each SDN controller has a dedicated transfer key received from blockchain and is applied to transfer and receive information. Importantly, scalability can be solved effectively by a blockchain-based hierarchical structure. If any SDN controller becomes down in a cell, the system will then manage this cell using another SDN controller in the network where consensus between SDN controller candidates can be achieved by blockchain ledgers. The integration of blockchain in SDN is thus promising to remove intermediaries for authentication, reduce transaction costs, and achieve global accessibility for all users. Meanwhile, the work in \cite{99} proposes a decentralized blockchain-based security framework for SDN-enabled vehicular ad-hoc networks (VANETs). The SDN controller is in charge of the global policies, including authentication, and mobility/traffic management, while the controller-defined policies are implemented at the data plane. With the immutable and decentralized features, blockchain helps record all vehicular messages and build trust for the SDN-based vehicular system to ensure reliable message transmissions and avoid fake messages from malicious vehicles. Further, in SDN, security also includes authentication in the control plane and data preservation in the data plane. Blockchain can be a solution for a decentralized security provisioning system in such scenarios \cite{100}. To improve throughput and ensure trust in vehicular SDN systems, the work in \cite{101} introduces a blockchain-based consensus protocol that interacts with the domain control layer in SDN, aiming to securely collect and synchronize the information received from different distributed SDN controllers. Specifically, in the area control layer, vehicles and link information is collected and sent to the domain control layer which operates in the distributed blockchain manner. Blockchain is able to share the model parameters of a domain controller to other domain controllers in a transactional manner to reach a consensus among multiple controllers in distributed software-defined VANET.  

Besides, blockchains also potentially address other security and networking issues caused by the centralized control concept of SDN. In fact, most network functions can be implemented by SDN applications and malicious software may cause severe damage to the SDN infrastructure. The lack of standards and guidelines for software development is also possible to pose security threats. For example, third party providers can access the network and modify control rules without the consent of SDN controllers, leading to serious data leakage risks. The work in \cite{102} uses immutable and incorruptible blockchain as a significant security mechanism for solving potential attacks in SDN such as unauthenticated access control, Denial-of-Service (DoS) attacks, SDN controller attacks and flooding attacks. Another work in \cite{103} builds a global trust assessment scheme using blockchain for SDN-based home network controllers. Users can assign a desired trust level to isolated network slices using a simplified risk assessment scale. The SDN controllers can update on the trust score of users and evaluate scores via reports which are then managed securely by blockchain in a tamper-resistant distributed manner. 

To achieve a high-efficiency fault tolerant control in SDN, the study \cite{104} employs blockchain on SDN controllers as depicted in Fig. 7. The data plane provides underlying data forwarding function which is software defined with OpenFlow protocol. In the control plane, all the controllers are connected via blockchain in a distributed manner within different control domains. At the software level, each controller in the control plane is loaded with the identical distributed ledger maintained by consensus plane, and smart contracts utilize the consistent data in the distributed ledger to provide the customized network function. The consensus plane performs multi-controller consensus for the pending-process services and inserts the results into a block data structure on a distributed ledger, while the contract plane contains smart contracts to perform automatic network functions. The blockchain-based solution is feasible to solve a number of security issues, including fault tolerance enabled by blockchain consensus, data consistency based on distributed ledger without the need of any third parties.

Moreover, the authors in \cite{105} propose a  Software-Defined Infrastructure (SDI) framework that leverages the blockchain technique along with abundant edge computing resources to manage secure data sharing and computing on sensitive data in healthcare. They focus on a blockchain-secured peer-to-peer network with SDI resources to make sure that every transaction on SDI is regulation compliant, while still providing high data interoperability. The proposed scheme is capable of performing effective authorized interactions between patients and medical applications, delivering patient data securely to a variety of organizations and devices, as well as improving the overall efficiency of medical applications. 

\subsection{	Blockchain for Network Function Virtualization (NFV)}
\begin{figure*}
	\centering
	\includegraphics [height=5.8cm, width=13cm]{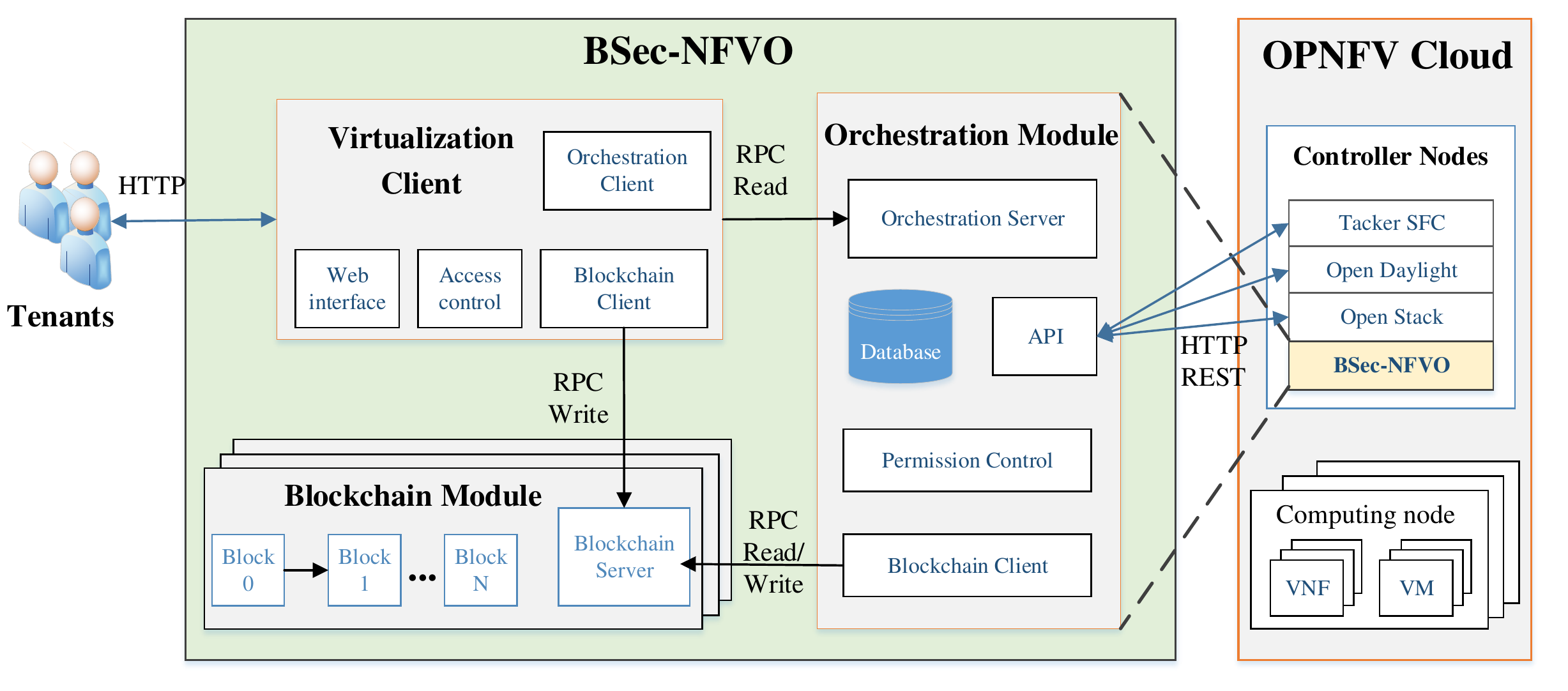}
	\caption{The conceptual blockchain-based NFV architecture.    }
	\vspace{-0.17in}
\end{figure*}
Network Functions Virtualization (NFV) is a network architecture concept, standardized by the European Telecommunications Standards Institute (ETSI) that employs standard hardware for hosting various independent and network software components \cite{106}. Basically, NFV includes three main architectural components, namely Network Function Virtualization Infrastructure (NFVI) which supports the execution of VNFs, Virtualized Network Functions (VNFs) that are the functions running on the NFVI, and Management and Network Orchestration (MANO) which cover the lifecycle management and orchestration of physical and software resources \cite{107}. NFV implements virtually Network Functions (NF) by decoupling hardware appliances (such as firewalls, gateways) from the functions that are running on them to provide virtualized gateways, virtualized firewalls and even virtualized components of the network, providing flexible network functions. In this way, the network operators can save significantly equipment costs and reduce operational expenditures as well as automate network operation tasks without concerning about hardware installation. Particularly, NFV envisions to provide a diverse number of benefits for 5G networks, including enhancing flexibility and scalability of NF deployments and connections thanks to the decoupling of software from hardware, optimizing resource provision of the VNFs for better cost and energy usage, and optimizing VNFs operations with maximum failure rate and tolerable unplanned packet loss \cite{108}. 

Network function virtualization and service function chaining, however, also incur new security challenges \cite{109}, \cite{110}. Since end-to-end service function chains may deploy NFVs in an environment involving multiple cloud providers, such data transmissions can be compromised by curious cloud entities, leading to data leakage concerns. Furthermore, in a virtualized scenario, tenants often share the same cloud infrastructure. In this context, the possibility of attacks inside the cloud can increase, which damage the transparency and accountability of service providers. In NFVs, virtualization servers can run on virtual machines (VM) to offer specific functions to execute distinct operating systems such as VM migration or resource allocation using orchestration protocols. However, the security for the communication between the orchestrator and the physical machines is a current challenge. In fact, these architectures are very sensitive to attacks that can come from different horizons. In fact, a VM can be created by an attacker to run in a server and leveraged to carry out external denial-of-service attacks. Besides, internal attacks from curious VMs are another concern which can adversely impact data integrity and confidentiality \cite{111}.  

In such a context, the blockchain technology has emerged as an efficient tool to help with these challenges. With the authenticity, integrity and non-repudiation natures, blockchain can facilitate NFV networks in three main aspects \cite{112}, \cite{113}. First, blockchain can enable reliable, easy and flexible orchestration of VNF services for better orchestration and network management. Second, blockchain can secure delivery of network functions and ensure system integrity against both insider attacks and external threats, i.e. malicious VM modifications and DoS attacks. Final, blockchain can perform data auditing and monitoring of system state during the network communication. We here review the latest advances in the use of blockchain to solve the above challenges for NFVs in 5G scenarios. 

The authors of \cite{114} propose a blockchain-based system called BSec-NFVO for secure management of service function chain orchestration operations in the Open Platform for Network Function Virtualization (OPNFV). A Practical Byzantine Fault Tolerance (PBFT) consensus protocol is employed to prevent collusion attacks without compromising latency and throughput. The architecture of BSec-NFVO is depicted in Fig. 8, consisting of three main modules: the visualization module, which provides an interface between tenants and the NFV and Service Function Chaining (SFC) services; the orchestration module, which executes instructions transmitted by tenants via the visualization module; and lastly the blockchain module that verifies and confirms transactions before execution by the orchestration module. By immutably logging all instructions that manipulate service chains enabled by blockchain, the proposed scheme can ensure authenticity, integrity and non-repudiation of instructions, which also provide data provenance and traceability in a multi-tenant and multi-domain NFV environment.

The work in \cite{115} builds a blockchain-based Virtual Machine Orchestration Authentication (VMOA) framework to secure NFV/cloud orchestration operations for better authentication of orchestration commands in the lifecycle of cloud services. Here, blockchain acts as a decentralized database ledger shared between the virtualization server, the orchestrator and VM agents.  The virtualization server is able to authenticate the orchestration command via blockchain VMOA ledger in an immutable and secure manner. Due to the removing of the requirement of third parties in the VMOA and using security features of blockchain, the proposed solution potentially achieves superior advantages such as records integrity, fault tolerance, and network trustworthiness, compared to its centralized counterparts. 

Additionally, to realize a faulty or compromised VNF configuration, the study in \cite{116} introduces a blockchain-based architecture to provide auditability to orchestration operations of network slices for securing VNF configuration updates. The prototype implements two smart contracts with specific transaction formats for safeguarding network slice management and VNF configuration operations. Especially, a Hyperledger Fabric blockchain platform associated with certificate authorities is integrated to manage digital certificates of every node, improving auditability and that only certified and authorized nodes participate in the blockchain-based NFV network.

The authors of \cite{117} introduce a scheme called BRAIN, a Blockchain-based Reverse Auction solution for Infrastructure supply in NFV scenarios for dealing with challenges of discovery and selection of infrastructures to host VNFs acquired by end users. Smart contracts are designed to achieve a trustworthy agreement between stakeholders such as users and infrastructure providers regarding resources contracted and configurations required. Meanwhile, to support efficiency and security in wireless virtualization, blockchain is proposed in \cite{118} to improve the trust and transparency among participants and stakeholders and enable more seamless and dynamic exchange of spectrum and computing resources in the 5G wireless networks. 

Another work \cite{119} presents a blockchain-based architecture for the secure configuration management of virtualized network functions (VNFs). Thanks to the immutability and the traceability features provided by blockchain and integrity and consistency of transactions ensured by a consensus protocol, the proposed solution can provide security for VNF configuration state migration, building a trust mechanism between different infrastructure providers (tenants) and VNF vendors. Asymmetric keys are employed to develop a transaction model for building anonymous authentication of tenants and VNFs and gaining confidentiality of configuration data through encryption. Such transactions are then appended in the blockchain data structure which also gives traceability and accountability of the VNF configuration updates. 

Meanwhile, to realize the orchestration/management capabilities and business support systems in the context of architectural NFV, the research in \cite{120} analyses blockchain-based Decentralized Applications (DApps) in support of multi-administrative domain networking. Blockchain can be an effective approach to establish an authentication layer for NFV Management and Orchestration (MANO) services across administrative domains. For example, blockchain can verify user access and grant access permission to resources between providers NFV-MANO components. In such a context, a smart contract can be leveraged to store access permission and assets information for MANO components as well as perform mappings of the structure of quotas, access grants and capacity of NFV users for efficient resource usage. 

\subsection{	Blockchain for network slicing}

5G offers a completely new vision of mobile networks to unify the management of IoT networks. In order to support various types of IoT applications, 5G relies on the concept of Network Slicing, which is the separation of multiple virtual networks operating on the same physical hardware \cite{121}. It enables telecom operators to portion their networks for specific services and applications, such as smart home, smart factory or vehicular network. Network slicing is well supported by Network Softwarization as the key technology enabler which consists of Virtual Network Functions (VNFs) running in the cloud inside virtual machines or containers. Each network slice contains a set of VNFs associated with physical network functions to enable network services based on the computing and storage capabilities of cloud infrastructure \cite{122}. Besides, network slicing also brings many unprecedented security challenges which consist of inter-slice security threats and the issues of resource harmonization between inter-domain slice segments \cite{123}, \cite{124}. For example, due to the design of network slice instances sharing on open cloud-based architectures, attackers may abuse the capacity elasticity of one slice to consume the resources of another target slice, which makes the target slice out of service. Further, since multiple slices have often common control plane functions, attackers can exploit this network weakness to compromise the data of the target slice by maliciously accessing the common functions from another slice, leading to serious data leakages and damage of the system integrity \cite{122}. 

In such contexts, blockchains can bring great opportunities for the security of 5G network slicing management. Blockchain can be exploited to build reliable end-to-end network slices and allow network slide providers to manage their resources. The work of \cite{125} uses blockchain for the dynamic control of the source reliability in vehicle-to-vehicle (V2V) and vehicle-to-everything (V2X) communications in vehicular network slices. In the V2X network slice operated with content-centric networking (CCN), vehicles can share securely messages (e.g., the specific messages for the management of the distributed ledger and the creation of new blockchains, including the list of trustable entities) with other nearby vehicles or roadside units via distributed blockchain ledgers. The blockchain acts as the middle-security layer between vehicles and network controllers (i.e. roadside equipment), which eliminates the need of installing additional hardware from the operator side. This not only solves trust issues thanks to no required external authorities but also improves significantly vehicular network performances with low latency and enhanced throughput. Further, the blockchain-based approach can allow for the dynamic control of resource reliability, and improved the integrity and validity of the information exchanged among vehicles in the untrusted vehicular environments. 

In order to guarantee secure and private transactions between the network slice provider and the resource provider for 5G services, blockchain is employed to build a brokering mechanism in network slicing \cite{126}. When a slice provider receives a request or query to establish an end-to-end slice, it submits this request to blockchain for tracking and sharing. To support the deployment of the sub-slice components, smart contracts are designed, called as slice smart contracts (SSCs), where each SSC specifies the essential resources needed by the sub-slice. In this way, the resource providers can perform resource trading on contracts with sub-slice components. All related information about the sub-slice deployment is immutably recorded and stored in a permissioned blockchain controlled by the slice provider. The proposed blockchain-based broker not only adds security abilities, but also supports privacy and accountability in network slicing.

The authors in \cite{127} consider a blockchain slice leasing ledger concept using the 5G network slice broker in a blockchain to reduce service creation time and enable autonomous and dynamic manufacturing process. Blockchain plays a significant role in the establishment of mutual trust relationships between the operators and management of virtual 5G network slices, enabling new end-to-end business models including the provision of connectivity or managed services for factories as well as IT infrastructure. In the same direction, the works \cite{128}, \cite{129} also present how the blockchain technology can support the resource configuration value creation micro-processes and the 5G network slice broker use case in industrial automation use and smart grid. Manufacturing equipment leases independently the network slice needed for operations on-demand, approve service-level agreement (SLA) and pay for the service fee based on actual usage. In this context, blockchain performs the network slice trading, while smart contract orders slice orchestration according to agreed SLA from a 5G network slice broker as shown in Fig. 9. 

\begin{figure}
	\centering
	\includegraphics [height=5.7cm,width=8cm]{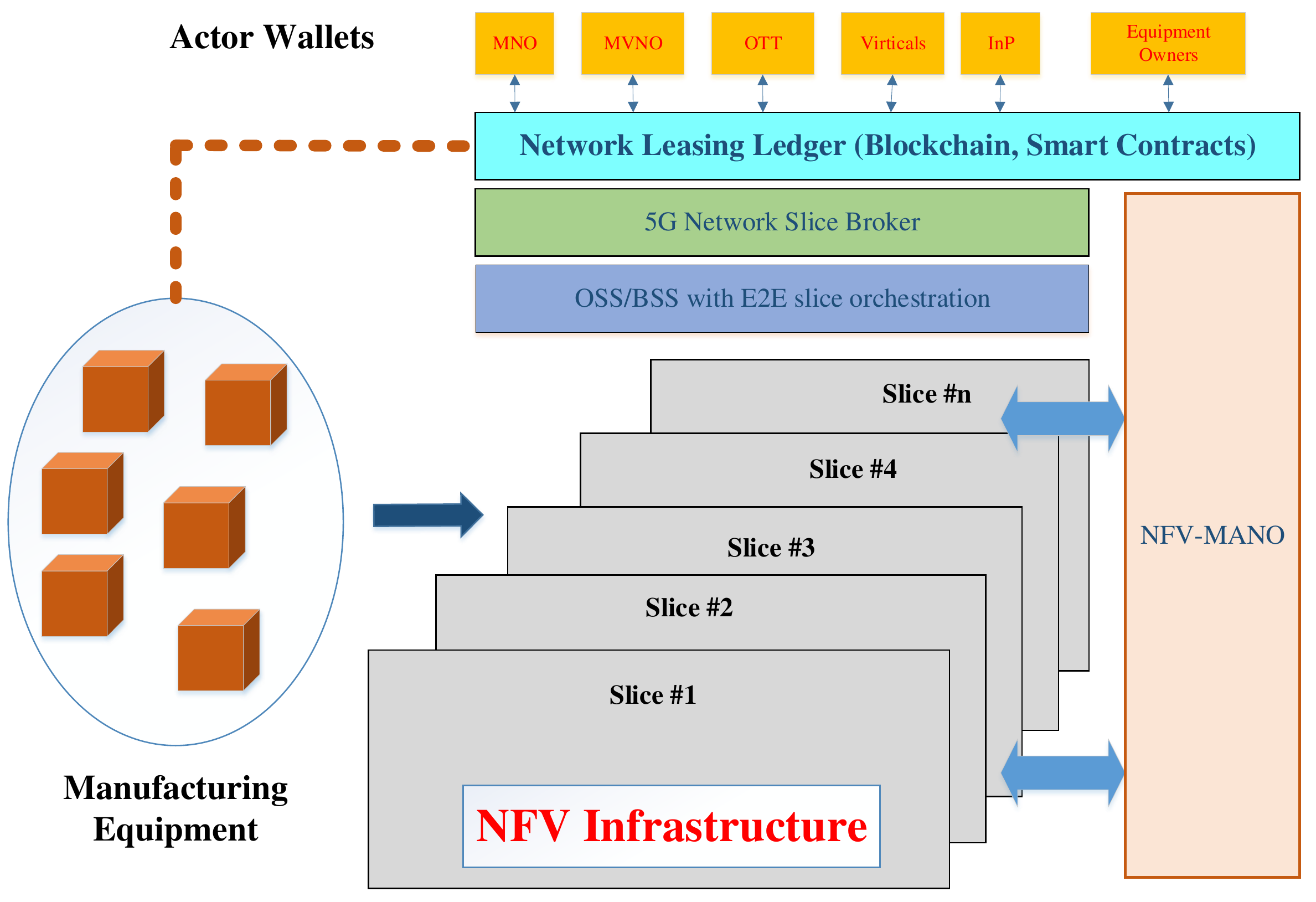}
	\caption{Blockchain for 5G Network Slice Brokering \cite{128}. }
	\label{fig11}
	\vspace{-0.15in}
\end{figure}
In an effort to virtualize the slicing network, the authors in \cite{130} propose a blockchain based wireless virtualization architecture where wireless resources such as RF channels are sliced into multiple (time/frequency) slices for mobile virtual network operators (MVNOs). Each transaction in blockchain for wireless virtualization contains information of bandwidth allocation, maximum channel power, and data rate which are used by the MVNOs when serving their users, and such a transaction is recorded immutably in the block for sharing. The blockchain based distributed scheme creates new MVNOs securely without revealing their private information to the public. Similarly, the work in \cite{131} also proposes a blockchain-based wireless network virtualization approach to optimally allocate wireless resources for wireless network virtualization where the blockchain technology helps Virtual Network Operators (MVNOs) to sublease the RF slices from trustworthy Wireless Infrastructure Providers (WIPs). Blockchain is mainly used to build a reputation-based scheme for RF allocation of network slices with the objective of minimizing the extra delay caused by double-spending attempts in NFVs. 

\subsection{	Blockchain for D2D communication}

The exponential growth of mobile 5G data traffic has given impetus to the demand for high network rate proximity services. Device-to-device (D2D) communication has been envisioned as an allied technology for such 5G scenarios \cite{132}. Conceptually, D2D communications refers to a type of technology that enables mobile devices (such as smartphone, tablet, etc.) to communicate directly with each other without the involvement of an access point or a core network of a cellular infrastructure. D2D takes advantage of the proximity of device communication for efficient utilization of available resources, enabling to improve the overall system throughput, mitigate communication delays and reduce energy consumption and traffic load \cite{133}. D2D communication thus can facilitate new peer-to-peer and location-based applications and services, making it well suitable for the next mobile 5G communication networks and services. 

However, direct communication between mobile devices also introduces new non-trivial challenges for D2D-based 5G networks in terms of security, network management and performance loss. Indeed, data sharing between devices may face risks of data leakage due to the malicious threats on the untrusted D2D environments. How to exchange mobile data to achieve low latency but ensure security is a critical challenge \cite{134}. Furthermore, D2D devices may not be trusted, and can obtain illegal access to resources on servers (i.e. edge/cloud servers) if there is no an authentication mechanism on the network. Besides, the existing D2D architectures rely on the external authorities to grant data permission and request authentication during the D2D communication, which can incur unnecessary communication latency and degrade the overall network performance \cite{135}.

Blockchain can be a good solution to help overcome such challenges to facilitate D2D communication in 5G networks. For example, the work in \cite{136} employs blockchain to build a secure content catching and sharing scheme among mobile devices for D2D networks. To mitigate the computation burden on devices, edge servers with high computing power are used to run mining puzzles for blockchain. In particular, blockchain demonstrates its efficiency in providing an incentive solution, which encourages caching-enabled users to store and share the contents with other mobile devices via D2D for better content sharing among mobile devices. The award policy empowered by blockchain stimulates the mining process in D2D devices, improving the robustness and security for the D2D network. 

In order to support the authenticity of channel state information (CSI) of mobile users in D2D underlying cellular network, blockchain is applied in \cite{137} to develop a secure mechanism using a consensus protocol. The blockchain consensus based D2D network is composed of mobile users and two blockchains, integrity chain (I-chain) and fraud chain (F-chain). The mobile users can verify and validate the received broadcast CSI messages through the consensus mechanism before signing and adding immutably to the decentralized ledgers for sharing and storage. The authors also suggest that the blockchain-based approach is potential to dramatically improve the spectral efficiency while providing efficient CSI authenticity services for D2D networks. 

Blockchain is also useful in another D2D scenario for supporting computation offloading \cite{138}. In this work, a decentralized computation offloading coordination platform is developed and empowered by the blockchain which is able to manage the computation offloading requests and perform user matching. Each mobile user can participate in the computation offloading process and submit offloading requests to the blockchain platform.  The other users in the D2D network and edge servers perform user matching to decide whether to participate in the offloading process to execute the requested computation tasks. The blockchain platform will incentivize COP which agrees to compute the task, and all request information is recorded and appended into blockchain for secure offloading management. 

The work in \cite{139} presents a delegated authorization architecture using blockchain-based smart contracts that enable users to use D2D communication to access IoT resources with respect to the preservation of the authorization information and network trust. Blockchains can immutably record hashes of the information exchanged during user authorization and payment events, while smart contracts can support for the concatenation of authorization requests. Here, smart contracts are placed on blockchain and run on all ledger nodes so that the resource access from D2D users can be handled automatically and quickly.  The authentication mechanism can also protect network resource against DoS attacks that involve a very high resource request rate. 

The authors in the works \cite{140}, \cite{141} integrate blockchain with D2D communication to support the computation and offloading of the mobile data tasks as Fig. 10. With the trust and traceability features of the blockchain, a decentralized incentive approach is introduced to foster the collaboration of content creators and D2D users without the intervention of any third party. Mobile data can be transferred securely over the D2D network via blockchain ledgers, and computation offloading and content caching can be performed by edger servers for efficient execution. 

\begin{figure}
	\centering
	\includegraphics[height=4.9cm,width=8cm]{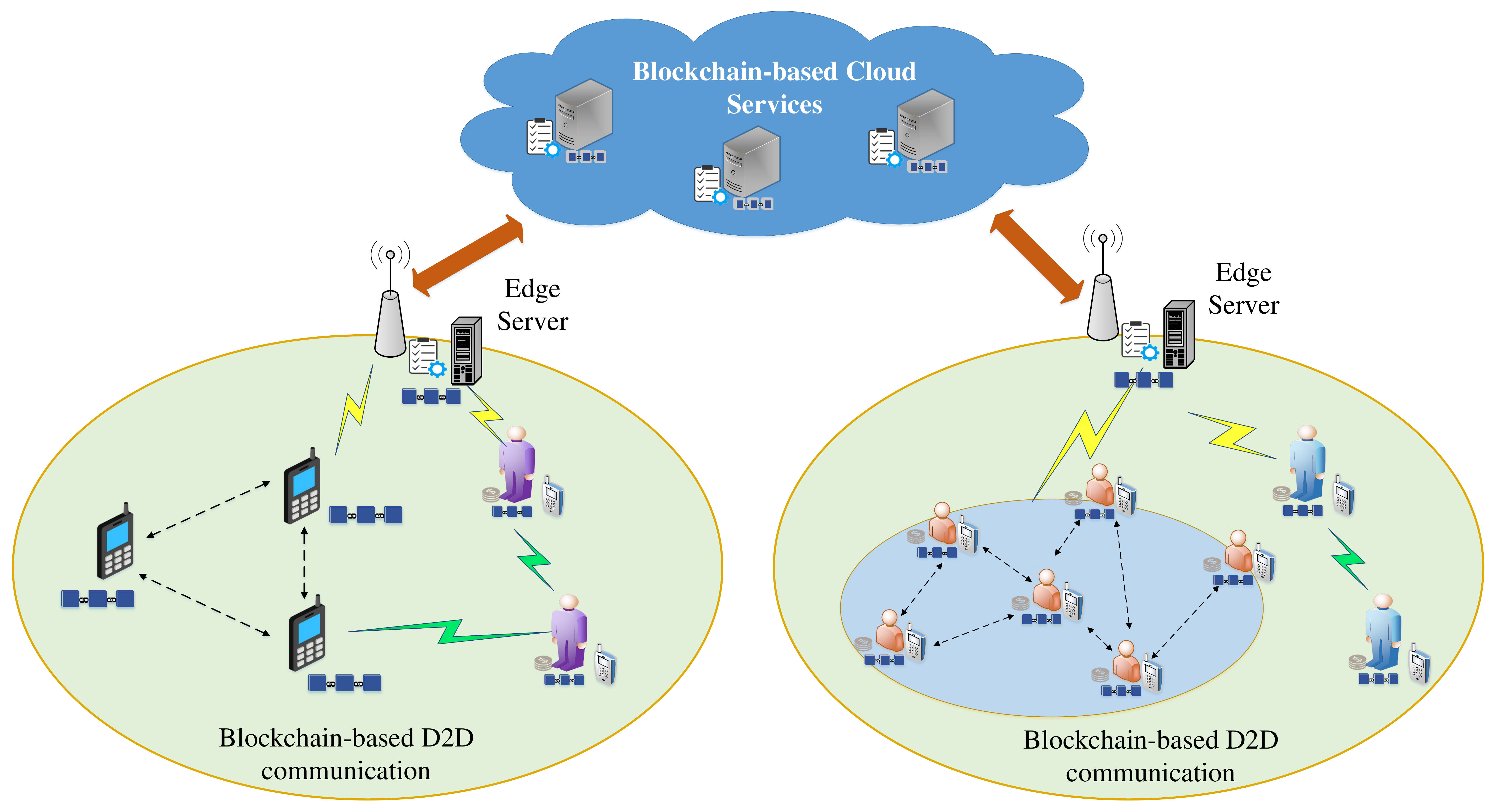}
	\caption{Blockchain for supporting D2D computation \cite{140}.}
	\label{fig11}
	\vspace{-0.15in}
\end{figure}
In \cite{142}, a consortium blockchain is considered for further security and efficiency in the feature extraction application for encrypted images in D2D systems.  Smart contracts are stored in blockchain, which solves the privacy leaking problem of image features (e.g. tempering, forging by the semi-trusted clouds). In a different direction, the study \cite{143} exploits blockchain and smart contracts for the design and implementation of a trading application between the seller and the buyer via D2D communication. The trading can be performed automatically on blockchain by triggering the contract, which ensures transparent and reliable data exchange among different users. Moreover, to build a distributed secure monitoring system in D2D systems, blockchain is also considered in \cite{144} to provide a high level of security with reduced computational and communication costs. In particular, a secure access control using blockchain is also integrated to support identity authentication in a lightweight and scalable manner.

In summary, blockchain brings numerous opportunities to support 5G technologies and provides emerging services for 5G systems. Reviewing the state of the art works, we find that blockchain can provide security, networking solutions to protect 5G services and improve the performance of 5G-based systems. In the next section, we will present an in-depth analysis and survey on the benefits of blockchain in a number of 5G services.

\section{Blockchain for 5G services}
Blockchains offer tremendous potential for improving existing 5G services and applications by supporting 5G technologies as discussed in the previous section. This vision can be achieved   by taking advantage of interesting features that blockchains offer such as decentralization, privacy, immutability, and traceability. Blockchain can be regarded as a natural choice to facilitate the next-generation mobile communication networks for better 5G services. In this section, we provide an extensive discussion on the use of blockchain for important 5G services, including spectrum management, data sharing, network virtualization, resource management, interference management, federated learning, privacy and security services. 

\subsection{Spectrum management}
With the increasing demand for bandwidth-hungry applications for 5G services such as mobile video streaming, big data processing, a foreseen network capacity shortage has become a key threat to mobile network operators (MNOs). Despite the technological achievements of 5G networks, the physical constraints such as spectrum limitations are still major limiting factors, which prevent operators from scaling their services properly. Spectrum scarcity in wireless networks hinders the fast improvement of throughput and service quality. Operators are forced to invest a large amount of money in their infrastructure to optimize the capacity by network densification and higher frequency reuse factors. Currently, MNOs have to face the challenges from the unavailability of usable frequency resources caused by spectrum fragmentation and the current fixed allocation policy, which prevents from meeting the requirements of the expanding market of wireless broadband and multimedia users \cite{145}. To deal with the desire of mobile users to be connected at all times, anywhere, and for any application, more spectrum bandwidth and/or more efficient usage of that bandwidth is urgently needed. Some solutions have been proposed, including the fixed spectrum allocation strategies, but these approaches are inefficient in terms of wasteful spectrum usage because the license holders (or primary users) do not continuously utilize their full spectrum allocation. One solution for addressing the spectrum scarcity problem in radio 5G networks is to introduce secondary users that opportunistically monitor the spectrum and then transmit their data whenever the spectrum is idle \cite{146}. However, spectrum sharing in such scenarios also raises security concerns and provides a central point of attack for malicious users. Another approach is to use certification authorities, providing provide certificates for cognitive radios inside each cell. This approach not only requires infrastructure to be implemented for each cell but also requires a protocol for defence against central-point attacks. Further, it requires greater calculation complexity and longer packet lengths, which increases overhead for spectrum sharing systems. Importantly, the use of such centralized architectures also adds single-of-failure bottlenecks when the authority is attacked or out of services, which leads to the disruption of the entire spectrum sharing network \cite{147}. 

In comparison to such conventional spectrum management schemes, blockchain can be a much better solution to overcome the security and performance issues for spectrum management in 5G. Since blockchain is a form of decentralized database where no single party has control, blockchain can be applied to build spectrum sharing and management models with improved security and better performances, i.e. low latency and enhanced throughput. Especially, blockchain envisions to support spectrum management by providing the following benefits \cite{148}.  
\begin{itemize}
	\item \textit{Decentralization:} The blockchain adoption eliminates the need of trusted external authorities such as spectrum licenses, band managers, and database managers. The inherent benefits are twofold: reducing unnecessary network overhead due to communicating with the authorities during the spectrum sharing, and improving system integrity and privacy due to no concerns about data leakage caused by curious third party intermediaries. 
	\item \textit{Transparency:} Since all transactions between spectrum users and service providers are reflected and recorded on distributed blockchain ledgers, the blockchain-based solution is able to provide better localized visibility into spectrum usage. Besides, blockchain can employ smart contracts, a self-executing platform, to perform auditability of spectrum sharing activities according to the pre-defined sharing policies. 
	\item \textit{Immutability:} The spectrum services, i.e spectrum sharing, monitoring or user payment is recorded to the only-appended blockchain in an immutable manner. By using consensus mechanisms empowered by blockchain miners, blockchain ledgers is well resistant to modifications caused by attacks or malicious users. This also ensures the reliability of the spectrum services and enhances the accuracy of the network implementation. 
	\item \textit{Availability:} Any network participants such as mobile users can access to spectrum resources managed by service providers to perform spectrum sharing and payment. Moreover, as blockchain broadcasts all service information to all entities, the spectrum sharing databases are also assessable to everyone in the network.  Furthermore, there is no central authority to verify or record the data and transactions, which potentially enables a more transparent system without a loss of security properties. 
	\item \textit{Permissionless:} Because there is no single trusted entity as the central authority to control the network, new users or applications can be added to the overall system without seeking the approval of other users, providing a flexible sharing environment. 
	\item \textit{Security:} Blockchains enable efficient communication between users and service providers with strong security capabilities against threats, DoS risks and insider attacks. 
	
\end{itemize}
In spectrum management, verification and access management is also of significant importance for enabling secure spectrum sharing \cite{149}. In this work, blockchain can secure distributed medium-access protocol for cognitive radios (CRs) to lease and access available wireless channels. Blockchain is responsible for verifying and authenticating each spectrum-leasing transactions between primary and secondary users. Here, primary users are defined as spectrum license holders and can lease their allocated spectrum to increase spectrum efficiency as well as gain profits via a spectrum coin protocol. The blockchain performs exchanging currency, mining and updating the transactions, and leasing available spectrum through an auction. The authors also demonstrated that the blockchain adoption is beneficial to spectrum management in terms of better scalability, power efficiency in spectrum usage, improved accessibility with high degree of security and better system protection capability against DoS attacks and threats.

The work presented in \cite{150} also describes a verification solution by taking advantage of blockchain for securing spectrum sharing in cognitive radio networks. The authors focus on building an auction protocol for spectrum payment services among primary users. Blockchain is regarded as a middle layer to perform spectrum trading, verify sharing transactions and lease securely the spectrum provided by a license holder. Besides, to solve the issues of privacy risks in spectrum sharing, a blockchain-based trustworthy framework called TrustSAS is presented in \cite{151} for a dynamic spectrum access system (SAS) to enable seamless spectrum sharing between secondary users (SUs) and incumbent users. The TrustSAS scheme relies on permissioned blockchains to monitor and control systems and cluster activities as well as tackle spectrum sharing events by using a Byzantine fault tolerant (BFT) consensus mechanism. All spectrum sharing transactions are validated by BFT and signed by blockchain miners for immutable recording on blocks. The experimental results show the superior advantages in terms of efficient auditability, improved privacy and lower end-to-end latency for spectrum access. 

In addition, a spectrum sensing platform empowered by blockchain has been proposed and referred to as Spectrum Sensing as a Service (Spass) \cite{152}, \cite{153}, which provide services of spectrum sensing trading and payment. Smart contract acts as the core component which is responsible for scheduling spectrum sensing among secondary users and helpers which are the nodes offering sensing service in the secondary user network. Based on operation rules defined in the contract, smart contracts also perform access verification by using a malicious helper detection mechanism to identify whether a helper is honest or malicious. The proposed solution not only maximizes the profits of MNOs to encourage spectrum provision for user applications but also guarantees security requirements in an untrusted and decentralized spectrum sharing setting. 

One of the biggest problems for unlicensed spectrum utilization is the unfair competition between MNOs for the utilization of unlicensed spectrum resources which are free to use and quite often available. To cope with this challenge, the authors of \cite{154} introduce a new unlicensed spectrum sharing among MNOs on blockchain. For this purpose, authors use smart contracts in conjunction with virtual cryptocurrency to develop a coalitional spectrum sharing game for optimizing spectrum allocation. The account balance of each MNO can be achieved fairly through a transparent sharing enabled by smart contracts, aiming to mitigate the conflict between MNOs during the sharing. To further improve spectrum sharing for sustainability in unlicensed frequency bands, the work in \cite{155} proposes to build a brokering platform to facilitate the collaboration between the network stakeholders. In this context, blockchain is feasible to establish a secure sharing to implement automatic negotiation processes for spectral resources between access point (AP) operators in a reliable manner. 

Meanwhile, in the spectrum sharing environment between the aerial and terrestrial communication systems, unmanned aerial vehicles (UAVs) has been used for facilitating communication on the sky. Currently, most UAVs in the market operate on the unlicensed spectrum (i.e., the industrial, scientific and medical bands) over the untrusted environment with significant security and privacy threats because of untrusted broadcast features and wireless transmission of UAV networks. To overcome such challenges, a spectrum blockchain architecture is considered in \cite{156} to improve the spectrum sharing. To avoid wasteful spectrum usage in UAV network, a pricing-based incentive mechanism is proposed to encourage MNOs to lease their idle spectrum to a secondary UAV network to obtain some revenue from the UAV operators. Then, a secure spectrum sharing framework is introduced where blockchain uses immutable distributed ledgers to implement spectrum exchange while protect the sharing system from threats. The authors focus on developing a Stackelberg game for an optimal spectrum sharing strategy, which can maximize the profits of MNOs while provide security services for UAV-based networks. 

\subsection{Data sharing}
One of the prominent characteristics of 5G is the strong data sharing capability in order to cope with the increasing content demands and data usage, especially in the 5G IoT scenarios. According to the latest release of Cisco \cite{157}, global mobile data traffic on the Internet will increase sevenfold between 2017 and 2022, reaching 77.5 exabytes per month by 2022. The rapid increase of content delivery over mobile 5G networks has revealed the need for new innovative data protection solutions to ensure secure and efficient data sharing over the untrusted environments \cite{158}. In fact, sharing data in mobile networks is highly vulnerable to serious data leakage risks and security threats due to data attacks \cite{159}. Mobile users tend to use information without caring about where it is located and the level of reliability of the information delivery, and the ability to control a large scale of information over the Internet is very weak. Blockchain may be an answer for such data sharing challenges. Indeed, blockchain can provide a wide range of features to improve the efficiency of data sharing in the 5G era such as traceability, security, privacy, transparency, immutability and tamper-resistance \cite{160}. To control the user access to data resources, blockchain miners can check whether the requester meets the corresponding access control policy. Due to the decentralized architecture which enables data processing for user requests over the distributed nodes, the overall system latency for data delivery is greatly reduced and the network congestion can be eliminated, which improves the performance of data sharing with blockchain. 

The problem of secure storage for data sharing is considered and discussed in \cite{161}. The authors leverage blockchain as an underlying mechanism to build a decentralized storage architecture called as Meta-key wherein data decryption keys are stored in a blockchain as part of the metadata and preserved by user private key.  Proxy re-encryption is integrated with blockchain to realize ciphertext transformation for security issues such as collusion-attack during the key-sharing under untrusted environments. In this context, the authors in \cite{162} study blockchain to develop a data storage and sharing scheme for decentralized storage systems on cloud. Shared data can be stored in cloud storage, while metadata such as hash values or user address information can be kept securely in blockchain for sharing. In fact, the cloud computing technology well supports data sharing services, such as off-chain storage to improve the throughput of blockchain-sharing \cite{163} or data distribution over the cloud federation \cite{164}.

In IoT networks, data transmission has faced various challenges in terms of low security, high management cost of data centre and supervision complexity due to the reliance on the external infrastructure \cite{165}. Blockchain can arrive to provide a much more flexible and efficient data delivery but still meet stringent security requirements.  A secure sharing scheme for industrial IoT is proposed in \cite{166}, which highlights the impact of blockchain for security and reliability of IoT data exchange under untrustworthy system settings. In comparison to traditional database such as SQL, blockchain can provide better sharing services with low-latency data retrieval and higher degrees of security, reliability, and stronger resistance to some malicious attacks (DoS, DDoS) for data sharing. Further, the privacy of data is well maintained by distributed blockchain ledgers, while data owners have full control on their data shared in the network, improving the data ownership capability of sharing models \cite{167}. 

The work in \cite{168} also introduces a sharing concept empowered by blockchain and fog computing. The proposed solution constitutes a first step towards a realization of blockchain adoption as a Function-as-a-Service system for data sharing. Fog nodes can collect IoT data arising from private IoT applications and securely share each other via a blockchain platform which can verify all data requests and monitor data sharing behaviours for any threat detection.  

Smart contracts running on blockchain have also demonstrated efficiency in data sharing services \cite{169}. Smart contracts can take the role of building a trusted execution environment so that we can establish a set of information exchange frameworks working on blockchain. For example, the study in \cite{170} leverages smart contracts to build a trustless data sharing in vehicular networks as depicted in Fig. 11. The roadside units (RSU) can set the constraints for data sharing by using smart contracts which define shared time, region scope, and objects to make sure the data coins is distributed fairly to all vehicles that participate in the contribution of data. In addition, the authors of \cite{171} introduce a smart contract-based architecture for consent-driven and double-blind data sharing in the Hyperledger Fabric blockchain platform. In the system, confidential customer data can be authorized and validated by smart contracts, and the service providers can execute the data tasks, add attributes and metadata, and submit it to the blockchain for validation and recording in a transparent manner.

\begin{figure}
	\centering
	\includegraphics[height=7.1cm,width=8cm]{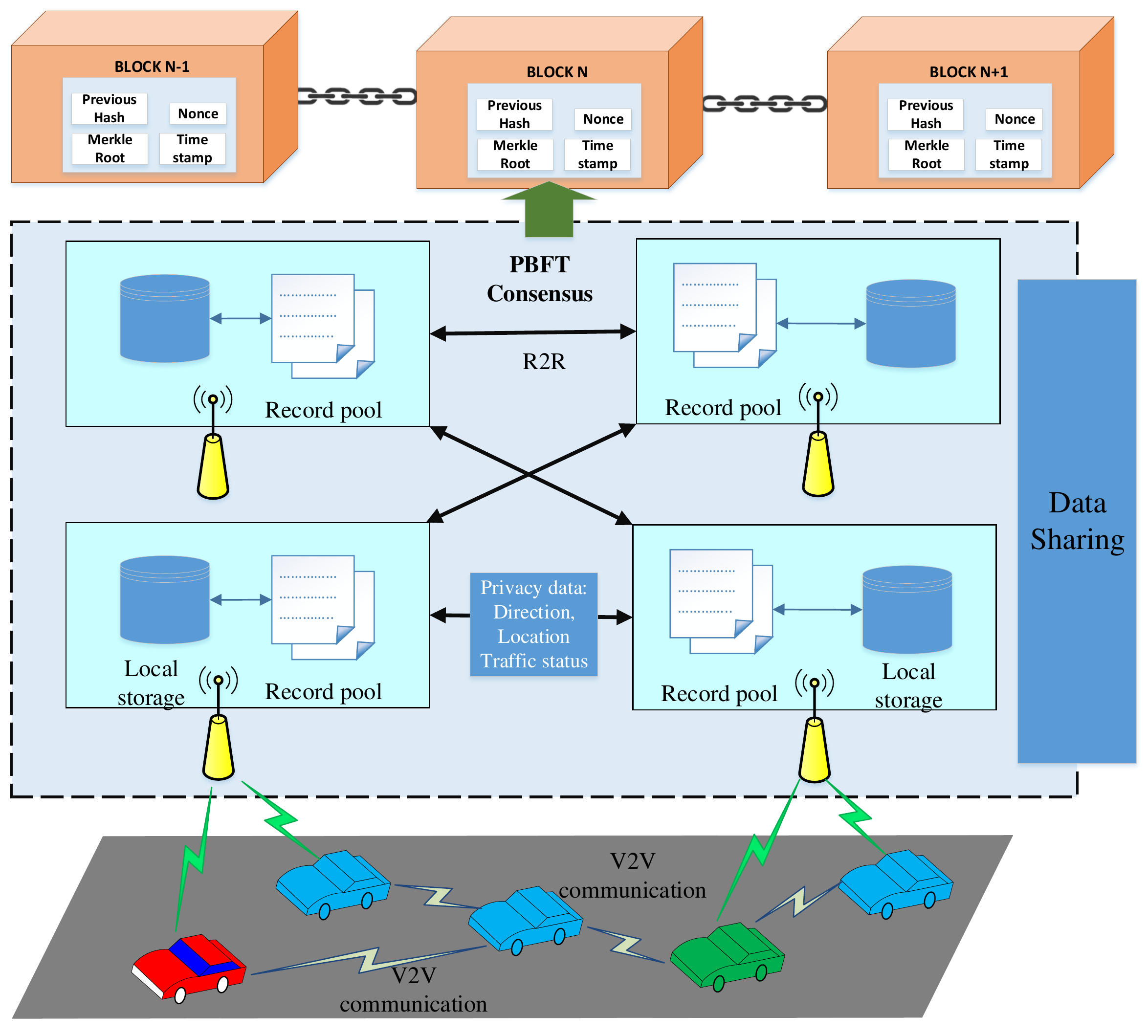}
	\caption{A data sharing model for vehicular IoT networks based on blockchain \cite{170}. }
	\label{fig11}
	\vspace{-0.15in}
\end{figure}
\subsection{Network virtualization }

Wireless network virtualization is considered as an emerging paradigm in 5G to build different virtual wireless networks (VWNs) through mobile virtual network operators (MVNOs) to support rapidly increasing data demands caused by emerging 5G IoT applications \cite{172}. Network virtualization is able to enhance wireless resource (RF slice) utilization, provide better coverage, and increase network capacity and energy efficiency \cite{173}. The blockchain technology can provide the required characteristics of nonrepudiation and immutability to overcome the shortcomings of the previous configuration models. More precisely, blockchain is capable of creating secure virtual wireless networks (VWNs) so that wireless resource-owners sublease their wireless resources (e.g., slice of RF spectrum, infrastructure) to mobile virtual network operators (MVNOs) \cite{130}. All participants of each virtual network slice is managed by a slice blockchain, which provides auditability of slice creation, monitors orchestration operations and data access of clients to the data centre. In such a decentralized virtual network, smart contracts can be very useful to provide automation and transparency in a distributed way instead of trusting a particular node or an authority process transactions. The solution of using blockchain and smart contracts can be an ideal solution to create secure end-to-end network slices for supporting virtual services with diverse requirements and resiliency \cite{116}.

Meanwhile, the work in \cite{115} proposes blockchain to secure virtual machine orchestration operations for cloud computing and network functions virtualization systems. The main objective is to protect and secure virtual machines and make virtual machine managers well resistant to be compromised by threats. In fact, the complexity of virtual networks with multiple physical machines and virtual machines raises security concerns to be solved. For instance, a virtual machine can be created virtually by an external attacker to run in a server and used to perform external DDOS attacks, and   internal attacks can act as legitimate entities to perform unauthorized data access which can impair the data integrity and confidentiality of the network. Therefore, the proposed work considers the authentication issues in virtualization using a blockchain system shared between the virtualization server, the orchestrator and VMM agents. The orchestration requests (create, destroy, resize, copy, migrate) to a virtualization server are recorded as a transaction which is then authenticated by smart contracts to grant permission for the orchestration command, avoid malicious access to the data centre. 

Moreover, in order to prevent from double-spending of same RF resources (frequency slices), the work in \cite{174} leverages a distributed blockchain based scheme to sublease the frequency slice to MVNOs through wireless network virtualization. The proposed wireless virtualization architecture contains three main entities: wireless service providers who participate in sharing or subleasing their wireless resources to MVNOs; data sharing services for wireless resources; and block managers that are trusted devices working to maintain the blockchain. Each transaction in blockchain for wireless virtualization includes the information of bandwidth allocation, allocated channel power, data rates which are utilized by the MVNOs while serving their users through virtual networks. Specially, the work pays special attention to addressing the double-spending issue which is the allocation of same wireless resources to multiple MVNOs with a hope that all MVNOs would not use their leased spectrum at the same time for obtain maximum revenues. Compared to traditional approaches which mainly rely on centralized trusted authorities to perform resource sharing, blockchain is much more efficient in verifying each transaction to ensure that the wireless resources are scheduled to a given MVNO, which not only solves double-spending problems but provides fairness and transparency for network virtualization. 

In an effort to secure management, configuration and migration of virtual networks services, the work in \cite{119} presents a blockchain-based architecture for network function virtualization (NFV) and service function chaining (SFC). The blockchain module designed mainly performs three main functions: verify the format of the transaction, validate the accuracy of the signature of the transaction, and check the duplication of transactions. The service requests sent from NFV clients would be verified by blockchain via VNF key pairs and blockchain module key pairs for authentication through a consensus mechanism. In the same direction, the authors of \cite{114} also analyse on how blockchain can support secure configuration and migration of NFVs. The consensus of Practical Byzantine Fault Tolerance (PBFT) is implemented on the Open Platform for Network Function Virtualization (OPNFV) to monitor and schedule the orchestration operations in virtualized networks. 

Furthermore, the security for SDN-based network virtualization is analysed in \cite{118}, and that is based on blockchain to enable privacy of spectrum resources. Here, blockchain is installed in the SDN controller of MVNOs to perform subleasing (or releasing) wireless recourses to virtual wireless network operators (VWNOs). Blockchain is able to offer auditability such that each spectrum assignment done by SDN controllers of PWROs is validated by other participants, with each allocation is recorded as a transaction in the blockchain with a timestamp. 

\subsection{	Resource management}
In 5G networks, mobile resource (i.e. computation, memory, bandwidth, channel and storage) is one of the most popular services. The growing variety of 5G services leads to unprecedented levels of complexity in mobile resource management \cite{175}. Edge/cloud computing in 5G needs to allocate its computation capacity to ensure efficient data execution while maintaining resources to serve the increasing demands of mobile users in the long term. In virtualized networks, the VNFs of a single slice may have heterogeneous resource requirements, i.e., CPU, memory, bandwidth and storage, depending on their functions and user requirements. The resource demands of slices of the same function type may be also different since they are serving different number of mobile users. For instance, a provider might run multiple Internet of Things (IoT) slices each one dedicated for a specific application. In such contexts, with heterogeneous resource capacities and heterogeneous resource requirements, implementing an optimal resource allocation to the mobile 5G network is a critical challenge. Importantly, the current resource management architectures mainly rely on a central authority to perform resource allocation and verification of user resource access, but such models obviously remain single point failure risks and security from the third party. Moreover, the traceability of the current resource sharing schemes is very weak, which makes shared resources being compromised by attacks or used illegally by malicious users. All of these issues need to be solved effectively before deploying 5G services in practical scenarios. 

Blockchains can be a highly efficient approach to solve the above remaining issues and improve the resource management. The use of blockchain enables the distributed resource allocation schemes as a strong alternative which is more preferable for both the service providers (edge/cloud, slice providers) and also mobile users/equipments. Blockchain would simplify the resource management concept, while remaining the important features of the core network and ensure strong security. For example, blockchain has been applied in VNFs in \cite{117} to implement reliable resource allocation corresponding to user requests from different aspects such as user demand, cost. More interesting, smart contracts are also integrated to build an auction scheme which enables to allocate optimally to the network of users in a transparent manner (due to the transparency and immutability of smart contracts) in dynamic mobile environments. 

Spurred by the power of blockchain, a resource management model is introduced in \cite{176} which proposes a new concept of blockchain radio access network (B-RAN). The main goal is to achieve a spectrum resource balance in the network of user equipment (UE), access points (AP), spectrum bands and blockchain. The resource access services between UE and AP can be implemented by a smart contract-enabled protocol which defines access rules in conjunction with certain resource constraints such as service time, service demand, and service fee. The service requestor, i.e. mobile user, can undertake resource trading with AP by triggering the smart contract so that spectrum access is authenticated and resource is released via blockchain. 

In the 5G networks, edge computing plays a significant role in improving QoS of mobile services thanks to its low latency and fast computing capabilities. Resource allocation for edge computing is of significant importance in edge-based mobile networks, such as IoT for better QoS and robustness of the system. A study in \cite{177} employs blockchain to develop a decentralized resource allocation scheme which overcomes the limitation of previous centralized schemes in terms of latency and service provision speed. To provide adaptive computation services for IoT data, resource allocation should be dynamically adjusted without any centralized controller to maintain the high QoS. Blockchain is well suitable for such scenarios by offering a distributed ledgers to update resource information in an automatic and trustworthy manner \cite{178}. In the case of resource scarcity in the network, a cooperative edge computing model can be necessary to support low-capability edge devices \cite{179}. In this regard, blockchain would be useful to provide a reliable resource sharing between edge nodes. Resource requests can be verified strictly by intelligent contracts with access policies without passing a centralized authority, which also reduces resource sharing latency.  

Another blockchain-based resource allocation architecture for edge computing is presented in \cite{180}. In this work, a three-stage auction scheme is introduced, including the blockchain miners act as the buyers, the edge servers which provide resources act as the sellers, and a trusted third party auctioneer that undertakes the resource trading. Blockchain is responsible to monitor resource trading and user payment between miners and edge servers. The experimental results also show that the blockchain-based solution is truthful, individual rational and computationally efficient, compared to the conventional approaches. 

In the multi-user network, a critical challenge is to allocate fairly the wireless resources among users with respect to their priorities (i.e., emergency levels). For example, a user who needs more resources for his service demand should be allocated more resources from the provider. Without authenticating the users’ priorities can lead to insufficient wireless resources to the users who are actually in high priorities. To provide a dynamic resource allocation solution for optimal resource usage, the work in \cite{181} presents a blockchain consensus protocol to check the authenticity of priorities. Each mobile user can take the role of a blockchain node to perform authenticity for a new message or request. The resource level is decided by an asynchronous Byzantine agreement among nodes, which guarantees trustworthiness and fairness for resource sharing. 

\subsection{Interference management}

The problem of interference management in the 5G infrastructural wireless networks is expected to become critical due to the unexpected data content traffic and numbers of 5G IoT devices. Although the telecom operators provide mobile services with the implementation of small size networks which can deliver various advantages such as high data rate and low signal delay, it is likely to suffer from various issues such as inter-cell, intra-cell, and inter-user interferences \cite{182}. In the data-intensive service scenarios where a huge amount of mobile data is required to be transmitted in cellular networks, D2D communication can be a good choice to implement low-latency data transmission. However, the coexistence of D2D devices and cellular users in the same spectrum for communication and the short distance between D2D devices and users in small cells can result in cross-tier interference (CTI). The possibility of collaborating communication and sharing service benefits between mobile devices can be infeasible in practice due to the interest conflict between them. Building a fair and trusted economic scheme can be a solution to this problem, and thus mitigate the network interference. Currently, electronic money transactions have received extensive attention in small cell deployments, but the transaction consensus is often reached by passing a central authority \cite{183}. This approach not only incurs additional costs of latency and transmission energy, but also raises security concerns from third parties. Distributed interference management with blockchain would be a feasible approach to cope with such challenges and facilitate interference management. 

For example, the authors in \cite{184} present a first example of using distributed blockchain to support a linear interference network. The main objective is to build a monetary mechanism using blockchain for optimal interference management. More precisely, a network model for a pair of two nodes including a transmitter and receiver is considered, wherein the transmitter (payer) may cause interference at the receiver (payee). A distributed interference avoidance transmission strategy is proposed so that a node has to pay in order to be active and then maximizes its monetary credit. The blockchain implementation realizes the monetary policies for cooperative interference management using a greedy algorithm. The proposed strategy also relieved that blockchain can help allocate economic benefits among users for interference avoidance \cite{185}.

In the D2D networks, interference may incur from the unfair resource allocation from the service providers to different user types. For example, users with higher spectral resource demands should be prioritized during resource scheduling. Motivated by this, a blockchain consensus method is proposed in \cite{137} to evaluate the amount of cross-tier interference (CTI) caused by each user. The authors pay special attention to building an access control mechanism using blockchain for the authenticity of channel state information (CSI) with a dynamic resource allocation. A user with higher CSI can be allocated a larger amount of wireless resource. A simulation implementation with an optimal user access algorithm is also presented, showing that the proposed scheme can improve the spectral efficiency for D2D users without interference effects. 

The study in \cite{186} utilizes power control with blockchain to support Quality-of-Service (QoS) provisioning for enabling efficient transmission of a macrocell user (MUE) and the time delay of femtocell users (FUEs) in blockchain-based femtocell networks. The macrocell base station (MBS) shares its spectrum resource to FUEs and the co-channel interference can be caused by the FUEs. In order to avoid excessive interference from FUEs, MBS can price the interference to the FUEs, the FUEs determine their transmission powers and payments with the constraint of time delay according to a modelled Stackelberg game. Blockchain is essential to build a decentralized femtocell network so that payment can be done in a reliable way without the involvement of a middle party. 

In another scenario, the interference between IoT transaction nodes (TNs) in the blockchain-enabled IoT network is also analysed in \cite{187}. In this work, the authors focus on investigating the performance of blockchain transaction throughput and communication throughput by deriving the probability density function (PDF) with respect to the interference of TNs, for a transmission from an IoT node to a blockchain full function node. The blockchain-based solution is able to ensure high successful rate and overall communication throughput and preserve the IoT network against security threats.

Despite great research efforts in the field, the use of blockchain for interference management in 5G mobile networks is still in its infancy with few investigated works. The preliminary findings from the literature works are expected to open the door for exploring blockchain in overcoming the challenges in network interference management in terms of network throughput and security. 

\subsection{	Federated learning }
Recent years, federated learning has emerged as a promising machine learning technique for large-scale mobile network scenarios \cite{306}, \cite{307}. Federated learning enables distributed model training using local datasets from distributed nodes such as IoT devices, edge servers but shares only model updates without revealing raw training data. More specific, it employs the on-device processing power and untapped private data by implementing the model training in a decentralized manner and keeping the data where it is generated. This emerging approach provides an ability to protect privacy of mobile devices while ensuring high learning performance and thus promises to play a significant role in supporting privacy-sensitive 5G mobile applications such as edge computing and catching, networking, and spectrum management \cite{307}.
In particular, the cooperation of blockchain and federated learning has been considered in recent works to solve complex issues in mobile 5G wireless networks. The authors in \cite{308} introduce a blockchained federated learning (BlockFL) architecture which enables on-device machine learning without any centralized training data or coordination by employing a consensus mechanism in blockchain. By relying on the decentralized blockchain ledger, the proposed model overcomes the single point of failure problem and enhances the network federation to untrustworthy devices in a public network due to federated validation on the local training results. Besides, the blockchain also accelerate the training process by a reward mechanism, which in return promotes the collaboration of ubiquitous devices. 

The study in \cite{309} considers a reputation scheme which selects reliable mobile devices (workers) for federated learning to defend against unreliable model updates in mobile networks. To ensure accurate reputation calculation, a consortium blockchain with the properties of non-repudiation and tamper-resistance is leveraged to create a secure decentralized model update network of edge servers and mobile devices, leading to the reliability of federated learning on mobile edge computing. Importantly, blockchain associated with contract theory enables an incentive mechanism, which stimulates high-reputation workers with high-quality data to join the model training for preventing the poisoning attacks in federated learning \cite{310}. 

Meanwhile, the authors in \cite{311} incorporate blockchain with federated learning in the determination of data relevance in mobile device networks. This can be done by encourage mobile users to aggregate relevant information belonging to a specific topic that they are seeking during the interaction process with other users. They also introduced a decentralized way of storing data which reduces the risk from centralized data storage. A consensus mechanism called the Proof of Common Interest is considered that provides data verification services to ensure that data that is added to the blockchain ledger is relevant. 

To provide a parallel computing architecture for big data analysis, especially for the precision medicine which data sets are owned by healthcare data users, an integrated blockchain-federated learning model is proposed in \cite{312}. Federated learning assists training large medical data sets from various distributed data sources owned and hosted by different hospitals, patients, and health service providers, while blockchain-empowered smart contract is used to enable a distributed parallel computing environment for distributed deep learning using heterogeneous and distributed data. Moreover, the blockchain adoption enables secure, transparent, and auditable data sharing to promote international collaboration.

The work in \cite{313} considers a blockchain empowered secure data sharing architecture for distributed devices in Industrial Internet of Things (IIoT). The key focus is on building a data sharing with privacy preservation by incorporating in federated learning. By using the power of federation of IoT devices, the data privacy is ensured via the federated learning model which allows to share the data model without revealing the actual data. Further, to enhance the data integrity of the data training, the federated learning is integrated with the consensus process of permissioned blockchain, which also ensures secure data retrieval and accurate model training. 

\subsection{	Privacy }

In addition to smart emerging services that 5G can provide to mobile users and stakeholders, the complex 5G mobile environments also raise many privacy issues to be investigated carefully. According to a survey work in \cite{49}, the privacy challenges in 5G come from various aspects, such as end-to-end data privacy, data sharing privacy, trust issues in information flows, and trust issues in centralized mobile data architectures with third parties. Blockchain with its decentralization, traceability, availability and trust capabilities has been demonstrated widely its great potential in solving privacy issues in 5G networks and services \cite{20}. As an example, blockchain is feasible to protect user data for decentralized personal data management \cite{188}, which enables to provide personalized services. Laws and regulations for data protection could be programmed into the blockchain so that they are enforced automatically. Interestingly, blockchain is capable of providing  full control of monitoring personal data when sharing on the network, which is unique from all traditional approaches which hinder users from tracking their data \cite{12}. 

To provide decentralized and trusted data provenance services on cloud computing, the work in \cite{189} uses blockchain to provide tamper-proof records and enable the transparency of data accountability. Blockchain can support in three steps, namely provenance data collection, provenance data storage, and provenance data validation. Data provenance record is published globally on the blockchain, where blockchain nodes (i.e. mobile users, data owners, and service providers) can participate in consensus for confirmation of every block.  During the data sharing between users and service providers, transmitted data can be highly vulnerable to malicious threats, i.e. data attacks, then privacy for shared data should be considered carefully. In this context, the authors in \cite{190} presented a blockchain-based solution for secure data exchange. Data can be recorded in blocks and signed by miners so that sharing is securely implemented. An automated access-control and audit mechanism is considered wherein blockchain enforces user data privacy policies when sharing their data across third parties for privacy preservation \cite{191}. 

In current IoT applications, the private information management often relies on centralized databases owned by third-party organizations for data services such as data processing, data storage, data sharing. However, it is easy to find that this architecture remains weaknesses in terms of data leakage coming from curious third parties and high communication latency due to such centralized models. A privacy architecture using blockchain for smart cities is presented in \cite{192}, focusing on solving the above issues. Blockchain has the potential to help mitigate privacy exposure while allowing users to benefit from trusted transactions and better data control. The records of data access are added to a transparent ledger so that blockchain with consensus mechanism can verify and validate the data requests from all users to detect any potential threats in a decentralized manner without the involvement of any third parties. In another research effort, the work in \cite{193} investigates how blockchain can support secure data storage and data availability in IoT health networks. With the combination of the cryptographically secured encryption and the common investment of the network peers via a consensus mechanism, blockchain empowers a decentralized and openly extendable network while protecting data on the network. 

A privacy-preserved scheme empowered by blockchain is also considered and discussed in \cite{194}. In this work, a consortium blockchain-oriented approach is designed to solve the problem of privacy leakage without restricting trading functions in energy networks. Both energy users and suppliers are verified by a trading smart contract so that all trading transactions are authenticated for trustworthiness. Moreover, to achieve good privacy in industrial IoT, the study \cite{195} introduces a decentralized blockchain architecture in conjunction with a hash algorithm and an asymmetric encryption algorithm. IoT data are still stored by the offline database (i.e. cloud storage), and the access record (storage, reading, and control) of each entity is stored in the block for tracking. Therefore, data storage on blockchain can be solved efficiently, and each operation will be strictly supervised via blocks.  

In dealing with privacy issues in vehicular networks, the authors of \cite{196} present a privacy-preserving authentication framework. The main goal of the proposed system is to preserve the identity privacy of the vehicles in the vehicular ad hoc networks. All the certificates and transactions are recorded immutably and securely in the blockchain to make the activities of vehicles (i.e. data sharing, energy trading) transparent and verifiable. In a similar direction, a model called CreditCoin for a novel privacy-preserving incentive announcement solution is presented in \cite{197}. On the one hand, by offering incentives to users, CreditCoin can promote data sharing for network expansion, and the transactions and account information of blockchain are also immutable and resistant to be modified by attacks. On the other hand, with a strongly linked ledger, the blockchain controller can be easy to trace user activities, including malicious behaviours, for data protection. 

In addition, the work in \cite{198} proposes to use private smart contracts to design a privacy-preserving business protocol in e-commerce. In the contract, the interaction policy is defined via a business logic that determines types of trade, counterparties, underlying assets, and price information of the online shopping. The transactions between the seller and the buyer can be implemented securely and transparently via the contract without the disclosure of private information. Recently, the blockchain benefit to privacy of machine learning algorithm implementation is investigated in \cite{199}. A privacy-preserving and secure decentralized Stochastic Gradient Descent (SGD) algorithm is established on blockchain, which enables computation in a decentralized manner in computing nodes. Computation parameters and information are kept in the block without revealing their own data and being compromised by data attacks. Obviously, the blockchain technology is promising to privacy preservation in the modern mobile networks and services, especially in 5G IoT systems, where data protection is becoming more important in the context of exponential mobile data growth in the 5G era \cite{200}.

\subsection{Security services}
The rapid increase of the 5G traffic and the explosive growth of valuable data produced by user equipment have led to strong demands for security mechanisms to protect mobile data against threats and attacks. With the important security properties, blockchain can provide a number of security services for 5G to improve the overall performance of future mobile systems.  Considering the state of the art literature \cite{20}, blockchain mainly offers three main security services, including access control, data integrity and authentication, which will be summarized as follows.

\subsubsection{	Access Control}

Access control refers to the ability of preventing the malicious use of network resource. Access control mechanisms guarantee that only legitimate users, devices or machines are granted permissions (e.g., read, write, etc.) the resources in a network, database, services and applications. Blockchain, especially smart contracts can offer access control capability to protect the involved system against any threats. As an example, a trustworthy access control scheme leveraging smart contracts is introduced in \cite{201} to implement access right validation for IoT networks. The access policy is predefined and stored in the contract, which runs on blockchain. The contract can verify the user request using such a policy in a dynamic and decentralized manner. Different from traditional access control architectures which always use external authority for verification, the blockchain-based approach can perform direct access control between the requestor and the data centre so that the access latency can be reduced and security is improved.

To achieve access control for user requests to data resources in fog cloud-based IoT networks, a privacy-oriented distributed key management scheme using blockchain is proposed in \cite{202} to achieve hierarchical access control. To receive a permission grant for data access, a subject needs to send a request with access information (i.e. identification, user address) to the security manager which checks the access and broadcast this request to other entities for verification via blockchain. The access is granted only when a consensus is achieved among all entities, which enhances reliability of the access control architecture. 

To overcome the challenges caused by complicated access management and the lack of credibility due to centralization of traditional access control models, the authors in \cite{203} introduce an attribute-based access control scheme. The ultimate goal is to simplify the access management by a distributed blockchain ledger while providing efficient access control ability to safeguard IoT data resources. Moreover, the work in \cite{204} introduces a combination of Ethereum blockchain and ciphertext-policy attribute-based encryption (CP-ABE) to realize fine-grained access control for cloud storage. An access control policy is programmed in a smart contract which verifies the request based on the access period time and the attributes of data users. All information of control functionality results is stored on the blockchain, so the access control is visible to all users. 

Meanwhile, a transaction-based access control scheme based on blockchain is proposed in \cite{205}. The access verification follows a four-step procedure: subject registration, object escrowing and publication, access request and grant. Each request of the subject is registered as a transaction that is then submitted to blockchain to be validated by the data owner on blockchain by suing a Bitcoin-type cryptographic script. The works in \cite{206}, \cite{207} also investigate the capability of blockchain for realizing access control services with Ethereum and Hyperledger Fabric platforms. To perform access control in the large-scale IoT networks, a platform called BlendCAC is considered in \cite{208} as a promising solution for securing data sharing and resource trading among devices, users and service providers. The proposed approach concentrates on an identity-based capability token management strategy which takes advantage of a smart contract for registration, propagation and revocation of the access authorization. 
\subsubsection{	Data integrity}

The integrity property ensures that the data is not modified in the transit or data is intact from its source to the destination. In recent years, distributed blockchain ledgers are starting to be used to verify data integrity for mobile services and networks, such as data management services or IoT applications, to overcome the limitations of the traditional models, which often rely on a third party auditor for integrity validation \cite{209}. A blockchain-based framework for data integrity service is also presented in \cite{210} which performs integrity verification based on blockchain for both data owners and data customers. To operate the data integrity service, a smart contract living on the blockchain is employed to audit transactions from all users. Upon the deployment of smart contract, participants can interact with it anytime, the integrity service cannot be terminated by any entities except the author. The blockchain store information of data history and database stored in blockchain is strong resistant to modifications, which improves data integrity. 

To provide data integrity services on resource-limited IoT devices, the authors in \cite{211} introduce a lightweight integrity verification model in Cyber-Physical Systems (CPS) by taking advantage of blockchain features. The key concept of the proposal is enabled by a three-level design, including the first level for running the Proof-of-Trust (PoT) mechanism among IoT devices, and two upper levels for data persistence and integrity verification by cloud. The implementation results reveal the efficiency of the blockchain-empowered model with good confidentiality, availability, integrity, and authenticity for IoT communication.

In an effort to deal with challenges caused by centralized traditional data integrity schemes such as symmetric key approaches and public key infrastructure (PKI) which often suffer from the single point of failure and network congestion, a decentralized stochastic blockchain-enabled data integrity framework is analysed and discussed in \cite{212}. The proposed stochastic blockchain design includes the chain structure and the consensus mechanism for the data integrity checking procedures. 

At present, with the popularity of cloud storage, how to guarantee data integrity on the cloud has become a challenging problem. The authors of \cite{213} describe a framework for data integrity verification in P2P cloud storage via blockchain which makes the verification process more open, transparent, and auditable to all data users. Moreover, a new solution for improving integrity on cloud is introduced in \cite{214}. In the system, blockchain constructs a semi-finished block on a candidate block arranged by data packages that is broadcast to all entities, while the consensus mechanism in blockchain, i.e Proof of Work, is able to generate tamper-resistant metadata associated with policy-based encryption method, leading to better data integrity. Besides, to tackle the issue of verification delay caused by procrastinating third-party auditors, the study \cite{215} implements a solution for cloud storage using blockchain which enables the auditors to record each verification result into a blockchain as a transaction with a stringent time requirement. The time stamp in conjunction with signature and hash values can provide a time-sensitive data integrity service with a high degree of system security. 
\subsubsection{	Authentication}

Recent years, blockchain has been also investigated to realize the authentication capability to improve the overall security levels of 5G networks \cite{65}. Mobile user access needs to be authenticated to detect and prevent any potential malicious behaviours to network resources (i.e. database, computing resources), which preserves the involved system and enhances the network robustness. In \cite{216}, a privacy-enhancing protocol is proposed by using the blockchain technology. The approach provides an ability to identify users by the evaluation on personal information which is extracted from the user request package. The smart contract is also integrated to perform authentication, aiming to prevent unauthorized access from attacks. 

In our recent works \cite{217}, \cite{218}, blockchain-based smart contracts are also leveraged to build an authentication mechanism for the cooperative edge IoT networks. By forcing an access control policy, smart contracts are able to identify and verify the user access for authentication. Only users with access grants can gain permission for their functionality, i.e. data offloading to edge servers.

The authors in \cite{219} consider an authentication scheme using blockchain for fog computing. The fog nodes running on Ethereum blockchain employ smart contracts to authenticate access from IoT users. The proposed scheme facilitates managing and accessing IoT devices on a large scale fog network while providing security features such as decentralization, privacy and authentication without the need of a trusted third party.

In order to achieve authentication in vehicular networks, a strategy working on the blockchain platform is proposed in \cite{220} which can undertake vehicle authentication and privacy preservation with seamless access control for vehicles. Blockchain can bring more advantages than conventional approaches using third party auditors in terms of high trust degree and transparency. Another blockchain application for privacy-awareness authentication is shown in \cite{221}, which allows both the server and the user to authenticate each other through this credential or certificate in a decentralized manner. All entities in the network achieve a consensus on an authentication task, and any potential threats can be detected and reflected on decentralized ledgers for necessary prevention. 

\section{	Blockchain for 5G IoT applications}
Nowadays, Internet of Things (IoT) have constituted a fundamental part of the future Internet and drawn increasing attention from academics and industries thanks to their great potential to deliver exciting services across various applications. IoT seamlessly interconnects heterogeneous devices and objects to create a physical environment where sensing, processing and communication processes are implemented automatically without human involvement. The evolution of the 5G networks would be the key enabler of the advancement of the IoT. A number of key enabling 5G technologies such as edge/cloud computing, SDN, NFV, D2D communication are developed to facilitate future IoT, giving birth to a new model as 5G IoT, which is expected to disrupt the global industry \cite{222}, \cite{223}. Especially, in recent years, blockchain has been investigated and integrated with 5G IoT networks to open up new opportunities to empower IoT services and applications \cite{224}. Reviewing the literature works, we find that blockchains mainly support some key IoT applications, namely smart healthcare, smart city, smart transportation, smart grid and UAVs, which will be highlighted as follows. 
\subsection{Smart healthcare}
Healthcare is an industrial sector where organizations and medical institutions provide healthcare services, medical equipment, health insurance to facilitate healthcare delivery to patients. The emerging 5G technologies are potential to support smart healthcare applications, which fulfill the new requirements for healthcare such as improved QoS, better density and ultra-high reliability \cite{225}. The integration of blockchain with 5G technologies can advance current healthcare systems and provide more performance benefits in terms of better decentralization, security, privacy \cite{226}, service efficiency and system simplification for lower operational costs \cite{168}.  Blockchain can incorporate with 5G technologies such as softwarization, cloud/edge computing for new smart healthcare services \cite{227} as depicted in Fig. 12.  The softwarized infrastructure can perform network functions through NFVs, which promote IoT communication, while cloud computing can support fast healthcare delivery services for early detection of patient health conditions. In such a 5G healthcare scenario, blockchain is employed to build a peer-to-peer database system which can validate and record all transactions (i.e. healthcare request, patient data) and store immutably them in decentralized ledgers. All transaction blocks are also visible to healthcare network members, including doctors, clinicians, and patients to accelerate data sharing during medications and treatment processes. 

Blockchain is also integrated with SDN-based healthcare networks \cite{228} for healthcare networking and computing. A software-defined infrastructure is designed to facilitate the specification of home-based healthcare services, and a cloud edge model is considered to provide a flexible heterogeneous health computation services. The role of blockchain in this work is to deal with health data interoperability and security issues, such as enabling effective authorized interactions between patients and healthcare providers (doctors, insurance companies), and delivering patient data securely to a variety of organizations and devices. Also, an access control mechanism empowered by smart contracts is integrated to support secure data sharing through user access verification, aiming to prohibit unauthorized users or threats from malicious access to health data resources. 

A healthcare architecture based on D2D communications can a notable solution for efficient information sharing and large-scale data sharing, but it also exists critical privacy issues due to untrusted sharing environments. An example is presented in \cite{229} wherein blockchain is incorporated with the D2D technology for large scale feature extraction applications on cloud. In healthcare, for example, image features extracted from health data collection contain important information of patients and thus need to be secured. Blockchain would ensure secure data storage by shifting the information to decentralized ledgers which are maintained by all participants. All stored data on blockchain is signed digitally and identified by hash values, which also solve privacy leaking issues from tampering or forging.  

Recently, blockchain is also considered and investigated in mobile edge computing (MEC)-empowered healthcare applications. The authors in \cite{181} consider an edge blockchain for telemedicine applications, with the main objective of providing secure transmission and computation of health data. The MEC-based cellular health network \cite{230} contains a base station and a set of mobile users. Here, mobile users can access the Internet via the cellular network, and they share the computation resources of a MEC server linked with a base station in a small cell. Blockchain provides a consensus protocol to verify the patient priority which is defined as the level of wireless resources that a user needs for their computation. As a result, the optimal resource allocation can be achieved to ensure the quality of data transmission of the whole network, and user information is secured due to storing on blockchain ledgers. Another blockchain approach in edge-based mass screening applications for disease detections is presented in \cite{231}. Due to a massive amount of captured multimedia IoT test data, an offline storage solution is considered and integrated with blockchain, which keeps cryptographic hashes of health data. This approach allows patients to take control of their information when performing clinical tests, visiting doctors or moving to other hospitals thanks to the transparency and availability of the blockchain protocol.

\begin{figure}
	\centering
	\includegraphics[height=6.63cm,width=8cm]{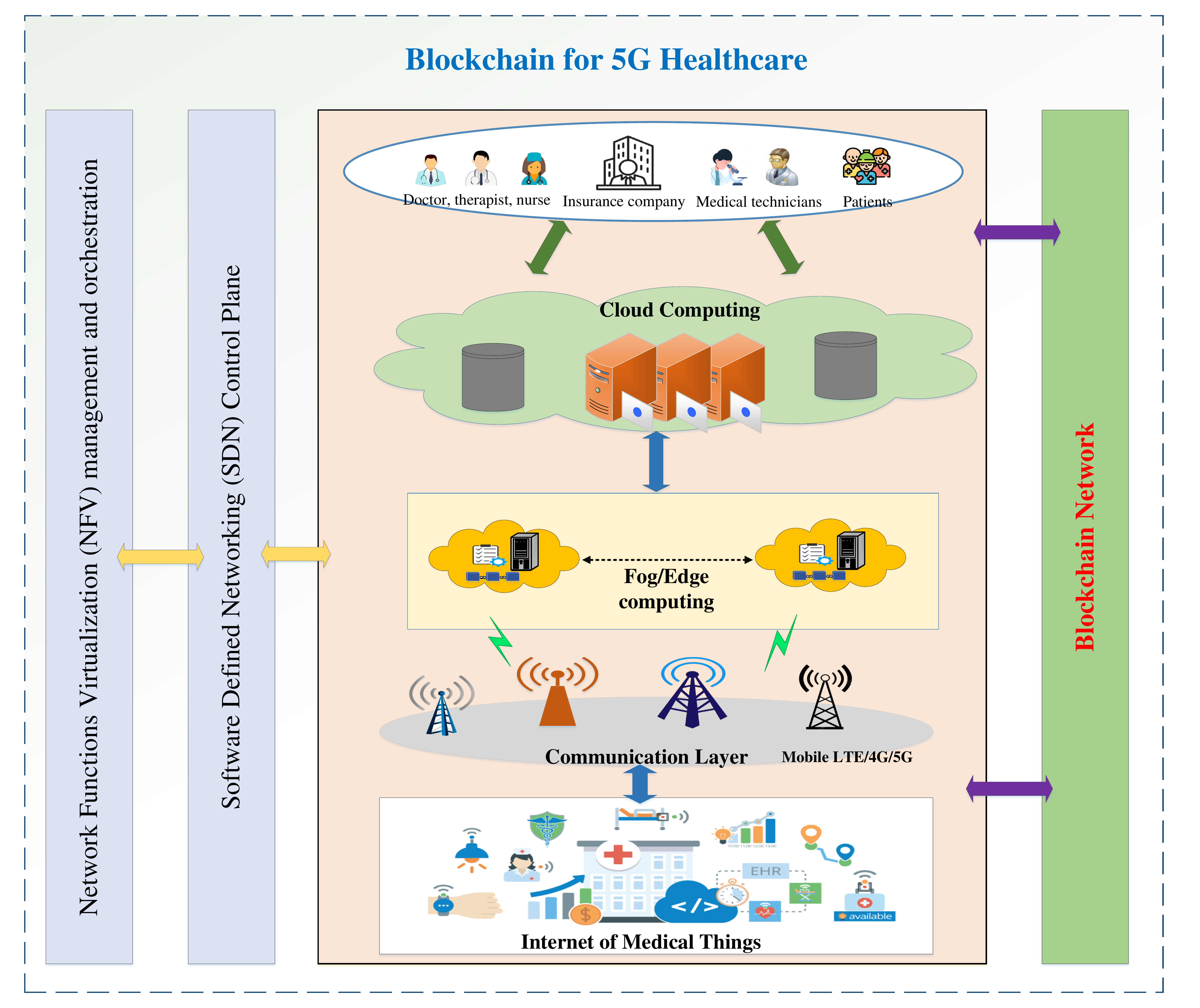}
	\caption{Blockchain for 5G healthcare \cite{227}. }
	\label{fig11}
	\vspace{-0.15in}
\end{figure}

Meanwhile, cloud computing, a key enabling technology of 5G networks, has also provided many notable solutions for healthcare services \cite{224}. Many research works have dedicated to use blockchain for cloud-based healthcare networks, such as \cite{232}. In this work, blockchain has proven its efficiency in improving the security of electronic health records (EHRs) sharing in cloud-assisted healthcare. The cloud computing is employed to store EHR ciphertext while the consortium blockchain keeps records of keyword ciphertext for data searching and sharing. In addition, to achieve secure data exchange between IoT health devices and cloud servers, a blockchain-enabled communication protocol is described in \cite{233}. All sensitive patient information and medical test results can be stored and managed by blockchain where a consensus mechanism is necessary for user verification when a medical test is performed. 

Very recently, we have also investigated and designed a blockchain architecture for cloud-based health management systems \cite{234}, \cite{235}. A mobile cloud blockchain platform is proposed to implement dynamic EHRs sharing among healthcare providers and patients. Blockchain is integrated with cloud computing to manage user transactions for data access enabled by smart contracts. In particular, a decentralized storage IPFS run by blockchain is combined with cloud computing to make data sharing more efficient in terms of low latency, easy data management and improved data privacy, compared to centralized cloud architectures. IoT users (i.e. doctors or patients) can perform data sharing transactions via their mobile devices such as smartphones, which offers flexible data sharing services with high security.

\subsection{	Smart city}
The evolution of 5G technologies has enabled enormous business opportunities and digital transformation initiatives for new smart city models, proving a wide range of services for city citizens \cite{236}. Smart cities involve a variety of components, including ubiquitous IoT devices, heterogeneous networks, largescale data storage, and powerful processing centres such as cloud computing for service provisions. Despite the potential vision of smart cities, how to provide smart city services with high efficiency and security remains unsolved. In this scenario, blockchain can be a promising candidate to solve critical security issues and empower smart city services \cite{237}, \cite{238}. To simplify the management of smart city services on a large scale, a city can be divided into small blocks called smart blocks. Each smart block consists of a variety of IoT devices, such as sensors, cameras, etc. of a certain area under the control of a block admin. A private blockchain using a ledger database is important to securely store all information generated from IoT devices during data exchange, data offloading and computation services. 

Another research in \cite{86} analyses a sustainable IoT architecture empowered by blockchain for a secure sharing economy services in mega smart cities. The proposed system employs cognitive fog nodes at the edge to gather and process offloaded multimedia payload and transactions from a mobile edge node and IoT devices. To extract significant information from the outsourced data, machine learning is used during the data analytic, and such results are then put in blockchain for secure sharing and storage. Furthermore, to solve data security issues in IoT for smart cities, blockchain is considered in \cite{239} to secure communication between the smart city and home devices and sensors. IoT data can be executed and computed at the edge layer for latency reduction, while user access information is recorded by blockchain, which works as a universal ledger. The key benefits of the proposed scheme include system transparency as well as the permissionless property which allows adding any new IoT devices without involving any authorities. 

In 5G smart cities, a prohibitively large amount of surveillance data will be generated continuously from ubiquitous video sensors. It is very challenging to immediately identify the objects of interest or detect malicious actions from thousands of video frames on the large scale. In such a context, building a distributed edge computing networks is highly efficient to achieve scalable data computation \cite{240}, \cite{241}. From the security perspective, blockchain would be a natural choice to establish decentralized security solutions by interconnecting edge nodes, IoT devices and city users, where data sharing, computation and business transactions can be performed on the blockchain ledger platform. It is also demonstrated that the use of distributed blockchain provides more benefits than the centralized architectures with a central cloud server in terms of lower latency, energy consumption, better service delivery, faster user response with security and privacy guarantees \cite{242}. 

Currently, most Mobility-as-a-Service (MaaS) which monitors the connections between transportation providers and passengers in smart cities is controlled by a central MaaS manager, which potentially introduces privacy leakage and system disruptions if this entity is attacked. By integrating with the blockchain, the MaaS model can be operated in a much more secure and decentralized manner \cite{243}. In this work, blockchain can help improve trust and transparency for all stakeholders and eliminate the need of centralized entity to make commercial agreements on MaaS. The mobility services, such as ticket purchase or payments for using transports, can be programmed by smart contracts, which enable automatic and reliable service trading and payment. 

Cloud computing is also a promising technology which can be incorporated to support strong computation and storage capabilities for smart city data, i.e big data from ubiquitous IoT devices. A cloud-smart city architecture is introduced in \cite{244}, wherein big data processing can be performed by cloud servers, while data auditing can be achieved by using the blockchain without third party auditors (TPAs). The proposed scheme focuses on building an optimized blockchain instantiation called data auditing blockchain (DAB) that collects auditing proofs and employs a consensus algorithm using a Practical Byzantine Fault Tolerance (PBFT) protocol. The simulation results reveal the potential of the blockchain adoption for big data in smart city with lower communication costs and better security. Furthermore, blockchain can enable interconnection cloud service providers to achieve a larger scale computation service \cite{245}. Any cloud server can be regarded as a blockchain node and cloud computing events are recorded on the ledgers, which effectively improves the system robustness and avoids the risks of single points of failures once the cloud server is compromised or attacked. 

\subsection{	Smart transportation}
With the rapid development of modern 5G communication and computation technologies, recent years have witnessed a tremendous growth in intelligent transportation systems (ITS), which create significant impacts on various aspects of our lives with smarter transport facilities and vehicles as well as better transport services \cite{246}, \cite{247}. Smart transportation is regarded as a key IoT application which refers to the integrated architectures of communication technologies and vehicular services in transportation systems. One critical issue in smart transportation is security risks resulted by dynamic vehicle-to-vehicle (V2V) communications in untrusted vehicular environments and reliance on centralized network authorities. Blockchain shed lights on several inherent features to implement distributed data storage, peer-to-peer communication, and transparently anonymous user systems, which envisions to build secure, decentralized ITS systems to facilitate customer transportations \cite{243}. One of the most significant services to realize intelligent transportation is data transmission among vehicles. How to provide efficient data exchange services in terms of low latency and increased network throughput while still ensure high degrees of security is a critical challenge. Blockchain would enhance QoS of the current ITS system by offering a decentralized management platform, wherein all vehicles and road side units (RSU) can perform data transmission and sharing on a peer-to-peer model to reduce end-to-end delay without using a vehicular authority \cite{248}. 

In order to adapt the large volumes of electric vehicle (EV) charging/discharging demand during transportation, the blockchain concept is introduced in \cite{249} that enables peer-to-peer transaction and decentralized storage to record all transaction data of EVs. In fact, EVs can be considered as a mobile power backup device to support the smart grid for load flattening, peak shaving and frequency regulation. This new energy trading paradigm is known as vehicle-to-grid (V2G), which is essential to build a safer and more sustainable energy platform for both EVs and the main power grid. Consumer power loads from smart city are all connected to the public blockchain power exchanging platform, where the electricity supply and user demand information are transmitted, encrypted and recorded in the blockchain platform. In such a context, the EV can publish and transmit the charging or discharging orders (for buying and selling) to the power blockchain platform which executes the EV request, performs energy trading and payment, and saves the transaction to the distributed ledger, which is also visible to every vehicle in the vehicular network. 

In the line of discussion, the authors in \cite{250} also analyse a V2G energy trading model with a combination of blockchain and edge computing. EVs can buy energy from local energy aggregators (LEAGs) via trading. The vehicular communication is secured by a consortium blockchain, in which all the transactions are created, propagated, and verified by authorized LEAGs. To further reduce latency and processing posing on burden blockchain, edge computing servers are employed to undertake block creation and mining. LEAGs can buy computation services from edge computing providers to finalize this process, and store mined blocks to the nearby edge nodes. The blockchain technology envisions a trustless network to eliminate the operation cost of the intermediary participation, which will realize a quicker, safer and cheaper way in ITS systems.  Moreover, authentication for vehicle access is of paramount importance for vehicular networks. In this regard, smart contract would be a notable approach which can authenticate and verify vehicular transactions by triggering the programmed logic functions \cite{251}. This enables direct authentication for registered vehicles without revealing device privacy and effectively prevents potential attacks from malicious vehicles.

Recently, blockchain has been incorporated with SDN to build secured and controlled vehicular ad hoc networks (VANETs) \cite{101}. With the increasing scale of the current VANETs, traditional VANET frameworks with centralized SDN control mechanisms obviously cannot match the diversification of VANET traffic requirements. Distributed SDN control can be an efficient solution to localize decision making to an individual controller, which thus minimizes the control plane response time to data plane requests. To achieve secure communications between SDN controllers as well as between SDN controllers and EVs, blockchain is leveraged to achieve agreement among different nodes in terms of traffic information and energy demands without using centralized trust management. 

Another aspect in VANETs is the security of power trading between EVs and V2G networks. In fact, it is very important to design a safe, efficient, transparent, information symmetrical trading model for VANETs to provide ubiquitous vehicular services (i.e. traffic transmission, vehicle cooperation, energy payment). Blockchain is introduced in \cite{252} for a reliable decentralized power trading platform where a V2G EV trading smart contract is integrated for trading authentication and a decentralized energy ledger is for data storage and sharing without relying on a trusted third party, eliminating the need for trusted third parties to address the high cost, inefficiency, and insecure data storage of traditional centralized organizations.

\subsection{	Smart grid}

The continuously growing power demand in modern society has been a critical challenge that needs significant attention in the present day of the smart grid era. The energy industry has witnessed a paradigm shift in power delivery from a centralized production and distribution energy system into a dynamic mode of decentralized operation thanks to the support of ubiquitous 5G technologies such as IoT, edge/cloud computing, SDN, network slice and D2D communication \cite{253}, \cite{254}. In this regard, blockchain, a decentralized database platform, enables completely new technological systems and business models for energy management with added features such as decentralization, security, privacy and transparency \cite{129}. In the 5G energy network slice, the electricity can be allocated to each power user in the housing society through a distributed blockchain platform where all users are interlinked with energy providers on secured and distributed ledgers. 

In smart grid, in order to monitor the electricity distribution and power usage of customers, a smart meter can be installed at each home to collect the real-time electricity consumption data for better smart home services. However, a critical drawback is that private user information such as home address, personal information may be disposed and adversaries can track users to obtain electricity consumption profile. To overcome this challenge, blockchain has been introduced in \cite{255} for a privacy-preserving and efficient data aggregation network. 
The power network has been divided into small groups, each group is controlled by a private blockchain. Instead of relying on a third party for data aggregation, a certain user is chosen to aggregate all user data within his network and record them to the blockchain for storage and monitoring. Such an aggregator only collects data and all other users share the equal right to verify and validate transactions to achieve consensus, which eliminates the risks of single points of failure and improves system trust accordingly. 

In order to achieve traceability of power delivery in smart grid, blockchain can be applied to provide transparency and provenance services \cite{256}. The customer can register their information on blockchain and perform energy trading and payment by uploading a transaction to blockchain. By creating an immutable data structure, data recorded and transferred onto the system cannot be altered. Smart contracts are also very useful to provide a transparent and fair energy trading between consumers and utility companies through an energy policy which defines all trading rules. Once the energy billing payment is completed, for example, both the user and the service provider receive a copy of the transaction, which allows users to keep track of their energy usage. 

At present, the sophistication of cyberattacks has posed a challenge to the current smart power systems. In recent years, cyber-attacks have caused power systems blackout due to data vulnerability, malicious events or market data manipulation \cite{257}. Therefore, the introduction of blockchain, a strong security mechanism, can help overcome such challenges. 
The interactions between the electricity market agent and the customer are reflected via transactions which contain electricity demands, electricity price, user information. All such transactions are signed by the private key of the sender (i.e. energy user) to perform energy trading with the agent. In such a context, an attacker can threaten the communication link between users and the agent, but it may be impossible to break the transaction due to the lack of user private key and such malicious access is detected and discarded by consensus mining. Additionally, the authors in \cite{258} also present a research effort in using blockchain to mitigate cyber-attacks on a smart grid. Every prosumer, consumer and substation are connected through a block chain based application under the control of a smart contract, which perform transaction verification when energy transmission occurs. The consensus is maintained by the computing power of all distributed energy servers and users, which also make the energy system well resistant to cyber-attacks \cite{259}. 

In a similar direction, the work in \cite{260} proposes a smart and scalable ledger framework for secure peer to peer energy trading in smart grid ecosystems. The energy network considered consists of a set of EVs which can participate in three operations, namely charging, discharging and staying idle, EV aggregator which works as an energy broker and provides access points to EVs for both charging and discharging operation, and energy cash as the currency for energy payment. To avoid the issue of spanning and Sybil attacks, instead of using PoW which remains high block generation latency, the authors suggest a proof of time concept. A client must collect a random token, i.e., random messages from neighbours, which makes the process costly for an attacker to achieve the throughput of honest transactions as each transaction contains associated timestamp with it. For security of energy transactions, another work in \cite{261} also builds a fully decentralised blockchain-based peer-to-peer trading scheme. The main goal is to present a pay-to-public-key-hash implementation with multiple signatures as a transaction standard to realise a more secure transaction and reduced storage burden of distributed prosumers. 

Recently, mobile edge computing (MEC), a significant 5G enabling technology, is also cooperated with smart grid. Although MEC can offer promising benefits such as low-latency computation, reduced network congestion for better energy delivery, the characteristics inherent of the MEC architecture such as heterogeneity, mobility, geo-distribution and location-awareness, can be exploited by attackers to perform nefarious activities. Thus, designing practical security solutions for MEC-based smart grid system is critical. In the work \cite{262}, a permissioned blockchain edge model is introduced with the main objectives of privacy protections and energy security. At the layer of distributed edge devices and power supply, smart devices and power supply facilities compose smart grid generating electricity trading transactions. Meanwhile, the smart contract running on blockchain assigns tasks to edge devices and records transaction on blockchain, which enables a secure and trustworthy trading environment. By integrating with distributed edge computing, blockchain can offer a larger number of services, such as device configuration and governance, sensor data storage and management, and trading payments.

Blockchain for edge-empowered smart grid has been considered in \cite{263}, in which a blockchain based mutual authentication and key agreement protocol is proposed without the need for other complex cryptographic primitives. The smart grid network model used consists of registration authority (RA), end users (EUs), edge servers (ESs) and blockchain. ESs are responsible to supply timely data analysis and service delivery, and each ES is linked with blockchain to prevent web spoofing attacks and guarantee smooth energy trading and user interactions. The authors in \cite{264} also present a blockchain implementation for smart grid to guarantee information privacy of energy user and energy trading. MEC servers act as active blockchain nodes with strong computation capabilities to enable fast data analytic services, i.e. processing large transaction graphs of energy trading, within the energy trading system among EVs. 

\subsection{	Unmanned Aerial Vehicles (UAVs)}
The rapid growth of drones or Unmanned Aerial Vehicles (UAVs) \cite{265} is creating numerous new business opportunities for service providers. UAVs can be regarded as flying IoT devices and have been employed widely in various areas, ranging from military, security, healthcare, and surveillance to vehicle monitoring applications \cite{266}. In the era of modern 5G communication networks, due to the rapidly growing IoT traffic, it is very challenging for static base stations (i.e. access point, router) to support data demands of billions of IoT devices in large scale IoT scenarios. Therefore, the adoption of UAV in IoT networks can be a future direction. Indeed, UAV can act as a flying base station to support unprecedented IoT services, i.e. dynamic data offloading, data sharing or service collaboration, due to its mobility and flexibility \cite{267}. However, the operation of UAVs in the sky is highly vulnerable to several privacy and security risks that target data accountability, data integrity, data authorization, and reliability \cite{268}. 

Recent years have also witnessed a new research trend on the combination of blockchain and UAVs for solving critical challenges in UAV networks and empowering new 5G IoT applications. For instance, the work in \cite{269} takes advantage of consortium blockchain for a spectrum sharing platform between the aerial and terrestrial communication systems for UAV-based cellular networks. The key idea is to establish the distributed shared database to perform secure spectrum trading and sharing between the mobile network operators (MNOs) and the UAV operators. The proposed model possibly addresses two key issues: security risks of UAV-based spectrum trading due to the unauthorized spectrum exploitations of malicious UAVs, and privacy leakages caused by the centralized sharing architecture with third parties. 

To support the security of UAV communication in ad hoc networks (UAANETs), permissioned blockchain has been adopted in \cite{270} to provide decentralized content storage services and detect internal attackers during efficient content dissemination. The key reason behind the blockchain adoption for UAANETs is the ability of blockchain to securely maintain a consistent and tamper-resistant ledger to record all the transactions of content sharing and storage in a decentralized environment without the need for any central authority, which is applicable to the complex and vulnerable network. Besides, to overcome the limitations of traditional blockchain models with low throughput and high resource consumption, an efficient and scalable Adaptive Delegate Consensus Algorithm (ADCA) is integrated to perform consensus without the mining procedures. Similarly, the work \cite{271} also proposes to use blockchain for secure data dissemination in UAV networks. Data collected from UAVs can be recorded and stored in decentralized database ledgers to mitigate the storage burden on UAVs. The use of blockchain allows any of the users in the UAVs network to participate in consensus processes and implement verification without any external authorities, such as cloud servers. The proposed model has the potential to solve various security issues, including spoofing, Denial-of-service (DoS), eavesdropping and data tampering. 

The authors in \cite{272} consider an autonomous economic system with UAVs where blockchain acts as a protocol of autonomous business activities in modern industrial and business processes. IoT devices, robots, UAVs in the multi-agent systems can exchange data each other to perform automatic collaborative works (i.e. in smart factory) and share collected data to users via a peer-to-peer ledger. Blockchain link all agents together to create a distributed network where any agent can join and perform block verification to maintain the correct operation and security of the system. To avoid the issues of data leakage or data loss during the transmission among UAVs, blockchain is also considered in \cite{273}. The data transfer process occurs within the blockchain which allows storing all user information and exchange records for security management.

More interesting, blockchain has been considered and incorporated with cloud/edge computing for enabling emerging UAV-based applications. The authors in \cite{274}, \cite{275} analyse a blockchain-enabled secure data acquisition scheme for UAV swarm networks in which data are collected from IoT devices employing UAV swarms. Each of the UAVs maintains its own shared key in order to expedite communication with IoT devices when performing the security mechanism (i.e., sign, verify, encrypt, and decrypt). A smart contract is also employed in order to handle the IoT devices and missions in data acquisition. The study in \cite{276} also explores a Hyperledger Fabric blockchain design for UAV swarm networks. Each communication request among UAVs is recorded as a transaction which is validated and verified by the mining process enabled by the computing power of all entities in the UAV network for maintaining the blockchain.

In an effort to enhance the security of edge-based UAV networks, the work in \cite{277} proposes a neural blockchain-based transport model as Fig. 13 to ensure ultra-reliability for UAV communication and enable intelligent transport during UAV caching through user equipment (UE) via MEC. The blockchain acts as a distributed database ledger which is shared among all the involved entities (UAVs, MEC servers, and users) identified by their public keys (IDs). The smart contract is responsible to monitor user access and perform verification, while blockchain provides a secure data sharing environment to facilitate content sharing and data delivery between the UEs and the caching servers.

\begin{figure}
	\centering
	\includegraphics[height=8.35cm,width=7.5cm]{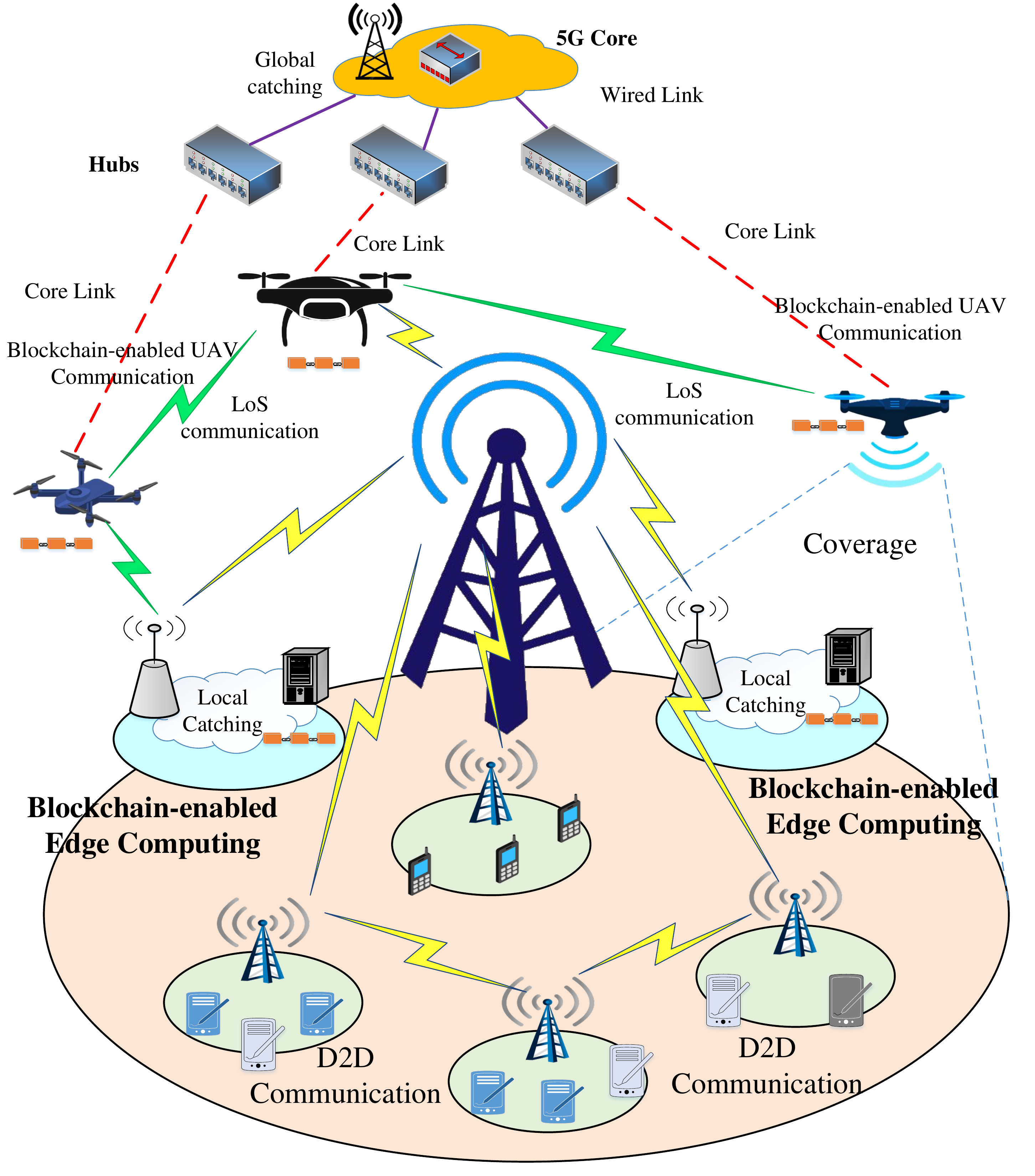}
	\caption{ Blockchain for secure 5G UAV networks \cite{277}.  }
	\label{fig11}
	\vspace{-0.15in}
\end{figure}

In addition, the authors in \cite{278} integrate blockchain in a cloud-assisted UAV network for surveillance services to investigate the safety condition of the dam infrastructure in real-time. Two blockchains are designed, a public bitcoin blockchain for payment trading, and a private blockchain for data storage on the network of UAV providers, users, and cloud providers. To join the blockchain, each entity, i.e. IoT sensor users should have certificates obtained from a certificate authority. Data gathered from cloud providers is considered as an object which is then hashed and anchored by the UAV provider into the blockchain network. The solution using blockchain bring various benefits, including reduced latency due to direct communication without passing a third party, and high data integrity and tampering resistance thanks to the hash function and consensus process.

\section{Main findings, Challenges and Future research directions }

Integrating blockchain in the 5G mobile networks is a hot research topic now. Many research efforts have been devoted to the development of blockchain technology for 5G mobile networks. In the previous sections, we have presented a state of the art review on the current achievements in the blockchain adoption in 5G networks. Specially, we have provided an extensive discussion on the convergence of blockchain into key 5G enabling technologies, namely cloud computing, edge computing, Software Defined Networks, Network Function Virtualization, Network Slicing, and D2D communication. The survey has also covered and highlighted the benefits of blockchain to empower fundamental 5G services such as spectrum management, data sharing, network virtualization, resource management, interference management, privacy and security services. We also analyse the integration of blockchain in a wide range of 5G IoT applications, ranging from smart healthcare, smart city, smart transportation to smart grid and UAVs. Based on the current great research efforts in the literature, in this section, we will summarize the key findings inherited from the integration of blockchain in 5G networks and services. We also identify possible research challenges and open issues in the field along with the future research directions that should be considered and investigated to encourage more innovative solutions and studies in this promising area. 

\subsection{	Main findings}

The comprehensive literature review on the integration of blockchain in 5G technologies, 5G services and IoT applications reveals many important findings, which would enable to open up numerous opportunities for the newly emerging 5G scenarios. This sub-section will highlight the key findings inherited from the convergence of these promising technologies. 

\subsubsection{Blockchain for 5G technologies}
Blockchain can offer many promising technical properties such as decentralization, privacy, immutability, traceability, and transparency to empower 5G technologies. Reviewing the literature works, we find that blockchain can support well 5G technologies mainly from three key aspects, including security, system performance, and resource management. The current 5G technology infrastructure is mainly enabled by the centralized network settings, such as edge/cloud computing, and SDN which obviously show security vulnerabilities due to the reliance of third parties. Blockchain can arrive to build decentralized network architectures for 5G technology platforms. For example, the concept of blockchain-based cloud computing enables decentralization of cloud/edge 5G networks \cite{64}, \cite{78} which gets rid of centralized control at the core network and offers a decentralized fair agreement with blockchain consensus platform. Even when an entity is compromised by malicious attacks or threats, the overall operation of the involved network is still maintained via consensus on distributed ledgers. More interesting, blockchain can help establish secure peer-to-peer communication among users (i.e. in D2D communication) using the computing power of all participants to operate the network instead of passing a third-party intermediary. This would potentially reduce communication latency, transaction costs, and provide the global accessibility for all users, all of which will enhance the overall system performance. 

Furthermore, blockchain is expected to improve the resource management for network function virtualization and network slicing. On the one hand, blockchain can boost the trust and transparency among participants and stakeholders and enable more seamless and dynamic exchange of computing resources in the cooperative. The secure spectrum resource provision can be achieved via blockchain which provides a decentralized sharing platform of the network of network servers, service providers and customers. Moreover, the network function resource can be shared at a faster speed, compared to conventional centralized schemes, which thus facilitates service delivery. Currently, the design of network slice instances is based on the open cloud-based architectures, and attackers may abuse the capacity elasticity of one slice to consume the resources of another target slice, which makes the target slice out of service. Blockchain can be exploited to build reliable end-to-end network slices and allow network slide providers to manage their resources, providing the dynamic control of resource reliability. 
\subsubsection{Blockchain for 5G services}
Blockchain is expected to facilitate the 5G services by adding security properties and simplification of service management. Blockchain is particularly useful to create secure sharing environments for spectrum or data exchange in the 5G mobile networks.  Blockchain is regarded as a middle layer to perform spectrum trading, verify sharing transactions and lease securely the spectrum provided by spectrum resource providers, i.e. license holders. Different from the conventional database management systems which often use a centralized server to perform access authentication, blockchain with smart contracts can implement decentralized user access validation by using the computing power of all legitimate network participants. This makes the sharing system strongly resistant to data modifications. Many research studies on blockchain \cite{150}, \cite{151}, \cite{152}, \cite{153} demonstrate that the blockchain adoption is beneficial to spectrum management in terms of better scalability, power efficiency in spectrum usage, improved accessibility with high degree of security and better system protection capability against DoS attacks and threats. 

Besides, blockchain can simplify the network virtualization in 5G networks with high degrees of security \cite{118}, \cite{119}. The blockchain technology can provide the required characteristics of nonrepudiation and immutability to overcome the shortcomings of the previous centralized configuration settings in virtual networks. More precisely, blockchain is capable of creating secure virtual wireless networks (VWNs) so that wireless resource-owners sublease their wireless resources (e.g., slice of RF spectrum, infrastructure) to mobile virtual network operators (MVNOs). In such a decentralized virtual network, smart contracts can be very useful to provide automation and transparency in a distributed way instead of trusting a particular node or an authority process transactions, which also enhances the trustworthiness of the resource management services. The building of a fair and trusted economic scheme empowered by blockchain can be a notable solution for network interference control, especially in small cell deployments \cite{184}. 

In addition to the above 5G services, blockchain also provides privacy and security benefits to 5G networks. By publishing user data to ledger where data is signed by hash functions and appended immutably to blocks, blockchain platforms ensure strong data protection. Blockchain is capable of providing full control of personal data when sharing on the network, which is unique from all traditional approaches which hinder users from tracking their data \cite{12}. Besides, blockchain is expected to offer a wide range of security merits such as access control enabled by smart contracts, data integrity thanks to the decentralized ledger and authentication from consensus process and smart contracts. 

\subsubsection{	Blockchain for 5G IoT applications}

Blockchain has been investigated and integrated into a number of key 5G IoT applications, such as smart healthcare, smart city, smart transportation, smart grid and UAVs. The integration of blockchain with 5G technologies can advance current IoT systems and provide more performance benefits in terms of better decentralization, security, privacy, service efficiency and system simplification for lower operational costs \cite{168}. For example, blockchain has been demonstrated its high efficiency in healthcare and smart city scenarios. By implementing a direct and secure interconnection in a network of users, service providers (i.e. hospital in healthcare or traffic control units in smart transportation) and network operators, the data sharing, resource sharing and cooperative communication can be achieved in a secure and low-latency manner. Importantly, the sharing of data over the untrusted environments is highly vulnerable to cyber-attacks, which can monitor and obtain the user information profile (patient information in healthcare of customer data in smart grid). Blockchain comes as a notable solution to address such challenges by securing the transaction and verifying the user access.

Recent years have also witnessed a new research trend on the combination of blockchain and UAVs for solving critical challenges in UAV networks and empowering new 5G IoT applications. UAV with its high mobility and flexibility can be a promising transmission solution for aerial and terrestrial communication systems, but it also remains critical challenges in terms of security due to adversaries and short battery life. Blockchain would be notable to solve such challenges. Recent studies show the feasibility of blockchain in UAV networks \cite{269}, \cite{270}, \cite{271}. UAV can collect data from the IoT devices and offload data to the blockchain, where data is hashed and recorded securely on the ledger. This would not only preserve IoT data against threats but also reduce the data storage burden on UAV, which is promising to prolong the duration of UAV operations for better service delivery. 

\subsection{	Challenges and Open issues}

At present, the amalgamation of blockchain and 5G networks has been received widespread research interests from academics and industries. The blockchain technology is promising to revolutionize 5G networks and services by offering the newly emerging features such as decentralization, privacy, and security. The arrival of this emerging technology is potential to change the current shape of 5G infrastructure and transform industrial network architectures with advanced blockchain-5G paradigms. However, the throughout survey on the use of blockchain for 5G networks also reveals several critical research challenges and open issues that should be considered carefully during the system design.  We analyse them from three main aspects: blockchain scalability, blockchain security, and QoS limitations, which will be analysed in details as follows.

\subsubsection{	Blockchain performance and scalability}

Despite the benefits of blockchain, scalability and performance issues of are major challenges in the integrated blockchain-5G ecosystems. Here, we analyse the scalability issues of blockchain from the perspectives of throughput, storage and networking.
\begin{itemize}
	\item \textit{Throughput: }In fact, blockchain has much lower throughput in comparison to non-blockchain applications. For instance, Bitcoin and Ethereum process only a maximum of 4 and 20 transactions per second respectively, while Visa and PayPal process 1667 and 193 transactions per second \cite{279} respectively. Obviously, the current blockchain systems have serious scalability bottlenecks regarding the number of replicas in the network as well the performance concerns such as constrained throughput (number of transactions per second) and latency (required time for adding a block of transactions in the blockchain) \cite{280}.  Many blockchains have long waiting time for transactions to be appended into the chain because of block size limitations. Therefore, the block generation time increases rapidly, which limits the overall system throughput. Therefore, in order to sustain a huge volume of real world transactions for 5G applications, proper solutions should be considered carefully to improve the throughput. 
\item	\textit{Storage:} When using blockchain in 5G networks, a huge quantity of data generated by ubiquitous IoT devices is processed by the blockchain for 5G services such as data sharing, resource management and user transaction monitoring. In the conventional blockchain systems, each blockchain node must process and store a copy of the complete transaction data. This can pose a storage and computation burden on resource-constrained IoT devices to participate in the blockchain network. Moreover, if all transaction data are stored on chain, the blockchain capacity will become very large to maintain on the chain over time \cite{281}.
\item	\textit{Networking:} Blockchain networking is another issue that also affects the scalability of blockchain systems. Blockchain is computationally expensive and requires significant bandwidth resources to perform computational mining puzzle. However, in the 5G scenarios, such as ultra-dense networks where resource is very limited due to the demands from IoT devices and service operators, it may be impossible to meet resource requirement for blockchain to achieve large scale transaction processing. Further, stemming from the property of blockchain consensus mechanisms which require multiple transaction transmissions among nodes to validate a block, the blockchain operation needs to consume much network resources (i.e. bandwidth, mining power, and transmission power), which also results in high network latency \cite{282}. 
	
\end{itemize}
Considering complex 5G IoT scenarios, i.e. smart cities, the IoT workload and data are enormous and thus will result in the rapid growth in the IoT blockchain size, making it difficult to process high volumes of data. The end-to-end latency in 5G networks is expected to achieve less than 1 millisecond \cite{2} for payload and data transmissions. This vision requires careful considerations in designing blockchain platforms before integrating into 5G systems. Many research efforts have been dedicated to improving the performance and scalability in blockchain from different design perspectives such as mining hardware design \cite{283}, hybrid consensus protocols \cite{284}, on-chain and off-chain solutions \cite{285}, \cite{286}. Very recently, a solution using 5G network virtualization is also considered \cite{287} to solve scalability of blockchain by decoupling the blockchain management from the transaction processing to improve QoS of blockchain operations. The preliminary results are expected to shed light on the blockchain research for solving scalability issues and improving the system performance in integrated blockchain 5G networks. 

\subsubsection{	Blockchain security and privacy}

Blockchain is considered as secure database platform to ensure safety and privacy for involved 5G networks. However, recent studies have revealed inherent security weaknesses in blockchain operations which are mostly related to 5G systems \cite{288}. A serious security bottleneck is 51\% attack which means that a group of miners controls more than 50\% of the network’s mining hash rate, or computing power, which prevents new transactions from gaining confirmations and halts payments between service providers and IoT users. Seriously, adversaries can exploit this vulnerability to perform attacks, for example, they can modify the ordering of transactions, hamper normal mining operations or initiate double-spending attack, all of which can degrade the blockchain network \cite{288}. In addition, the security aspect of smart contract, which is regarded as core software on blockchain, is also very important since a small bug or attack can result in significant issues like privacy leakage or system logic modifications \cite{289}, \cite{290}. Some of the critical security vulnerabilities can include timestamp dependence, mishandled exceptions, reentrancy attacks on smart contracts in 5G applications.

In addition to that, in current 5G IoT systems, data can be stored off-chain in cloud computing to reduce the burden on blockchain. However, this storage architecture can arise new privacy concerns. Specifically, an autonomous entity can act as a network member to honestly perform the cloud data processing, but meanwhile obtains personal information without the consent of users, which leads to serious information leakage issues. External attacks can also gain malicious access to retrieve cloud data, or even alter and modify illegally outsourced IoT records on cloud. Besides, privacy leakage on blockchain transactions is another significant problem. Although blockchain uses encryption and digital signature to preserve transactions, recent measurement results \cite{291} show that a certain amount of transaction is leaked during blockchain operations and data protection of blockchain is not very robust in practice. Furthermore, criminals can leverage smart contracts for illegal purposes, facilitating the leakage of confidential information, theft of cryptographic keys. Importantly, privacy of IoT users cannot be ensured once they join the network. Indeed, by participating in the blockchain network, all information of users such as address of sender and receiver, amount values is publicly available on the network due to the transparency of blockchain. Consequently, curious users or attacks can analyse such information and keep track of activities of participants, which can lead to leakage of information secrets such as personal data. 

Security problems in blockchain in 5G networks can be solved by recent security improvements. For example, a mining pool system called SmartPool \cite{292} was proposed to improve transaction verification in blockchain mining to mitigate security bottlenecks, such as 51\% vulnerability, ensuring that the ledger cannot be hacked by increasingly sophisticated attackers. Particularly, recent works \cite{293}, \cite{294} introduced efficient security analysis tools to investigate and prevent threat potential in order to ensure trustful smart contract execution on blockchain. Such research efforts make contributions to addressing security issues in blockchain 5G environments and improving the overall performance of the system.

\subsubsection{QoS limitations}
With the advances of mobile 5G technologies, blockchain now can be implemented in mobile devices to provide more flexible blockchain-based solutions for 5G IoT applications. The foundation of the efficient and secure operation of blockchain is a computation process known as mining. In order to append a new transaction to the blockchain, a blockchain user, or a miner, needs to run a mining puzzle, i.e. Proof of Work (PoW) or Proof of Stake (PoS) which is generally complicated and requires vast computing and storage resources. Further, blockchain also requires network bandwidth resources to perform its consensus process. Without a careful design, the blockchain implementation to operate involved IoT applications may lead to Quality of Service (QoS) degradation with long latency, high energy consumption, high bandwidth demands, and high network congestion. Obviously, the integration of blockchain can introduce new QoS challenges that would negatively impact the overall performance of blockchain-5G networks. It is noting that one of the most important goals of future 5G is to provide user-centric values with high QoS to satisfy the growing demands of user traffic and emerging services \cite{2}. Therefore, it is vitally important to develop efficient solutions that can enhance service qualities of blockchain ecosystems to empower the future blockchain-5G networks. 

Recently, some strategies have been proposed to solve the above issues from different perspectives. On the one hand, the design of lightweight blockchain platforms can be a notable solution to enhance the QoS, by eliminating computation consensus mechanisms of blockchain \cite{295}, compressing consensus storage \cite{296}, or designing lightweight block validation techniques \cite{297}, \cite{298}, \cite{299}. These solutions potentially simplify the blockchain mining process for lower energy consumption and better latency efficiency, which make greats contributions to the QoS improvements in blockchain-5G applications. On the other hand, computation offloading is also another feasible approach to solve the low QoS issues of blockchain \cite{217}. With the development of 5G technologies such as edge/cloud computing, SDN, D2D communication, blockchain computation tasks (i.e. consensus puzzle) can be offloaded to resourceful servers such as edge/cloud servers \cite{300}, \cite{301}  by combining SDN \cite{302} and D2D communication \cite{138} to bridge the gap between constrained resources of local mobile devices and growing demands of executing the computation tasks. By using offloading solutions, the performance of blockchain-5G systems would be improved significantly, such as saving system energy, reducing computation latency and improving the quality of computation experience for mobile devices. As a result, the system QoS will be enhanced while blockchain features are ensured for high level network security. The offloading optimization solutions should be explored further to balance both blockchain and the core 5G networks for future mobile blockchain-5G applications. 

\subsection{	Future research directions}
Motivated by our detailed survey on research studies on the convergence of blockchain and 5G networks, we point out possible research directions which should be considered in the future works.
\subsubsection{	Integrating machine learning with blockchain for 5G}
The rapid developments in blockchain technology are creating new opportunities for artificial intelligence applications. The revolution of machine learning (ML) technology transforms current 5G services by enabling its ability to learn from data and provide data-driven insights, decision support, and predictions. These advantages of machine learning would transform the way data analytics are performed to assist intelligent services in the age of 5G. For example, ML has the ability to interact with the wireless environment to facilitate resource management and user communication \cite{217}. ML also exhibits great potential on data feature discovery to predict data usage behaviour for developing control algorithms, such as data traffic estimation for network congestion avoidance or user access tracking for privacy preservation \cite{218}. Recent years, there is a growing trend of integrating machine learning with blockchain for 5G use case domains. For example, deep reinforcement learning (DRL) \cite{23} has been investigated and combined with blockchain to enable secure and intelligent resource management and orchestration in 5G networks. An advanced DRL algorithm is proposed to accurately analyze the topology, channel assignment, and interference of the current wireless network, and then select the most appropriate wireless access mode (i.e., cellular network, V2V, or D2D) to improve communication rate, reduce energy consumption, or enhance user experience. Meanwhile, blockchain provides a secure decentralized environment where operating reports and network configurations can be replicated and synchronized among edge servers, which can facilitate network diagnosis and enable reliable orchestration. Other significant works also propose the integrated blockchain-DRL architectures for flexible and secure computation offloading \cite{303}, reliable network channel selection \cite{304}, and networking optimization \cite{305}. 

\subsubsection{Blockchain for big data in 5G}

In the age of data explosion, big data becomes a hot research topic in 5G \cite{314}. A large amount of multimedia data generated from ubiquitous 5G IoT devices can be exploited to enable data-related applications, for example, data analytics, data extraction empowered by artificial intelligence solutions \cite{315}. Cloud computing services can offer high storage capabilities to cope with the expansion of quantity and diversity of digital IoT data. However, big data technologies can face various challenges, ranging from data privacy leakage, access control to security vulnerabilities due to highly sophisticated data thefts \cite{316}. Further, big data analytics on cloud/edge computing are also highly vulnerable to cyberattacks in the complex operational and business environments. 

In such contexts, blockchain appears as the ideal candidate to solve big data-related issues \cite{317}. Indeed, the decentralized management associated with authentication and reliability of blockchain can provide high-security guarantees to big data resources. Specifically, blockchain can offer transparency and trustworthiness for the sharing of big data among service providers and data owners. By eliminating the fear of security bottlenecks, blockchain can enable universal data exchange which empowers large-scale 5G big data deployments. Recently, some big data models enabled by blockchain are proposed, such as data sharing with smart contracts \cite{318}, access control for big data security \cite{319}, or privacy preservation for big data analytics \cite{320}. Such preliminary results show that blockchain can bring various advantages in terms of security and performance enhancement to big data applications in the age of 5G. 

\subsubsection{Blockchain for 6G}
Beyond the fifth-generation (B5G) networks, or so-called 6G, will emerge to provide superior performance to 5G and meet the increasingly high requirements of future mobile services and applications in the 2030s. The key drivers of 6G will be the convergence of all the past features, such as network densification, high throughput, high reliability, low energy consumption, and massive connectivity \cite{321}. According to \cite{322}, 6G wireless networks are expected to support massive user connectivity and multi-gigabits data transmissions with super-high throughput, extremely low-latency communications (approximately 10 µs), and support underwater and space communications. The 6G networks are also envisioned to create new human-centric values \cite{323} enabled by numerous innovative services with the addition of new technologies. The new services may include smart wearables, implants, fully autonomous vehicles, computing reality devices, 3D mapping, smart living, space travel, Internet of Nano-Things, deep-sea sightseeing and space travel \cite{324}. 

To satisfy such applications for the 2030 intelligent information society, 6G will have to meet a number of stringent technical requirements. Following this rationale, high security and privacy are the all-important features of 6G, which shall be paid special attention from the wireless research community \cite{325}. With the promising security capability, blockchain is expected to play a pivotal role in the successful development of the future 6G networks. Blockchain potentially provides a wide range of security services, from decentralization, privacy, transparency to privacy and traceability without needing any third parties, which will not only enhance the security of 6G networks but also promise to promote the transformation of future mobile services \cite{326}. The Federal Communications Commission (FCC) also suggests that blockchain will be a key technology for 6G services. For example, it is believed that blockchain-based spectrum sharing \cite{327} is a promising technology for 6G to provide secure, smarter, low-cost, and highly efficient decentralized spectrum sharing. Blockchain can also enable security and privacy of quantum communications and computing, molecular communications, and the Internet of Nano-Things via secure decentralized ledgers. 

In summary, blockchain has provided enormous opportunities to 5G mobile networks thanks to its exceptional security properties. The convergence of blockchain and 5G technologies has reshaped and transformed the current 5G service provision models with minimal management effort, high system performance with high degrees of security. This detailed survey is expected to pay a way for new innovative researches and solutions for empowering the future blockchain-5G networks.  

\section{Conclusions}

Blockchain is an emerging technology that has drawn significant attention recently and is recognized as one of the key enablers for 5G networks thanks to its unique role to security assurance and network performance improvements. In this paper, we have explored the opportunities brought by blockchain to empower the 5G systems and services through a state-of-art survey and extensive discussions based on the existing literature in the field. This work is motivated by the lack of a comprehensive review on the integration of blockchain and 5G networks. In this article, we have presented a comprehensive survey focusing on the current state-of-the-art achievements in the integration of blockchain into 5G wireless networks. Particularly, we have first provided a brief overview on the background knowledge of blockchain and 5G networks and highlighted the motivation of the integration. We have then explored and analysed in detail the potential of blockchain for enabling key 5G technologies, such as cloud computing, edge computing, Software Defined Networks, Network Function Virtualization, Network Slicing, and D2D communication. A comprehensive discussion on the use of blockchain in a wide range of popular 5G services has been provided, with a prime focus on spectrum management, data sharing, network virtualization, resource management, interference management, federated learning, privacy and security services. Our survey has also covered a holistic investigation on the applications of blockchain in 5G IoT networks and reviews the latest developments of the cooperated blockchain-5G IoT services in various significant use-case domains, ranging from smart healthcare, smart city, smart transportation to smart grid and UAVs. Through the comprehensive survey on the related articles, we have summarized the main findings derived from the integrations of blockchain in 5G networks and services. Finally, we have pointed out several research challenges and outlined potential research directions toward 6G networks.

Research on blockchain for 5G wireless networks is still in its infancy. But it is obvious that blockchain will significantly uplift the shape and experience of future mobile services and applications. We believe our timely study will shed valuable light on the research problems associated with the blockchain-5G integration as well as motivate the interested researchers and practitioners to put more research efforts into this promising area. 
\bibliography{Ref}
\bibliographystyle{IEEEtran}
\end{document}